\documentclass{aa}
\nolinenumbers

\usepackage{graphicx}
\usepackage{txfonts}
\usepackage[colorlinks=true, citecolor=blue, linkcolor=blue, urlcolor=blue]{hyperref}
\usepackage{natbib,twoopt}
\bibpunct{(}{)}{;}{a}{}{,} 
\makeatletter
\newcommandtwoopt{\citeads}[3][][]{\href{http://adsabs.harvard.edu/abs/#3}%
{\def\hyper@linkstart##1##2{}%
\let\hyper@linkend\@empty\citealp[#1][#2]{#3}}}
\newcommandtwoopt{\citepads}[3][][]{\href{http://adsabs.harvard.edu/abs/#3}%
{\def\hyper@linkstart##1##2{}%
\let\hyper@linkend\@empty\citep[#1][#2]{#3}}}
\newcommandtwoopt{\citetads}[3][][]{\href{http://adsabs.harvard.edu/abs/#3}%
{\def\hyper@linkstart##1##2{}%
\let\hyper@linkend\@empty\citet[#1][#2]{#3}}}
\newcommandtwoopt{\citeyearads}[3][][]%
{\href{http://adsabs.harvard.edu/abs/#3}
{\def\hyper@linkstart##1##2{}%
\let\hyper@linkend\@empty\citeyear[#1][#2]{#3}}}
\makeatother

\begin{document}

   \title{Revisiting the dynamical masses of the transiting planets in the young AU\,Mic system: Potential AU\,Mic b inflation at $\sim$20 Myr\thanks{Full Tables \ref{tab:rv_CARMV}, \ref{tab:rv_CARMN}, \ref{tab:rv_HARPS}, and \ref{tab:rv_SPIRou} are only available in electronic form at the CDS via anonymous ftp to cdsarc.u-strasbg.fr (130.79.128.5) or via \url{http://cdsweb.u-strasbg.fr/cgi-bin/qcat?J/A+A/}}}

   \author{
 M.\,Mallorqu\'in\inst{\ref{i:iac},\ref{i:ull}}, 
 V.\,J.\,S.\,B\'ejar\inst{\ref{i:iac},\ref{i:ull}},
 N.\,Lodieu\inst{\ref{i:iac},\ref{i:ull}},
 M.\,R.\,Zapatero Osorio\inst{\ref{i:cab}}, 
 H.\,Yu\inst{\ref{i:oxf}}, 
 A.\,Su\'arez Mascareño\inst{\ref{i:iac},\ref{i:ull}}, 
 M.\,Damasso\inst{\ref{i:inaf}}, 
 J.\,Sanz-Forcada\inst{\ref{i:cab}}, 
 I.\,Ribas\inst{\ref{i:ieec},\ref{i:ice}}, 
 A.\,Reiners\inst{\ref{i:got}}, 
 A.\,Quirrenbach\inst{\ref{i:hei}}, 
 P.\,J.\,Amado\inst{\ref{i:iaa}}, 
 J.\,A.\,Caballero\inst{\ref{i:cab}}, 
 S.\,Aigrain\inst{\ref{i:oxf}}, 
 O.\,Barrag\'an\inst{\ref{i:oxf}}, 
 S.\,Dreizler\inst{\ref{i:got}}, %
 A.\,Fern\'andez-Mart\'in\inst{\ref{i:caha}}, 
 E.\,Goffo\inst{\ref{i:tls},\ref{i:unito}}, 
 Th.\,Henning\inst{\ref{i:max}}, 
 A.\,Kaminski\inst{\ref{i:hei}}, 
 B.\,Klein\inst{\ref{i:oxf}}, 
 R.\,Luque\inst{\ref{i:chi}, \ref{i:iaa}}, 
 D.\,Montes\inst{\ref{i:ucm}}, 
 J.\,C.\,Morales\inst{\ref{i:ieec}, \ref{i:ice}}, 
 E.\,Nagel\inst{\ref{i:got}}, 
 E.\,Pall\'e\inst{\ref{i:iac},\ref{i:ull}}, 
 S.\,Reffert\inst{\ref{i:hei}}, 
 M.\,Schlecker\inst{\ref{i:ari}} 
 A.\,Schweitzer\inst{\ref{i:ham}} 
           }
   \authorrunning{Mallorqu\'in, M., et al.}
   \titlerunning{Revisiting the dynamical masses of the transiting planets in the young AU\,Mic star}
  \institute{
        \label{i:iac} Instituto de Astrof\'isica de Canarias (IAC), Calle V\'ia L\'actea s/n, 38205 La Laguna, Tenerife, Spain,
             \email{mmd@iac.es}
        \and 
        \label{i:ull} Departamento de Astrof\'isica, Universidad de La Laguna (ULL), 38206 La Laguna, Tenerife, Spain
        \and
        \label{i:cab} Centro de Astrobiolog\'ia (CAB), CSIC-INTA, ESAC Campus, Camino bajo del castillo s/n, 28692, Villanueva de la Cañada, Madrid, Spain
        \and
        \label{i:oxf} Sub-department of Astrophysics, Department of Physics, University of Oxford, Oxford, OX1 3RH, UK
        \and
        \label{i:inaf} Osservatorio Astrofisico di Torino, Via Osservatorio 20, 10025 Pino Torinese, Italy
        \and
        \label{i:ieec} Institut d'Estudis Espacials de Catalunya (IEEC), Calle Gran Capital 2-4, 08034, Barcelona, Spain
        \and
        \label{i:ice} Institut de Ci\'encies de l’Espai (CSIC-IEEC), UAB Campus, Calle de Can Magrans s/n, 08193 Bellaterra, Barcelona, Spain
        \and
        \label{i:got} Institut f\"ur Astrophysik und Geophysik, Georg-August-Universit\"at G\"ottingen, Friedrich-Hund-Platz 1, 37077 G\"ottingen, Germany
        \and
        \label{i:hei} Landessternwarte, Zentrum f\"ur Astronomie der Universit\"at Heidelberg, K\"onigstuhl 12, 69117, Heidelberg, Germany
        \and
        \label{i:iaa} Instituto de Astrof\'isica de Andaluc\'ia (IAA-CSIC), Glorieta de la Astronom\'ia s/n, 18008, Granada, Spain
        \and
        \label{i:caha} Centro Astron\'omico Hispano en Andaluc\'ia (CAHA), Observatorio de Calar Alto, Sierra de los Filabres, 04550, G\'ergal, Spain
        \and
        \label{i:tls} Th\"uringer Landessternwarte Tautenburg, 07778 Tautenburg, Germany
        \and 
        \label{i:unito} Dipartimento di Fisica, Universita degli Studi di Torino, via Pietro Giuria 1, I-10125, Torino, Italy
        \and
        \label{i:max} Max-Planck-Institut f\"ur Astronomie, K\"onigstuhl 17, 69117, Heidelberg, Germany
        \and
        \label{i:chi} Department of Astronomy \& Astrophysics, University of Chicago, Chicago, IL 60637, USA
        \and
        \label{i:ucm} Departamento de F\'isica de la Tierra y Astrof\'isica and IPARCOS-UCM (Instituto de F\'isica de Part\'iculas y del Cosmos de la UCM), Facultad de Ciencias F\'isicas, Universidad Complutense de Madrid, 28040, Madrid, Spain
        \and
        \label{i:ham} Hamburger Sternwarte, Gojenbergsweg 112, 21029 Hamburg, Germany
        \and
        \label{i:ari} Steward Observatory and Department of Astronomy, The University of Arizona, Tucson, AZ 85721, USA
             }

   \date{Received 20 March, 2024; Accepted 05 July, 2024}








 
  \abstract
   {Understanding planet formation is important in the context of the origin of planetary systems in general and of the Solar System in particular, as well as to predict the likelihood of finding Jupiter, Neptune, and Earth analogues around other stars.}
   {We aim to precisely determine the radii and dynamical masses of transiting planets orbiting the young M star AU\,Mic using public photometric and spectroscopic datasets.}
   {We performed a joint fit analysis of the TESS and CHEOPS light curves and more than 400 high-resolution spectra collected with several telescopes and instruments. We characterise the stellar activity and physical properties (radius, mass, density) of the transiting planets in the young AU\,Mic system through joint transit and radial velocity fits with Gaussian processes.}
   {We determine a radius of $R_{p}^{b}$=\,4.79\,$\pm$\,0.29 R$_\oplus$, a mass of $M_{p}^{b}$=\,9.0\,$\pm$\,2.7 M$_\oplus$, and a bulk density of $\rho_{p}^{b}$\,=\,0.49\,$\pm$\,0.16 g\,cm$^{-3}$ for the innermost transiting planet AU\,Mic\,b. For the second known transiting planet, AU\,Mic\,c, we infer a radius of $R_{p}^{c}$=\,2.79\,$\pm$\,0.18 R$_\oplus$, a mass of $M_{p}^{c}$=\,14.5\,$\pm$\,3.4 M$_\oplus$, and a bulk density of $\rho_{p}^{c}$\,=\,3.90\,$\pm$\,1.17 g\,cm$^{-3}$. According to theoretical models, AU\,Mic\,b may harbour an H$_{2}$ envelope larger than 5\% by mass, with a fraction of rock and a fraction of water. AU\,Mic\,c could be made of rock and/or water and may have an H$_{2}$ atmosphere comprising at most 5\% of its mass. AU\,Mic\,b has retained most of its atmosphere but might lose it over tens of millions of years due to the strong stellar radiation, while AU\,Mic\,c likely suffers much less photo-evaporation because it lies at a larger separation from its host. Using all the datasets in hand, we determine a 3$\sigma$ upper mass limit of $M_{p}^{[d]}\sin{i}$\,=\,8.6\,M$_{\oplus}$ for the AU\,Mic 'd' TTV-candidate. In addition, we do not confirm the recently proposed existence of the planet candidate AU\,Mic\,'e' with an orbital period of 33.4 days. We investigated the level of the radial velocity variations and show that it is lower at longer wavelength with smaller changes from one observational campaign to another.}
   {}


   \maketitle
%

\section{Introduction}
\label{sec:intro}
\nolinenumbers

Young systems ($<$ 1 Gyr) offer a unique opportunity to understand planet formation and evolution. At early ages, planets can still preserve information about their original radius and mass or their initial composition and internal structure. According to models, the timescales for these formation and evolution processes can range from a few million years to 1 Gyr, with the planets evolving most rapidly in the first tens of millions of years \citep{Baruteau2016, Jin2018, gupta20, Lopez2018, Raymond2018}. Although today there are thousands of transiting exoplanets known mostly thanks to space missions such as Kepler \citep{Kepler}, K2 \citep{K2}, or TESS \citep{TESS}, only a few tens of young transiting planetary systems have been discovered \citep[][and references therein]{Mallorquin2023, Mallorquin2024}. One of the main difficulties is identifying planets around young stars whose age determination is accurate enough to constrain the timescales of the formation and evolution models. In addition, these stars are very active, sometimes with variations that could be orders of magnitude larger than the signals attributed to planets, which makes the discovery and characterisation of these planets challenging. The high level of stellar activity and strong magnetic fields of young stars is mainly due to the rapid rotation and relatively large area of the active regions, which produces quasi-periodic and sporadic variations in light curves and radial velocity (RV) curves. These variations can confuse transits or mimic Keplerian signals produced by planets \citep{Damasso2020, Simpson2022}. The most promising candidates to characterise their physical properties are transiting planets, because the RV follow-up allows us to derive their masses and bulk densities when combined with radii obtained from the transit depth. While space missions with simultaneous photometric observations in a large field allow us to observe stars continuously for weeks or months, spectroscopic follow-up from the ground is much more complicated. In addition to the visibility of the target from the ground, only the brightest stars are amenable to achieve sufficient RV precision to infer the mass and density of the planets. Moreover, as the period of the stellar activity due to the rotation may be similar to the orbital periods of the transiting planets (of the order of days), an adequate sampling of these signals is key to being able to model and separate them properly. Currently, the most widely used technique for modelling the stellar activity in young stars is Gaussian processes regression \citep[GP;][]{Rasmussen2006}, which generates models that are flexible enough to reproduce quasi-periodic variations, but that also strongly depend on the temporal cadence of the data. Therefore, measuring the masses of young planets requires a combination of several ingredients: ($i$) high-precision spectrographs to reach a precision level of a few meters per second at each epoch; ($ii$) an observational strategy that makes it possible to obtain 3--5 RV points per stellar rotation period, that is 3--5 RV points every few days to adequately model the activity with the GP; and ($iii$) a minimum number of 100 RV epochs. This number is in fact typical for results published in the literature in the mass measurement of young planets.

Only twelve transiting young planetary systems similar in size to Neptune or smaller have measured densities thanks to RV follow-up: AU\,Mic\,b and c \citep[$\sim$20\,Myr;][]{klein21, Cale2021, zicher22, klein22, Donati2023}, TOI-1807\,b \citep[$\sim$300\,Myr;][]{nardiello22}, TOI-179\,b \citep[$\sim$300\,Myr;][]{vines23, Desidera2023}, K2-233\,d \citep[$\sim$360\,Myr;][]{barr23}, HD\,63433\,c \citep[$\sim$400\,Myr;][]{Mallorquin2023}, TOI-560\,b and c \citep[$\sim$490\,Myr;][]{barr22, elmufti2023}, TOI-1099\,b \citep[$\sim$520\,Myr;][]{barros23}, TOI-5398\,c \citep[$\sim$650\,Myr;][]{Mantovan2024}, K2-25\,b \citep[$\sim$725\,Myr;][]{stef20}, K2-100\,b \citep[$\sim$750\,Myr;][]{barr19}, TOI-1201\,b \citep[600--800\,Myr;][]{kossak21}, and TOI-1801\,b \citep[600--800\,Myr;][]{Mallorquin2023b}, where AU\,Mic is the only star younger than 100 Myr in this sample. In general, AU\,Mic and V1298\,Tau \citep[$\sim$20 Myr; whose star hosts four planets larger than Neptune;][]{david18b, damasso23} are the youngest systems with transiting planets whose masses have been determined by several groups \citep{Suarez-Mascareno2022, Sikora2023, Finociety2023}. Only three other systems with transiting planets are younger than AU\,Mic and V1298\,Tau: HIP67522 \citep[$\sim$15\,Myr;][]{Rizzuto2020}, TOI-1227 \citep[$\sim$12\,Myr;][]{Mann2022}, and K2-33 \citep[$\sim$9.3\,Myr;][]{Trevor2016, Mann2016}. For the former, its fast stellar rotation period ($P_{\text{rot}}$\,$\sim$\,1.4 d) makes modelling the activity a huge challenge, while TOI-1227 and K2-33 are comparatively faint stars ($G$\,$\sim$\,15.2, 14.1 mag, respectively) currently within the reach of only a few high-resolution spectrographs.

The star AU\,Microscopii (AU\,Mic, HD\,197481, TOI-2221) is a very young \citep[22\,$\pm$\,3\,Myr;][]{Mamajek2014} M1 pre-main-sequence star located at a distance of 9.7 pc \citep{gaia2016, gaiadr3}. It is a member of the $\beta$ Pictoris moving group \citep{Zuckerman2001, Gagne2018}, whose age is estimated in the range of 10--30\,Myr \citep{Miret-roig2020, Lee2024}. The star hosts a spatially resolved debris disc \citep{Kalas2004}, shows very intense magnetic activity, and is orbited by at least two planets. Based on the first analysis of a TESS light curve, \cite{plav20} published the detection of AU\,Mic\,b, a Neptune-sized transiting planet that orbits the star every 8.46 d. Using RV measurements from several spectrographs, they estimated a 3$\sigma$ upper mass limit of 57.2 M$_\oplus$ for the planet. In addition, they observed a mono-transit event, indicating the possible presence of an additional planet in the system. Shortly after, \cite{Hirano2020}, \cite{Martioli2020}, and \cite{Palle2020} measured the Rossiter-McLaughlin (RM) effect of AU\,Mic\,b with the InfraRed Doppler \citep[IRD;][]{Tamura2012, Kotani2018}, the SPectropolarim\`etre InfraROUge \citep[SPIRou;][]{Donati2020}, and the \'Echelle SPectrograph for Rocky Exoplanets and Stable Spectroscopic Observations \citep[ESPRESSO;][]{Pepe2021} spectrographs, respectively, inferring that the perpendicular to the orbital plane of the planet is aligned with the stellar rotation axis. Subsequently, \cite{klein21} obtained a mass measurement of 17.1$^{+4.7}_{-4.5}$ M$_\oplus$ for AU\,Mic\,b using RV measurements collected with the SPIRou near-infrared (NIR) spectrograph. With the observation of the second (and last) TESS sector of the star, \cite{Martioli2021} published the detection of the second planet in the system, AU\,Mic\,c, with an orbital period of 18.86 d and a slightly smaller size than planet b. Subsequently, \cite{Cale2021}, \cite{zicher22}, and \cite{klein22} attempted to infer the mass of both planets using slightly different methodologies and datasets to characterise the stellar activity. \cite{Cale2021} proposed a chromatic GP using data from different spectrographs to model the stellar activity of AU\,Mic, obtaining a mass of 20.1$^{+1.7}_{-1.6}$ M$_\oplus$ for planet b and a 5$\sigma$ upper mass limit of 20.1\,M$_\oplus$ for planet c. \cite{zicher22} used 85 HARPS measurements during two observing campaigns and a multidimensional GP framework \citep{Rajpaul2015} to infer masses of 11.7\,$\pm$\,5.0 and 22.2\,$\pm$\,6.7 M$_\oplus$ for AU\,Mic\,b and c, respectively. Using the same dataset as \cite{zicher22}, \cite{klein22} estimated a mass of 14.3\,$\pm$\,7.7 M$_\oplus$ for AU\,Mic\,b and 34.9\,$\pm$\,10.8 M$_\oplus$ for AU\,Mic\,c with a Doppler imaging technique designed to simultaneously model the activity-induced distortions and the planet-induced shifts in the line profiles. The radii of both planets have been measured using light curves from TESS, CHEOPS, Spitzer, and tens of transits from ground-based facilities \citep{plav20, Martioli2021, Szabo2021, Gilbert2022, Szabo2022, Wittrock2022, Wittrock2023}. The values of the radii range from 3.6 to 4.4 R$_{\oplus}$ and from 2.4 to 3.2 R$_{\oplus}$ for AU\,Mic\,b and c, respectively. These transit observations have enabled the study of transit time variations (TTVs) in these planets. \cite{Szabo2021, Szabo2022}, \cite{Gilbert2022}, and \cite{Wittrock2022, Wittrock2023} found TTVs of several minutes that could be explained by a non-transiting candidate (AU\,Mic\,'d') with an orbital period of 12.73596 d and a mass similar to that of Earth (1.1\,$\pm$\,0.5 M$_{\oplus}$). Recently, \cite{Donati2023} published new RV measurements with the SPIRou spectrograph and identified an additional non-transiting planet candidate with an orbital period of 33.39 d. These authors derived masses of 10.2$^{+3.9}_{-2.7}$, 14.2$^{+4.8}_{-3.5}$, 2.9$^{+2.9}_{-1.3}$, and 35.2$^{+6.7}_{-5.4}$ M$_\oplus$ for AU\,Mic\,b, c, and the candidates 'd', and 'e', respectively. AU\,Mic has also been observed with the NIRCam instrument aboard the JWST, using the coronagraphy technique. No massive companions ($>$0.1\,M$_{\mathrm{Jup}}$) beyond 20\,AU were found with a 5$\sigma$ level of confidence \citep{Lawson2023}.

Our paper introduces the first joint analysis of photometry and RV where a measure of the bulk density, with a significance of at least 3$\sigma$, of both transiting planets of the AU\,Mic system is obtained. The paper is structured as follows. In Sects.\,\ref{sec:ph} and \ref{sec:sp}, we present the photometric and RV time series of the system, respectively. In Sect.\,\ref{sec:star}, we revise the stellar properties of the star. In Sect.\,\ref{sec:analysis}, we outline a full photometric and RV analysis. In Sect.\,\ref{sec:disc}, we discuss the composition of the planets, their evolutionary status, and key implications. Finally, in Sect.\,\ref{sec:concl}, we summarise the main results of our study and place them in the wider context of planet formation.

\section{Transit photometry}
\label{sec:ph}

\renewcommand{\arraystretch}{1.2} 
\begin{table}
\caption{Main characteristics of photometric transit datasets analysed in this work.}\label{tab:PHdatasets}
\centering
\begin{tabular}{l c c c c c}
\hline\hline
Mission & Transit date & $\lambda$ range & $t_{\mathrm{exp}}$ & Planet & Ref\\
 & [UT] & [nm] & [s] &  & \\
\hline
TESS S1 & 2018-07-30 & 600--1000 & 120 & b & 1\\
TESS S1 & 2018-08-11 & 600--1000 & 120 & c & 1\\
TESS S1 & 2018-08-16 & 600--1000 & 120 & b & 1\\
TESS S20 & 2020-07-10 & 600--1000 & 20 & b & 1\\
CHEOPS & 2020-07-10 & 400--1100 & 15 & b & 2\\
TESS S20 & 2020-07-09 & 600--1000 & 20 & c & 1\\
TESS S20 & 2020-07-19 & 600--1000 & 20 & b & 1\\
TESS S20 & 2020-07-27 & 600--1000 & 20 & b & 1\\
TESS S20 & 2020-07-28 & 600--1000 & 20 & c & 1\\
CHEOPS & 2020-08-22 & 400--1100 & 15 & b & 2\\
CHEOPS & 2020-09-24 & 400--1100 & 3 & b & 2\\
CHEOPS & 2021-07-26 & 400--1100 & 15 & b & 3\\
CHEOPS & 2021-08-09 & 400--1100 & 15 & c & 3\\
CHEOPS & 2021-08-12 & 400--1100 & 15 & b & 3\\
CHEOPS & 2021-08-28 & 400--1100 & 15 & c & 3\\
CHEOPS & 2021-08-29 & 400--1100 & 15 & b & 3\\
CHEOPS & 2021-09-06 & 400--1100 & 15 & b & 3\\
\\
\hline
\end{tabular}
\tablebib{
(1): \href{https://archive.stsci.edu/}{MAST}; 
(2): \citet{Szabo2021};
(3): \citet{Szabo2022}.
}
\end{table}

\subsection{CHEOPS}

We used the transit photometry from the CHaracterising ExOPlanet Satellite (CHEOPS) previously analysed by \cite{Szabo2021, Szabo2022}. These data cover seven transits of planet b and two transits of planet c (Table\,\ref{tab:PHdatasets}). The data were collected between July and September, 2020 and between July and September, 2021. The data are publicly available, and the detailed explanation of their reduction can be found in \citet[][Sect.\,2.1]{Szabo2021} and \citet[][Sect.\,2.1]{Szabo2022}. The transit photometry presents several flares with an amplitude comparable to or larger than the depth of the transits. For this reason, we decided to visually search and mask the flares identified by eye. Therefore, we use this photometry free of flares in the subsequent analysis. Since the published CHEOPS data do not include error uncertainties, we adopted a common error bar of 0.31 ppt, which corresponds to the dispersion of the out-of-transit data.

\subsection{TESS}

\begin{figure*}
\includegraphics[width=1\linewidth]{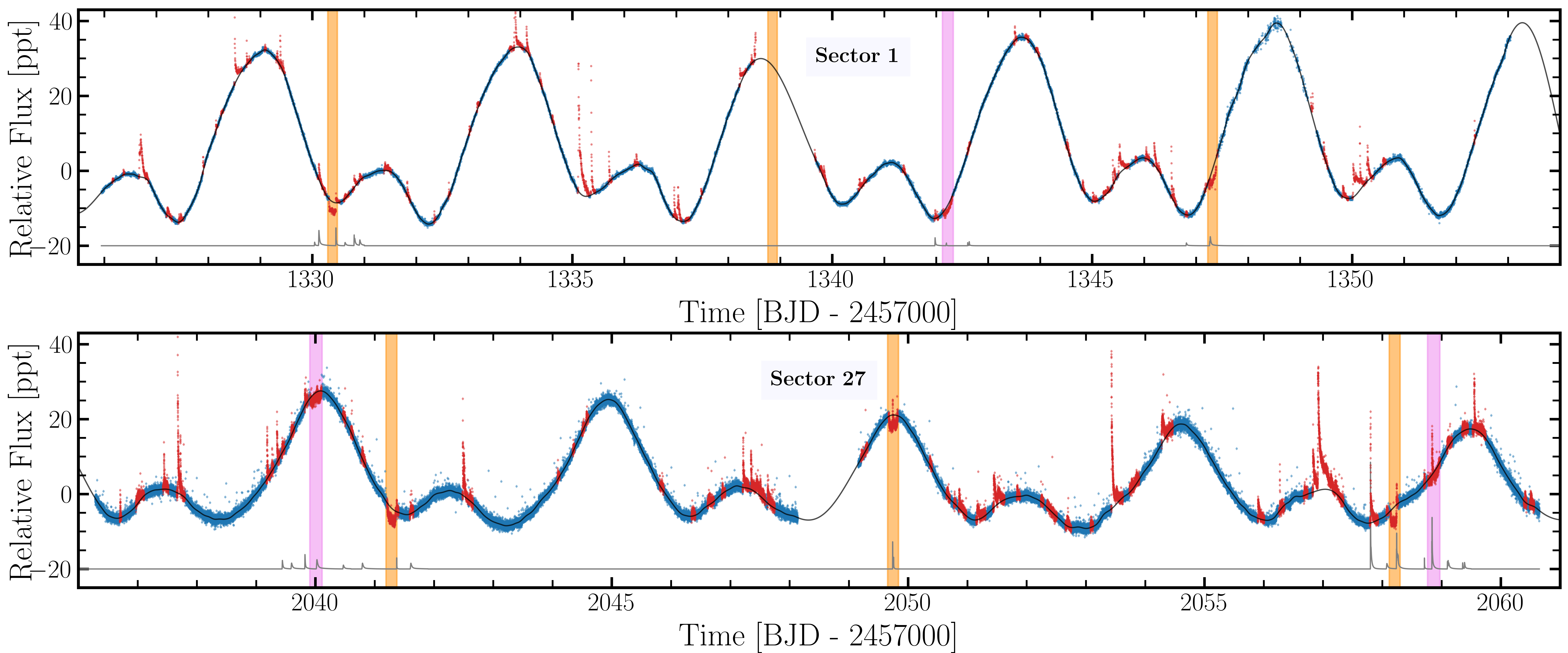}
\caption{TESS light curve of AU\,Mic. The PDCSAP flux used in the analysis is shown as blue dots. The data considered as flares or transits are shown as red dots. The black and grey lines represent the stellar activity models for the stellar rotation and for the flares, respectively. The flare model was vertical shifted for a better visualisation. The vertical orange and purple bands indicate the timing of the planetary transits for AU\,Mic\,b and c, respectively.}
\label{fig:LC_TESS}
\end{figure*}

The star AU\,Mic was observed by TESS in sector 1 in August, 2018 and sector 27 in July, 2020 with a 2-min and 20-s cadence, respectively (Table\,\ref{tab:PHdatasets}). Both sectors were processed by the Science Processing Operations Center \citep[SPOC;][]{SPOC} photometry and transit search pipeline at the NASA Ames Research Center and were downloaded from the Mikulski Archive for Space Telescopes\footnote{\url{https://archive.stsci.edu/}} (MAST). TESS is scheduled to be pointed at AU\,Mic again in August, 2025. The PDCSAP light curve of AU\,Mic is characteristic of a young and active star, with numerous flares (more than six flares per day), as reported by \citet{Martioli2021}, and a quasi-periodic double-peaked pattern whose shape is repeated during the two sectors. The light curve of the first sector had a peak-to-peak amplitude of up to 60 ppt, while the amplitude decreased to around 40 ppt in the second sector. Due to the large number and intensity of the flares present in the light curve, and for homogeneity with the CHEOPS light curve, we decided to identify and remove them by visual inspection. For the rest of our analyses, we used the TESS photometric light curve after the removal of flares (Fig.\,\ref{fig:LC_TESS}).

\section{Spectroscopic observations}
\label{sec:sp}

The spectroscopic and/or RV data presented in this paper are available in public databases and/or have been published previously. The details of the observations, instrumentation, and reduction are given in the respective papers. The main characteristics of each dataset such as wavelength range, number of epochs, rms, precision, and pipeline are listed in Table\,\ref{tab:RVdatasets}. The variations due to stellar activity are larger than the expected Keplerian amplitudes, making the modelling of the activity a major challenge. For this reason, we only included the most stable spectrographs with the highest quality data in our RV analysis (i.e. CARMENES, HARPS, HIRES, iSHELL, SPIRou, and TRES). Additionally, because the variations are quasi-periodic, i.e. they cannot be modelled with a simple periodic function, the cadence of the data is essential to properly model the activity and, consequently, determine the masses of the planets. Therefore, we excluded isolated observations or low-cadence datasets that do not help to constrain the activity model. Finally, because of the significant impact that the Earth's atmosphere can have on RV calculations, all spectra from spectrographs covering wavelengths beyond 700\,nm were corrected for telluric absorption lines.

\begin{table*}
\caption{Main characteristics of RV datasets analysed in this work.}\label{tab:RVdatasets}
\centering
\begin{tabular}{l c c c c c c l}
\hline\hline
Spectrograph & $\lambda$ range & $N_{\mathrm{RV}}$\tablefootmark{a} & rms$_{\mathrm{RV}}$ & Median $\sigma_{\mathrm{RV}}$ & $N_{\mathrm{RV}}$/$P_{\mathrm{rot}}/T$\tablefootmark{b} & TAC\tablefootmark{c} & Pipeline\\
 & [nm] & & [m\,s$^{-1}]$ & [m\,s$^{-1}]$ & & & \\
\hline
\textbf{CARMENES VIS} & \textbf{520--960} & \textbf{88 (86)} & \textbf{93.7} & \textbf{7.8} & \textbf{3.1} & \textbf{Yes} & \tt{serval}\tablefootmark{c} \\
\textbf{CARMENES NIR} & \textbf{960--1710} & \textbf{89 (87)} & \textbf{84.9} & \textbf{23.4} & \textbf{3.1} & \textbf{Yes} & \tt{serval}\tablefootmark{c} \\
\textbf{HARPS} & \textbf{380--690} & \textbf{122 (117)} & \textbf{148.6} & \textbf{3.2} & \textbf{3.0} & \textbf{No} & \tt{serval}\tablefootmark{d} \\
HIRES & 480-600 & 40 & 130.9 & 2.6 & 1.1 & No & \cite{Howard2010}\\
iSHELL & 2180--2470 & 28 & 39.6 & 4.8 & 1.0 & Yes & \tt{pychell}\tablefootmark{e} \\
\textbf{SPIRou} & \textbf{950--2350} & \textbf{179 (176)} & \textbf{32.5} & \textbf{3.8} & \textbf{2.1} & \textbf{Yes} & \tt{LBL}\tablefootmark{f} \\
TRES & 390--910 & 43 & 106.0 & 22.6 & 1.5 & No & \cite{Furesz2008} \\
\hline
\textbf{Full dataset} & ... & \textbf{478 (466)} & ... & ... & \textbf{3.9} & ... & ... \\
\hline
\end{tabular}
\tablefoot{
\tablefoottext{a}{Total numbers of RV epochs. The number of RV epochs used in the final analysis is provided in parentheses.}
\tablefoottext{b}{Number of RV epochs used in the final analysis per stellar rotation period and per total time range. The instruments used to derive the planetary parameters of the system are represented in bold face.}
\tablefoottext{c}{Telluric absorption corrected.}
\tablefoottext{d}{\cite{SERVAL}.}
\tablefoottext{e}{\cite{Cale2019}.}
\tablefoottext{f}{\cite{Artigau2022}.}
}
\end{table*}

\subsection{CARMENES}

The star AU\,Mic was observed 98 times with the Calar Alto high-Resolution search for M dwarfs with Exoearths with Near-infrared and optical \'Echelle Spectrographs \citep[CARMENES;][]{CARMENES, CARMENES18} with the visible (VIS) and NIR channels, in two observing campaigns during 2019 and 2020 \citep{Ribas2023}. Unlike the data presented by \cite{Cale2021}, we used spectra that were corrected for telluric absorption lines (TAC) following \cite{Nagel2023}. We examined the relative intensity of spectral lines associated with chromospheric activity, including H$\alpha$, \ion{Ca}{II}\,IRT, \ion{Na}{I}, \ion{K}{I}, and \ion{He}{I}, because RVs can be affected by the presence of strong flares due to the young age of AU\,Mic \citep{Fuhrmeister2023}. We discarded five and four spectra in the VIS and NIR channels, respectively, suspected to be affected by flares due to the broader of emission lines. Moreover, the last five spectra were obtained separated by 70 d from the main RV follow-up, and therefore they do not have the required cadence to properly sample the stellar activity, leaving a total of 88 and 89 spectra for subsequent analysis. We calculated the RVs using {\tt serval}\footnote{\url{https://github.com/mzechmeister/serval}} \citep{SERVAL}, applying a smoothing factor to the template (called {\tt{ofac}} in the software) to account for the relatively high $v\sin{i}$ of the star (Sect.\,\ref{sec:star}).

\subsection{HARPS}

There are 197 public spectra of AU\,Mic in the ESO archive\footnote{\url{https://archive.eso.org/wdb/wdb/adp/phase3_spectral/form}} collected with the High Accuracy Radial velocity Planet Searcher \citep[HARPS;][]{Pepe2000}. We discarded 63 spectra obtained between 2003 and 2019 because of the poor cadence, which is not optimal for our scientific goal. The remaining 134 spectra correspond to the 91 spectra analysed by \cite{zicher22}, and one campaign of 43 spectra collected in 2022 (under program 110.241K.001; PI Yu). We proceeded as we did for the CARMENES spectra and found 12 spectra that we rejected because they were affected by flares, leaving a sample of 122 spectra for our analysis. We extracted the RV from reduced spectra with {\tt serval}, in the same manner as those from CARMENES data.

\subsection{SPIRou RVs}

A total of 185 RVs secured with the SPectropolarim\`etre InfraROUge \citep[SPIRou;][]{Donati2020} during four campaigns over a three year baseline (2019-2022) were analysed and published by \cite{Donati2023}. Of those 181 measurements, we discarded the first six because of poor cadence, resulting in 175 RV measurements to include in our analysis.

\subsection{Other RVs}
There are 60, 46, and 73 RVs collected with the HIgh Resolution \'Echelle Spectrometer \citep[HIRES;][]{Vogt1994}, the Immersion Grating \'Echelle Spectrograph \citep[iSHELL;][]{Rayner2016}, and the Tillinghast Reflector \'Echelle Spectrograph (TRES), respectively, which were published by \cite{Cale2021}. Most of these measurements are distributed over a baseline of thousands of days. This means that they do not have a sufficient cadence to model the activity using our methodology. Therefore, we were left with 40 HIRES RVs, 28 iSHELL RVs, and 43 TRES RVs for subsequent analysis.

\section{Stellar properties}
\label{sec:star}

\begin{figure}
\includegraphics[width=1\linewidth]{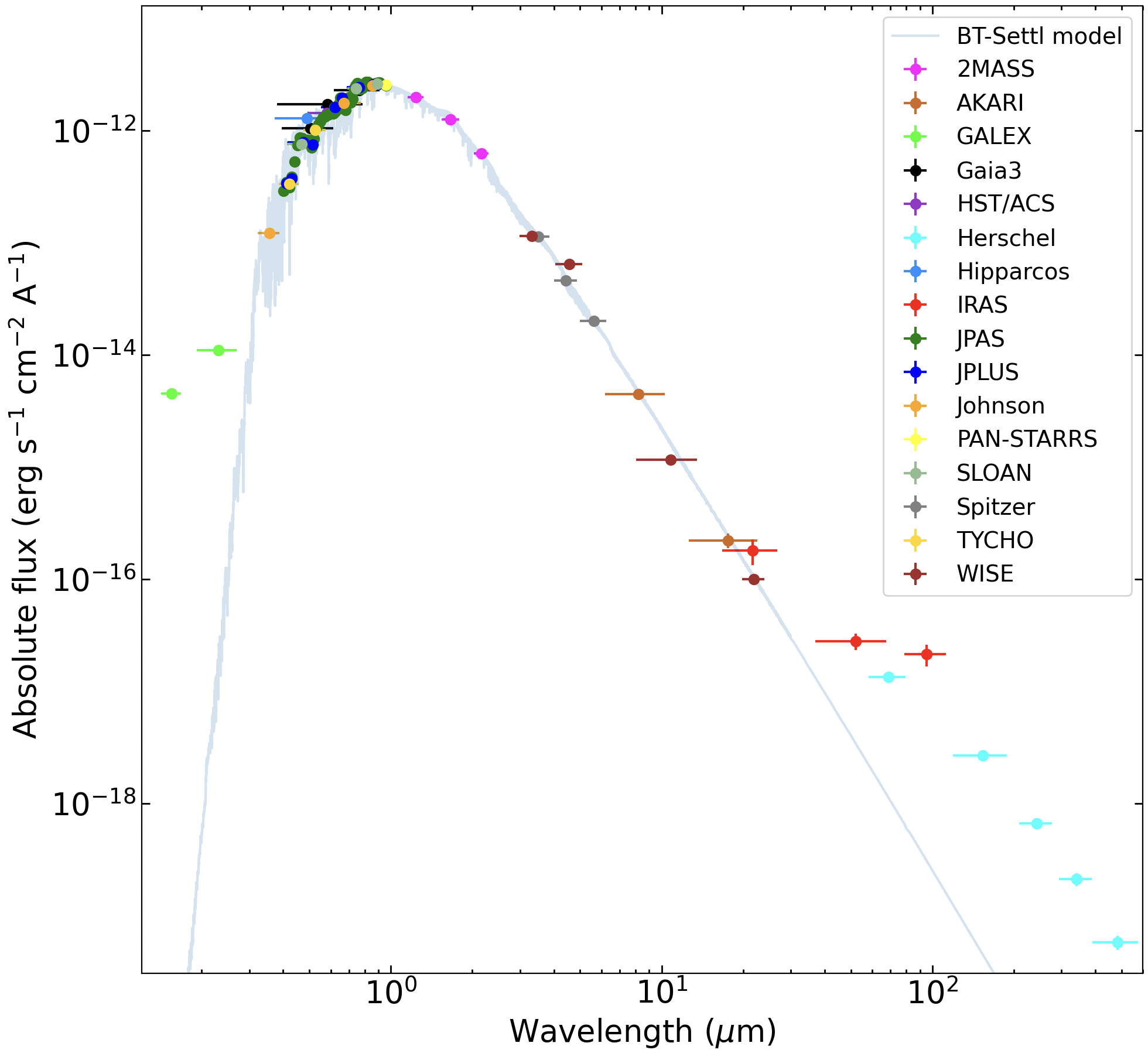}
\caption{Photometric SED of AU\,Mic (coloured dots) covers 0.13\,$\mu$m to 670\,$\mu$m. The photospheric emission of a 3600\,K dwarf with solar metallicity is shown by the blue line \citep[BT-Settl model;][]{Allard2012}. AU\,Mic shows significant flux excesses at wavelengths shorter than 0.3 $\mu$m and longer than 40 $\mu$m, which are compatible with stellar activity and the presence of a circumstellar debris disc \citep{Kalas2004}, respectively. Horizontal error bars account for the width of the filter passbands. Effective wavelengths and widths of the passbands are taken from the Virtual Observatory filter database \citep{bayo2008}.}
\label{fig:SED}
\end{figure}

To determine the stellar parameters of AU\,Mic, we first built its photometric spectral energy distribution (SED) using broad- and narrow-band photometry from various optical and infrared archives. The stellar SED is shown in Fig.\,\ref{fig:SED}, including the far- and near-ultraviolet data of the Galaxy Evolution Explorer satellite \citep[GALEX;][]{bianchi17}; Johnson $UBVR$ photometry \citep{ducati02}; the $ugriz$ magnitudes from the Sloan Digital Sky Survey catalogue \citep[SDSS;][]{york00}; the optical multi-band photometry of the Observatorio Astrof\'isico de Javalambre (OAJ) Physics of the Accelerating Universe Astrophysical Survey \citep[J-PAS;][]{Cenarro2019} and Photometric Local Universe Survey \citep[J-PLUS;][]{Dupke2019} catalogues; {\sl Gaia} Early Data Release 3 photometry \citep{gaia2016,gaiadr3}; the $y$-band magnitude from the Panoramic Survey Telescope \& Rapid Response System \citep[PAN-STARRS;][]{flewelling20}; the Two Micron All Sky Survey near-infrared $JHK_s$ photometry \citep[2MASS;][]{skrutskie2006}; the Wide-field Infrared Survey Explorer $W1$, $W2$, $W3$, and $W4$ data \citep[{\sl WISE};][]{wright2010}; the {\sl AKARI} S9W and L18W fluxes \citep{matsuhara06}; the Infrared Astronomical Satellite 60-$\mu$m and 100-$\mu$m data \citep[{\sl IRAS};][]{moshir90}; and the observations using the Photodetector Array Camera and Spectrometer (PACS) and SPIRE camera from the {\sl Herschel Space Observatory} \citep{matthews15}. AU\,Mic's SED is well covered from 0.13 $\mu$m up to 850 $\mu$m, although in Fig.\,\ref{fig:SED} we show data only up to 670 $\mu$m for clarity. The SED has three distinguishable components: photospheric emission dominant between 0.3 and 21 $\mu$m and well reproduced by the BT-Settl model of $T_{\rm eff}$\,=\,3600 K, log\,$g$\,=\,4.5 (cgs), and solar metallicity \citep{Allard2012}; flux excess at blue wavelengths below 0.3 $\mu$m due to stellar chromospheric activity; and far-infrared flux excess at wavelengths greater than 40 $\mu$m indicative of circumstellar dust discovered by {\sl IRAS}. The disc emission continues through submillimetre wavelengths \citep{liu04}, but we did not extend the SED of AU\,Mic because we are mostly interested in the photospheric fluxes for deriving stellar parameters. 

We integrated the photospheric emission of the SED over the wavelength range 0.35--21 $\mu$m using the trapezoidal rule and the {\sl Gaia} Data Release 3 trigonometric distance \citep{gaiadr3} to derive the bolometric luminosity of AU\,Mic, obtaining $L_{\star}$\,=\,0.1053\,$\pm$\,0.0014 L$_\odot$. The green and red optical broad-band filters (e.g.\ $Gaia$, Johnson $VRI$, SDSS) were not used in the integration because their passbands encompasse various redder and bluer narrower filters (e.g.\ J-PAS). We then applied $M_{\rm bol}\,$=\,$-2.5~{\rm log}\,F_{\rm bol} - 18.988$ \citep{cushing2005}, where $F_{\rm bol}$ is in units of W\,m$^{-2}$, to derive an absolute bolometric magnitude $M_{\rm bol}$\,=\,7.184$\pm$0.014 mag for AU\,Mic. The quoted error bar accounts for the photometric uncertainties in all observed bands and the trigonometric distance error. Our bolometric luminosity is slightly larger than the values available in the literature \citet[e.g.][]{plavchan09} and indicate that AU\,Mic is more luminous than stars with the same temperature and spectral type on the main sequence (see below).

The radius of AU\,Mic was determined to be $R_\star$\,=\,0.862\,$\pm$\,0.052 R$_\odot$ by \citet{Gallenne2022} using interferometric observations and the same {\sl Gaia} distance as used above for the stellar luminosity computation. Therefore, there was no need to apply any correction to the conversion from angular diameter to linear radius by \citet{Gallenne2022}. We did not employ the smaller stellar radius of \citet{White2015}, because these authors reported the uniform disc diameter without any limb-darkening calculation. By combining the measured stellar bolometric luminosity and radius using the Stefan-Boltzman equation,
\begin{equation}
L_\star = R_\star^2 \times (T_{\rm eff}/5772)^4,
\end{equation}
where $L_\star$ and $R_\star$ are in solar units and 5772\,K is the solar effective temperature, we derived $T_{\rm eff}$\,=\,3540\,$^{+120}_{-110}$ K for the AU\,Mic star. The temperature uncertainty accounts for the errors in luminosity and radius, although the radius uncertainty mainly dominates it. This result is derived only from parameters obtained through observations. It is in good agreement with the temperature obtained from the SED after finding the best BT-Settl model reproducing the observed fluxes. It is also compatible with the expected temperature of an M1 dwarf star at the 1$\sigma$ level \citep[e.g.][]{pecaut13,cifuentes20}, although it appears systematically $\approx$100 K cooler than the mean values tabulated in the literature. Our temperature deviates from the hotter determination of \citet{Maldonado2020}, \citet{Marfil2021}, and \citet{Donati2023} by 100--250 K.

\begin{figure}
\includegraphics[width=1\linewidth]{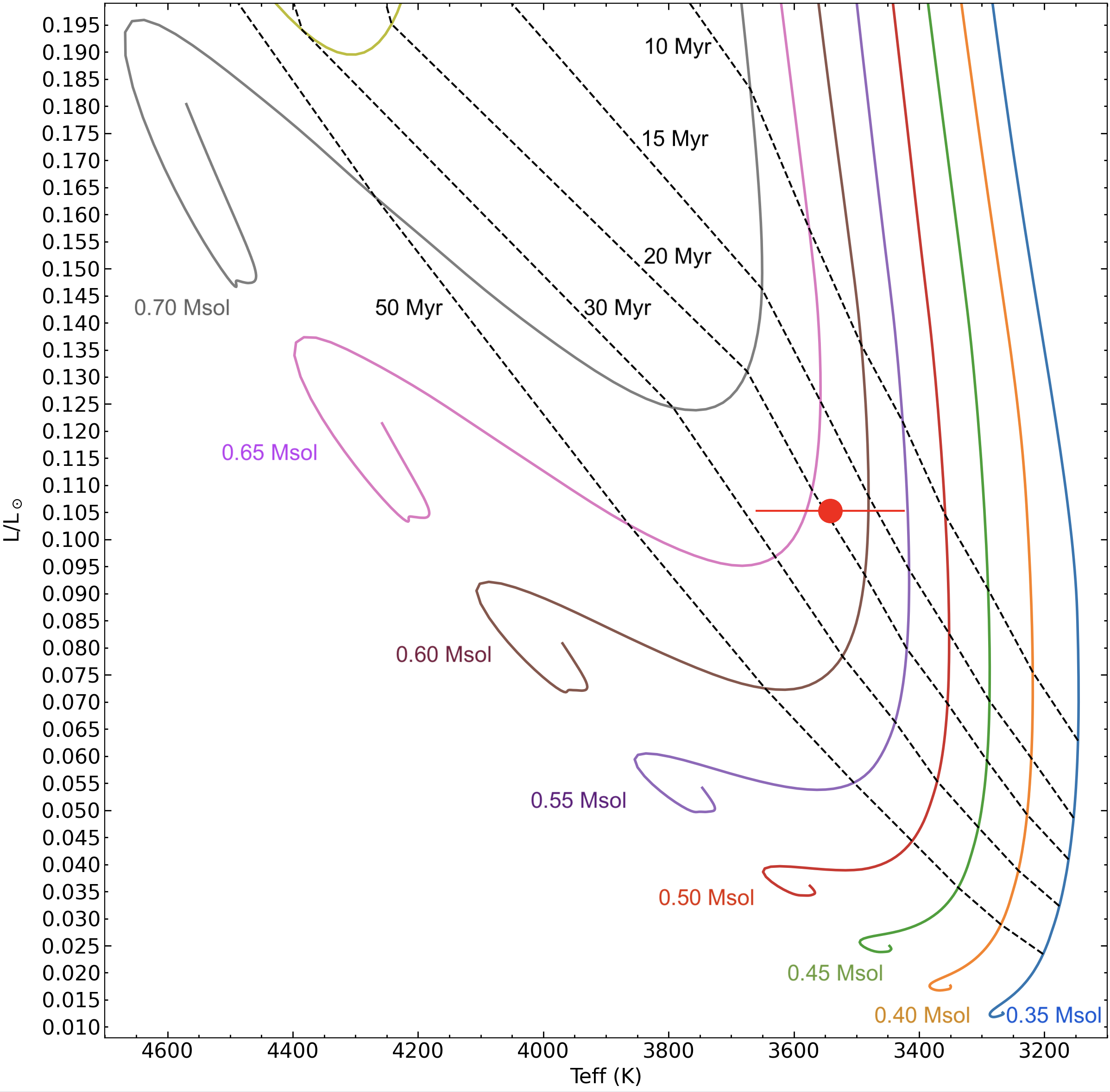}
\caption{AU\,Mic (red dot) in the Hertzsprung-Russell diagram. The solid coloured lines correspond to the evolutionary tracks of the \texttt{PARSEC} v1.25 solar metallicity models \citep{Chen2015}: masses in solar units are labelled. Isochrones with ages in the 10--50 Myr interval are shown by the dashed lines. The tracks are terminated at the age of 8.5\,Gyr. The temperature and luminosity error bars of AU\,Mic are plotted; the luminosity error bar is smaller than the size of the symbol.}
\label{fig:HR}
\end{figure}

Because of its young age, the stellar mass of AU\,Mic cannot be derived from mass-luminosity or mass-radius relations valid for main-sequence stars. The stellar mass and age can be derived; however, by placing AU\,Mic in the Hertzsprung-Russell diagram and comparing it against evolutionary models. Figure~\ref{fig:HR} depicts the luminosity and temperature of AU\,Mic together with the PARSEC v1.25 solar metallicity tracks and isochrones of different masses and ages \citep{Chen2015}. We obtained a mass of $M_\star$\,=\,0.635$^{+0.040}_{-0.070}$ M$_\odot$ and an age of $19 ^{+10}_{-6}$ Myr for AU\,Mic. The quoted error bars account for the $T_{\rm eff}$ and luminosity uncertainties, although the temperature error contributes most to the mass and age error budgets. Interestingly, our age derivation is fully consistent with the estimated age based on the membership of AU\,Mic to the $\beta$\,Pictoris young moving group, which has a dynamical traceback age of $18.5 ^{+2.0}_{-2.4}$ Myr \citep{Miret-roig2020}, which is a lower limit on the age of the group. Using different methods, the age of the $\beta$\,Pic Moving Group is estimated in a wider range of 10--30\,Myr (\citealp{Miret-roig2020}, and references therein), consistently with our estimated age uncertainties. We note that our mass determination for AU\,Mic yields a value up to 27\,\% larger than the masses used in the literature, except for the mass derived by \cite{Donati2023}, which is consistent with our determination.

With the derived mass and the interferometric radius, we computed the surface gravity of AU\,Mic to be log\,$g$\,=\,4.37$^{+0.08}_{-0.10}$ (cgs). All stellar parameters are summarised in Table\,\ref{tab:stellar_parameters}.

\renewcommand{\arraystretch}{1.2} 
\begin{table}
\caption{Stellar parameters of AU\,Mic.}\label{tab:stellar_parameters}
\centering
\begin{tabular}{lcr}
\hline\hline
Parameter & Value & Reference\\
\hline
Name & HD\,197481 & Can18\\
     & CD-31\,17815 & Tho92\\
     & AU\,Mic & Kuk72\\
\noalign{\smallskip}
$\alpha$ (J2016) & 20:45:09.9 & \textit{Gaia} DR3\\
$\delta$ (J2016) & --31:20:33.0 & \textit{Gaia} DR3\\
$G$ [mag] & 7.843\,$\pm$\,0.003 & \textit{Gaia} DR3\\
$J$ [mag] & 5.436\,$\pm$\,0.017 & 2MASS\\
Sp. type & M1V & Ken89 \\
\noalign{\smallskip}
$\varpi$ [mas] & 102.94\,$\pm$\,0.02 & \textit{Gaia} DR3\\
$d$ [pc] & 9.7144\,$\pm$\,0.0018& \textit{Gaia} DR3\\
\noalign{\smallskip}
$L_{\star}$ [$L_{\odot}$] & 0.1053\,$\pm$\,0.0014 & This work \\ 
$R_{\star}$ [$R_{\odot}$] & 0.862\,$\pm$\,0.052 & Gal22 \\
$T_{\text{eff}}$ [K] & 3540$_{-110}^{+120}$ & This work \\
$M_{\star}$ [$M_{\odot}$] & 0.635$_{-0.070}^{+0.040}$ & This work \\
$\log{g}$ [cgs] & 4.37\,$_{-0.10}^{+0.08}$ & This work \\
{[Fe/H]} [dex] & 0.01\,$\pm$\,0.06 & Mar21 \\
$v\sin i$ [km\,s$^{-1}$] & 9.2\,$\pm$\,0.2 & Koc20 \\
$P_{\text{rot}}$ [d] & 4.86\,$\pm$\,0.01 & Pla20 \\
Kinematic group & $\beta$ Pictoris & Bar99\\
Age [Myr] & 19$_{-6}^{+10}$ & This work \\
\noalign{\smallskip}
\hline
\end{tabular}
\tablebib{
Can18: \citet{Cannon1918}; 
Tho92: \citet{Thome1892};
Kuk72: \citet{Kukarkin1972};
\textit{Gaia} DR3: \cite{gaia2016, gaiadr3}; 
2MASS: \citet{skrutskie2006};
Ken89: \citet{Keenan1989};
Gal22: \citet{Gallenne2022};
Mar21: \citet{Marfil2021};
Koc20: \citet{Kochukhov2020};
Pla20: \citet{plav20};
Bar99: \citet{Barrado1999}.
}
\end{table}

\section{Analysis}
\label{sec:analysis}

\subsection{Transit analysis}
\label{sec:tr}

Heterogeneities on stellar surfaces produce photometric variations. In young stars such as AU\,Mic, these photometric variations can produce peak-to-peak amplitudes of tens of ppt or even larger. In the particular case of AU\,Mic, the star shows a photometric pattern due to stellar rotation with a periodicity of $\sim$4.86 days \citep{plav20}. This double-peaked, quasi-periodic pattern was relatively stable in amplitude in both TESS sectors (Fig.\,\ref{fig:LC_TESS}). Additionally, the star shows a large amount of intense flaring activity throughout the light curve, including during transits (as occurred in the ESPRESSO observations of \citealt{Palle2020}, which prevented the detection of the planet's atmosphere). These flares are also observed in the CHEOPS light curve. Hence, modelling the transits requires prior modelling and subtraction of the stellar activity.

We modelled the quasi-periodic variations due to stellar rotation using GPs as in \cite{plav20}. We used the double simple harmonic oscillator (dSHO) kernel:
\begin{equation}
\begin{aligned}
k_{\mathrm{dSHO}}(\tau) = & \  k_{\mathrm{SHO}}(\tau; \eta_{\sigma_{1}}, \eta_{L_{1}}, \eta_{P}) + k_{\mathrm{SHO}}(\tau; \eta_{\sigma_{2}}, \eta_{L_{2}}, \eta_{P}/2) \\
               = & \ \eta_{\sigma_{1}}^{2}  e^{-\frac{\tau}{\eta_{L_{1}}}}  \left[ \cos \left(\eta_{1} \frac{2\pi\tau}{\eta_{P}} \right) + \eta_{1} \frac{\eta_{P}}{2\pi \eta_{L_{1}}} \sin \left(\eta_{1} \frac{2\pi\tau}{\eta_{P}} \right) \right] \\
             & + \eta_{\sigma_{2}}^{2}  e^{-\frac{\tau}{\eta_{L_{2}}}}  \left[ \cos \left(\eta_{2} \frac{4\pi\tau}{\eta_{P}} \right) + \eta_{2} \frac{\eta_{P}}{4\pi \eta_{L_{2}}} \sin \left(\eta_{2} \frac{4\pi\tau}{\eta_{P}} \right) \right],
\end{aligned}
\label{eq:dsho}
\end{equation}
implemented in the \texttt{celerite} package \citep{celerite}, with the goal of creating a smooth function that is flexible enough to model the stellar activity without affecting the depth of the transits (black line in Fig.\,\ref{fig:LC_TESS}). In Eq.\,\ref{eq:dsho} (valid only if $\eta_{P}$\,$<$\,$2\pi \eta_{L}$), the variable $\tau$\,$\equiv$\,$|t_i - t_j|$ represents the time-lag between two data points, and $\eta$ is defined as $|1 - (2\pi \eta_{L}/\eta_{P})^{-2} |)^{1/2}$. The hyperparameters of the GP $\eta_{\sigma_{i}}$, $\eta_{L_{i}}$, and $\eta_{P}$ represent the amplitude of the covariance, the decay timescale, and the period of the fundamental signal, respectively. To design a smooth function that fits the temporal and amplitude scales of the TESS light curve, we imposed a normal prior on $\eta_{P}$ centred on the rotation period, uniform priors larger than the rotation period on the temporal variation scales ($\eta_{L_{i}}$), and normal priors on $\eta_{\sigma_{i}}$ centred on the rms of the light curve. In the CHEOPS transit photometry, which had already been flattened (as explained by \citealp{Szabo2021, Szabo2022}), we only used a jitter term added in quadrature to the error bars ($\sigma_{\mathrm{jit,CHEOPS}}$) to take into account the activity not related to stellar rotation or other possible instrumental systematics, as we did for the TESS light curve ($\sigma_{\mathrm{jit,TESS}}$). Since there were data with different exposure times in both CHEOPS and TESS photometry, a different jitter term was considered for each cadence (Table\,\ref{tab:PHdatasets}). We also incorporated an offset parameter for each instrument ($\gamma_{\mathrm{TESS}}$, $\gamma_{\mathrm{CHEOPS}}$).

To subtract stellar activity due to flares, we inspected the TESS and CHEOPS light curves looking for local maxima above the noise level that have the shape of a flare or a combination of several flares. Once the flares were located, we proceeded in two different ways. If the flare is located at a distance larger than 0.5 d from the central time of any transits, we masked it, because we considered that modelling the flare would not recover useful information about the transit fits. On the other hand, flares identified within a 0.5 d margin of the transits were modelled following the approach of \cite{Mendoza2022}. This is an analytical flare model based on that proposed by \cite{Davenport2014}, which convolves two exponential functions and a Gaussian function. This is fully characterised by three parameters: the central temporal position of the maximum, its full width at half maximum, and its amplitude. Therefore, for each transit, we modelled each individual flare by means of an iterative fit to each local maximum that we found (grey line in Fig.\,\ref{fig:LC_TESS}). This process was applied in all the subsequent analyses.

To generate transit models, we used the \texttt{PyTransit} package \citep{Parviainen2015}\footnote{\url{https://github.com/hpparvi/PyTransit}}, whose input parameters are the central time of the transit ($T_c$), orbital period ($P$), planet-star radius ratio ($R_p/R_{\star}$), the orbital semi-major axis divided by the stellar radius ($a/R_\star$), the orbital inclination ($i$), the eccentricity ($e$), and the argument of periastron ($\omega$). The parameter $a/R_\star$ is reparameterised using Kepler's third law; thus, it only depends on the orbital period of the planet and the radius and mass of the star (under the assumption that $M_p/M_\star$\,$\ll$\,1). In addition, we took into account the variations in the shape of the transit due to limb darkening ($u_1$, $u_2$), adopting the parameterisation proposed by \cite{kipping02013} ($q_1$, $q_2$), with different coefficient values for TESS and CHEOPS instruments since they depend on wavelength. The initial values for the stellar limb-darkening coefficients were calculated using the \texttt{LDTk}\footnote{\url{https://github.com/hpparvi/ldtk}} package \citep{Parviainen2015b, Husser2013}.

\begin{figure*}
\includegraphics[width=1\linewidth]{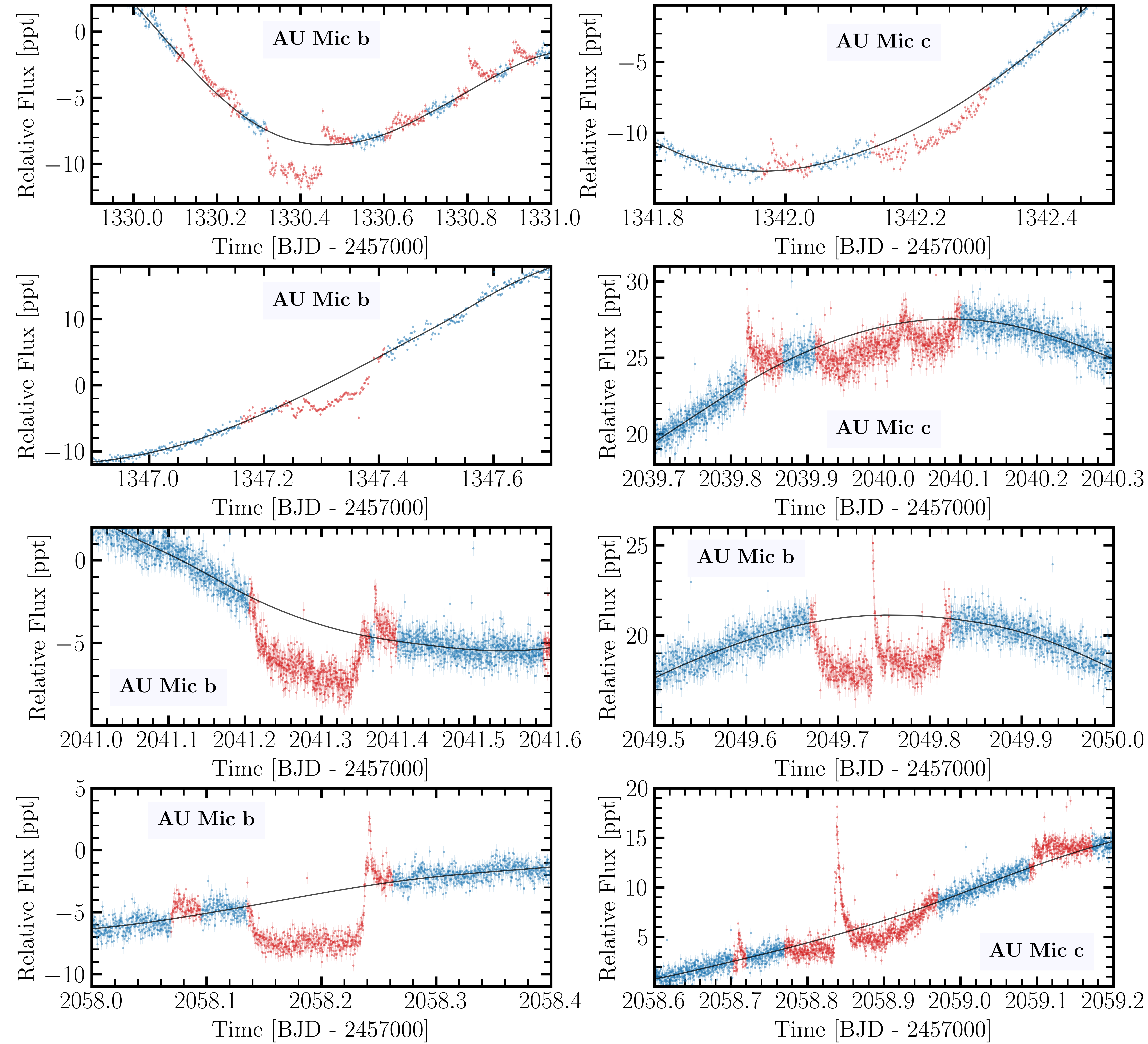}
\caption{TESS light curve of transits in AU\,Mic. The PDCSAP flux used to create a smooth GP model is shown as blue dots. The data considered as flares or transits were masked (red dots) to create the GP model (black line).
\label{fig:LC_GPmodel}}
\end{figure*}

Given the differences in the radii of AU\,Mic\,b and AU\,Mic\,c published in the literature (Sect.\,\ref{sec:intro}), we also created an alternative transit model where each transit (12 transits of planet b and 5 transits of planet c, Table\,\ref{tab:PHdatasets}) is modelled, sharing all the parameters except for the transit depth, which was left as a free parameter for each transit. Since the planets of AU\,Mic also present TTVs of tens of minutes (Sect.\,\ref{sec:pl_d}), the $T_c$ is also a free parameter in each transit, so that we ensure the correct fit of the depth in each transit. In this alternative transit model, we prevented the activity model of the TESS light curve from altering the transit depths by flattening the curve with a very smooth GP model. First, we masked the transits (5 transits of planet b and 3 of planet c in the TESS light curve). Then, we modelled the activity with a GP whose hyperparameters were introduced ad hoc ($\eta_{\sigma_{1}}$\,$=$\,$\eta_{\sigma_{2}}$\,$=$\,$\sigma_{\mathrm{TESS}}$; $\eta_{L_{1}}$\,$=$\,$\eta_{L_{2}}$\,$=$\,100 d; $\eta_{P}$\,$=$\,4.86\,d), visually inspecting that the model is smooth in a region around the transits (Fig.\,\ref{fig:LC_GPmodel}). Finally, we fitted the transit model on the flattened light curves in these close regions around the transit. These prior and posterior results are listed in Table\,\ref{tab:tr_depth} and are discussed in Sect.\,\ref{sec:tr_depth}.

\subsection{RV analysis}
\label{sec:rv}

Unlike photometric observations, for which cadence is very good for sampling transits (of the order of minutes), RV data are usually taken with a daily or weekly cadence. In young and active stars such as AU\,Mic, the cadence of the observations is critical to disentangle the Keplerian signal from the stellar activity. First, because the stellar activity produces prominent variability on timescales close to the orbital periods of the planets that are being measured, and second, because the ratio in amplitude between the stellar activity and the Keplerian amplitudes can be as large as one or two orders of magnitude (with a clear dependence on wavelength; see Sect.\,\ref{sec:chr} for details). All studies dedicated to the mass determination of AU\,Mic\,b and c made use of GPs \citep{klein21, Cale2021, zicher22, Donati2023}, except \cite{klein22}. GP models are often too flexible, even with restricted hyperparameters. The best way to limit their flexibility is to have data with a better observing cadence than the periodicities to model. Table\,\ref{tab:RVdatasets} shows the number of points observed per stellar rotation period, where the CARMENES VIS, CARMENES NIR, HARPS, and SPIRou datasets have more than two points per stellar rotation period. Due to the cadence and the number of points per dataset, these are the most adequate datasets for measuring the masses of transiting planets in the AU\,Mic system. Therefore, we only used these, and we excluded the HIRES, iSHELL, and TRES data from our RV analysis. In total, we have 478 RV measurements, with an average of 4 RV points per stellar rotation.

\subsubsection{Periodogram analysis}
\label{sec:rv_GLS}

\begin{figure*}
\begin{center}
\includegraphics[width=0.8\linewidth]{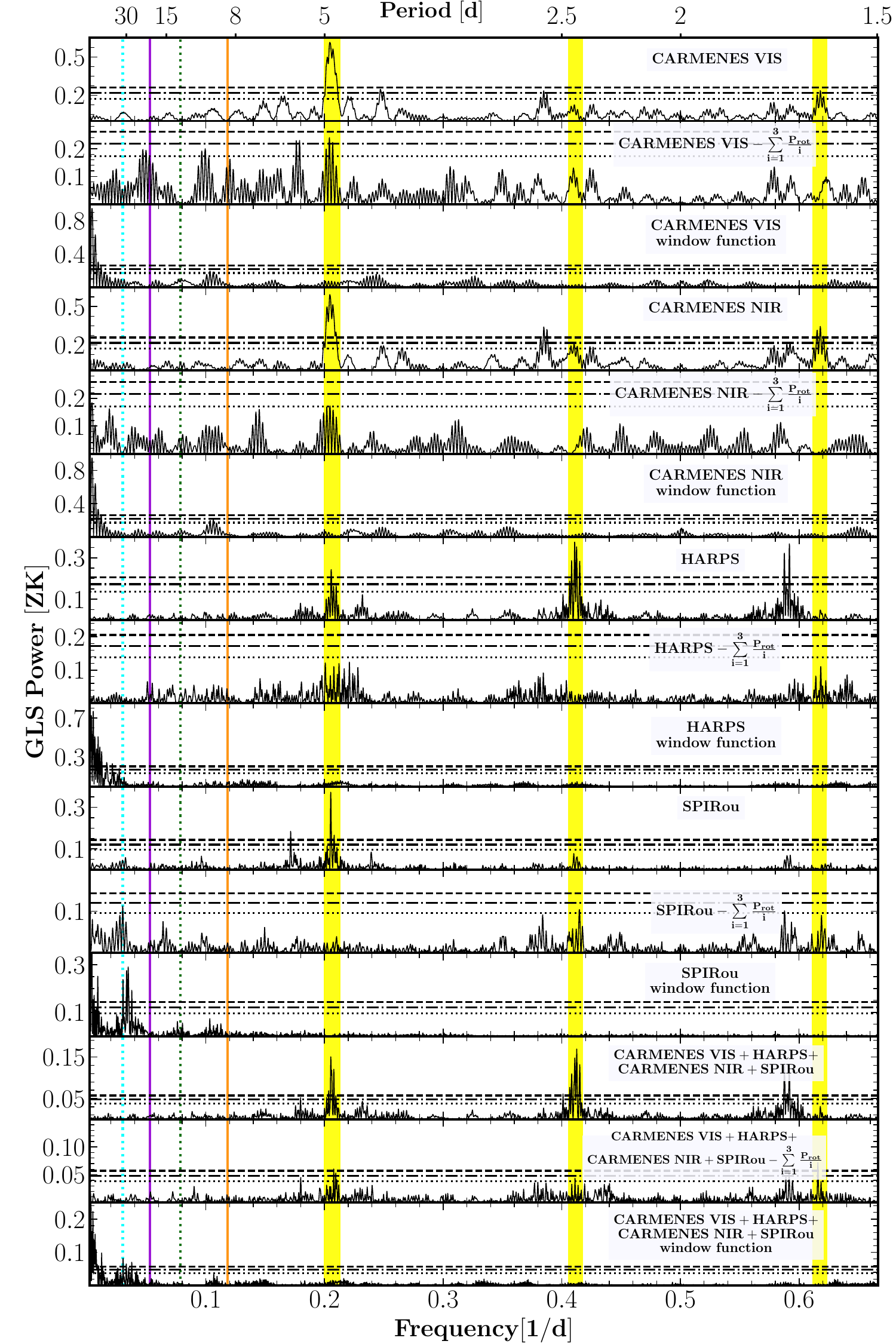}
\caption{GLS periodograms for CARMENES VIS, CARMENES NIR, HARPS, SPIRou datasets, and the combination of all of them. For each dataset, the GLS periodogram, the GLS periodogram of the residuals after applying the pre-whitening, and the GLS periodogram of the window function are shown. The stellar rotation period and its second and third harmonics are shown as vertical yellow bands, centred at 0.206\,d$^{-1}$ (4.9 d), 0.410\,d$^{-1}$ (2.5 d), and 0.617\,d$^{-1}$ (1.6 d). The vertical orange and purple lines indicate the orbital periods of planets b and c. The vertical green and cyan dotted lines indicate the orbital periods of the candidates d and e, respectively. The dashed horizontal black lines of each panel correspond to the FAP levels of 10\%, 1\%, and 0.1\% (from top to bottom).
\label{fig:GLS_RV}}
\end{center}
\end{figure*}

To identify periodic signals in the RV data, we computed the generalised Lomb-Scargle (GLS; \citealp{Zechmeister2009}) periodograms for CARMENES VIS, CARMENES NIR, HARPS, and SPIRou separately, as well as the combined one (Fig.\,\ref{fig:GLS_RV}). In panels one, four, seven, and ten (from top to bottom), we observe that the dominant signal is at $\sim$4.9 days (CARMENES VIS, CARMENES NIR, and SPIRou) or $\sim$2.5 d (HARPS). These first signals correspond to the stellar rotation period and its second harmonic. Panels two, five, eight, eleven, and fourteen (from top to bottom) show the periodograms of the residuals after applying the pre-whitening method, where we iteratively subtracted the signals, by order of significance, related to the rotation period and its second and third harmonics. The signals related to the stellar rotation are thus reduced in significance, although they do not completely disappear. In addition, the window functions show significant peaks at $\sim$256 d (CARMENES VIS, CARMENES NIR), $\sim$180 d (HARPS), or $\sim$30 d (SPIRou). A significant signal is observed solely in the case of the SPIRou dataset, corresponding to the AU\,Mic\,“e” candidate (Sect.\,\ref{sec:pl_e}), while no significant signal of that planet candidate is seen in other periodograms. No signal is observed close to the periods of the transiting planets, AU\,Mic\,b or c, nor to the planet candidate AU\,Mic\, “d”. As the amplitude of the expected Keplerian signals is at least an order of magnitude smaller than the activity, and given the nature of this variability, pre-whitening to subtract the activity may not be the best model method. It is reasonable to assume that one does not find significant signals associated with the planets.

\subsubsection{RV modelling}
\label{sec:rv_pl}

When a planet transits its star, it hides a specific region of the star, producing variations in RV, also known as the RM effect \citep{rossiter, mclaughlin}. AU\,Mic\,b produces RM variations with amplitudes of $\sim$20 m\,s$^{-1}$ at optical wavelengths \citep{Palle2020}, larger than the Keplerian amplitudes measured for that planet (4--10 m\,s$^{-1}$). For AU\,Mic\,c, the expected RM amplitude is $\sim$10 m\,s$^{-1}$ \citep{gaudi07}, comparable to the measurements for the Keplerian signature of that planet (4--13\,m\,s$^{-1}$). We discarded observations collected during transits to avoid RM effects, namely BJD\,=\,2459379.86239, 2459506.73618, 2459887.61698, and 2459904.56330 for transits by AU\,Mic\,b; and 2458700.47479, 2458700.47939, 2458700.47501, 2458700.47901, 2459040.04526, 2459058.95264, 2459888.61540, and 2459907.56879 for transits by AU\,Mic\,c. Therefore, the final datasets contain 86, 87, 117, and 176 RV measurements for CARMENES VIS, CARMENES NIR, HARPS, and SPIRou, respectively, resulting in a total of 466 data points (Table\,\ref{tab:RVdatasets}).

We used GPs to model stellar activity in RV, specifically the quasi-periodic (QP) kernel \citep{aigrain12}, widely used in the field of exoplanets and in previous works on AU\,Mic \citep{plav20, klein21, zicher22, Donati2020}. The hyperparameters of the QP kernel represent the scale of variations in amplitude ($\eta_{\sigma}$), the scale of temporal variations ($\eta_{L}$), the period of the fundamental signal ($\eta_{P}$), and a balance factor ($\eta_{\omega}$) between periodicity and non-periodicity: 

\begin{equation}
k_{\mathrm{QP}}(\tau)  =  \eta_{\sigma}^2 \exp \left[ -\frac{\tau^2}{2\eta_{L}^2} -\frac{\sin^2{ \left( \frac{\pi \tau}{\eta_{P}} \right)}}{2\eta_{\omega}^2} \right]. \\
\label{eq:qp}
\end{equation}

As for the photometric analysis, we constrained the hyperparameters of the GPs according to the type of stellar activity we wanted to model; that is, the variations on timescales of the rotation period. Another higher frequency activity and/or systematics is captured by a jitter term ($\sigma_{\mathrm{jit,RV}}$) added to the error bars, independent for each dataset. To generate Keplerian models, we used \texttt{RadVel}\footnote{\url{https://github.com/California-Planet-Search/radvel}} \citep{fultonradvel}, which includes $T_c$, $P$, $e$, $\omega$, and the stellar RV amplitude due to the planet ($K$). To find the best model, we applied the Markov chain Monte Carlo (MCMC) method, sampling the parameter space with an affine-invariant ensemble sampler \citep{goodman10}, which is implemented in the \texttt{emcee} code \citep{emcee}. However, our RV analysis was also tested with a nested sampler and the Bayesian inference tool \texttt{MultiNest} \citep{Feroz2008} with the \texttt{pyMultiNest} wrapper \citep{Buchner2014}. Therefore, our RV model includes an activity model generated with GP (QP kernel) plus a jitter term and a circular Keplerian model that assumes the existence of AU\,Mic\,b and c. In this model, we sampled the hyperparameter $\eta_{\sigma, \mathrm{RV}}$ with a Gaussian prior centred on the rms of each dataset, while we employed uniform priors for $\eta_{L}$, $\eta_{P}$, and $\eta_{\omega}$. In addition, we considered an offset term ($\gamma_{\mathrm{RV}}$) for each dataset. We used normal priors for $T_c$ and $P$ and a uniform positive prior for $K$.

We first studied the amplitudes of the planets individually for each dataset (Table\,\ref{tab:rv_datasets}). The CARMENES VIS, HARPS, and SPIRou data show similar amplitudes for planet b ($\sim$2--4 m\,s$^{-1}$), while CARMENES NIR gives a value of the amplitude that is significantly larger for the planet ($\sim$14 m\,s$^{-1}$). For planet c, we obtain amplitudes between 3--8 m\,s$^{-1}$, all of which are compatible within the error bars. Therefore, the amplitude of planet b seems to be around 2--4 m\,s$^{-1}$, while the amplitude of planet c is comparable to planet b or higher. In none of the cases did we obtain a 3$\sigma$ detection for the planets, including the SPIRou data, which is in disagreement with the recent findings by \cite{Donati2023}. Their amplitudes for an equivalent analysis of the SPIRou data alone yielded 4.1$^{+1.8}_{-1.2}$ and 4.0$^{+1.7}_{-1.2}$ m\,s$^{-1}$; i.e.\ barely a $>$3$\sigma$ detection. The difference between our results and theirs likely comes from the use of different priors. \cite{Donati2023} used a modified Jeffreys prior on the Keplerian amplitudes, leading to an asymmetry in their posterior distribution and a difference in significance. Instead, we considered a uniform prior and yielded a symmetric posterior. Since the SPIRou dataset has the largest number of RVs and the lowest rms, in order to check that the results are not biased by these data, we analysed the rest of the data together to check whether the planet signals were consistent with the SPIRou results alone. To combine the datasets, all parameters are shared, except $\eta_{\sigma, \mathrm{RV}}$, $\sigma_{\mathrm{jit}, \mathrm{RV}}$, and $\gamma_{\mathrm{RV}}$, which are independent for each instrument. Combining the datasets from CARMENES VIS, CARMENES NIR, and HARPS, we obtained amplitudes of 3.2$^{+1.4}_{-1.4}$ and 4.7$^{+1.3}_{-1.3}$ m\,s$^{-1}$ for the planets AU\,Mic\,b and c, respectively (Table\,\ref{tab:rv_datasets}). These amplitudes were consistent with those obtained by us using only the SPIRou data and with those of \citet{Donati2023}. Finally, the combination of the four datasets yielded planetary amplitudes of 3.6$^{+1.1}_{-1.1}$ for AU\,Mic\,b and 4.3$^{+1.0}_{-1.0}$ m\,s$^{-1}$ for AU\,Mic\,c, which are similar to those obtained when using only the SPIRou dataset, but with an error bar that is about 30\% smaller (Table\,\ref{tab:rv_datasets}).

\renewcommand{\arraystretch}{1.25} 
\begin{table*}
\caption{Comparison of amplitudes of transiting planets of AU\,Mic derived from the RV analysis.}
\label{tab:rv_datasets}
\begin{center}
\begin{tabular}{l l c c}
\hline
\hline
Dataset & Activity model & $K^{b}$[m\,s$^{-1}$] & $K^{c}$[m\,s$^{-1}$] \\
\hline
CARMENES VIS & GP$_{\mathrm{QP}}$ + $\sigma_{\mathrm{jit,}}$ & 2.3$^{+2.0}_{-1.5}$ & 3.2$^{+1.8}_{-1.7}$\\
CARMENES NIR & GP$_{\mathrm{QP}}$ + $\sigma_{\mathrm{jit}}$ & 14.2$^{+4.8}_{-5.0}$ & 7.8$^{+4.4}_{-4.0}$\\
HARPS & GP$_{\mathrm{QP}}$ + $\sigma_{\mathrm{jit}}$ & 2.6$^{+2.1}_{-1.7}$ & 5.7$^{+2.3}_{-2.3}$\\
SPIRou & GP$_{\mathrm{QP}}$ + $\sigma_{\mathrm{jit}}$ & 3.7$^{+1.7}_{-1.7}$ & 3.8$^{+1.6}_{-1.6}$\\
\hline
CARMENES VIS + CARMENES NIR + HARPS & GP$_{\mathrm{QP}}$ + $\sigma_{\mathrm{jit}}$ & 3.2$^{+1.4}_{-1.4}$ & 4.7$^{+1.3}_{-1.3}$\\
\hline
CARMENES VIS + CARMENES NIR + HARPS + SPIRou & GP$_{\mathrm{QP}}$ + $\sigma_{\mathrm{jit}}$ & 3.6$^{+1.1}_{-1.1}$ & 4.3$^{+1.0}_{-1.0}$\\
\hline
\end{tabular}
\tablefoot{In the model name, $\sigma_{\mathrm{jit}}$ refers to a jitter term added in quadrature to the RV error bars. All models assumed circular orbits. The amplitudes ($K$) are given with their 1$\sigma$ uncertainties.}\\
\end{center}
\end{table*}

\subsubsection{Dependence of Keplerian amplitudes on the activity models}
\label{sec:rv_models}

Although in our analysis we modelled the stellar activity using GPs, we investigated how a different activity model may affect the planetary amplitudes and the significance of the planet signal detections. To address this issue, we compared three different activity models together with the Keplerian model for the planets. First, a model that only includes a jitter term. Second, a model that includes three sinusoidal functions whose periods are centred on the stellar rotation period and its second and third harmonics (based on the periodograms analysed in Sect.\,\ref{sec:rv_GLS}), and third, the aforementioned GP model. In addition, these activity models were also contrasted with a model that only includes activity, with the aim of determining whether these planets are detected significantly in our RV data. We used the Bayesian evidence ($\ln \mathcal{Z}$) calculated following the method by \citet{diaz16} to compare these models. We used the criteria proposed by \cite{trotta2008}, where the model with the higher log-evidence is strongly favoured if the absolute difference $|\Delta \ln \mathcal{Z}|$ is greater than five. If 2.5\,$<$\,$|\Delta \ln \mathcal{Z}|$\,$<$\,5, the evidence in favour of a given model is moderate. If 1\,$<$\,$|\Delta \ln \mathcal{Z}|$\,$<$\,2.5, the evidence is weak, and indistinguishable if $|\Delta \ln \mathcal{Z}|$\,$<$\,1. This comparison is shown in Table\,\ref{tab:logZ}. Modelling the activity with GPs is strongly favoured over modelling with sinusoidal functions or only with a jitter term. This best activity model with GP also produces the most precise determination of the planet amplitudes. The other two activity models, although they do not result in a 3$\sigma$ detection, give similar amplitudes that agree within the error bar.

\renewcommand{\arraystretch}{1.3} 
\begin{table*}
\caption{Comparison of activity models from the RV analysis of the AU\,Mic system using the difference between Bayesian log-evidences ($\Delta \ln \mathcal{Z}$).}
\begin{center}
\begin{tabular}{cccc|ccc|ccc}
\hline
\hline
 & \multicolumn{9}{c}{Activity model}\\
\cline{2-10}
 & \multicolumn{3}{c}{$\sigma_{\mathrm{jit}}$} & \multicolumn{3}{|c}{3 sin ($P_{1}$, $P_{2}$, $P_{3}$\,$\sim$4.9, 2.4, 1.6 d) + $\sigma_{\mathrm{jit}}$} & \multicolumn{3}{|c}{GP$_{\mathrm{QP}}$ + $\sigma_{\mathrm{jit}}$}\\
\cline{2-10}
 Planets & $K^b$[m\,s$^{-1}$] & $K^c$[m\,s$^{-1}$] & $\Delta \ln \mathcal{Z} $ & $K^b$[m\,s$^{-1}$] & $K^c$[m\,s$^{-1}$] & $\Delta \ln \mathcal{Z}$ & $K^b$[m\,s$^{-1}$] & $K^c$[m\,s$^{-1}$] & $\Delta \ln \mathcal{Z}$\\
\hline
0 & ... & ... & --2660.2 & ... & ... & --2522.0 & ... & ... & --2226.5\\
2 & 5.3$^{+3.3}_{-3.0}$ & 3.6$^{+2.8}_{-2.3}$ & --2662.7 & 5.5$^{+2.7}_{-2.6}$ & 4.4$^{+2.2}_{-2.3}$ & --2543.5 & \textbf{3.6$^{+1.1}_{-1.1}$} & \textbf{4.3$^{+1.0}_{-1.0}$} & \textbf{--2218.9}\\
\hline
\end{tabular}
\tablefoot{In the model name, $\sigma_{\mathrm{jit}}$ refers to a jitter term added in quadrature to the RV error bars. All models assume circular orbits. The amplitudes are given with their 1$\sigma$ uncertainty. '3 sin' refers to three sinusoidal functions with their periods. The data used for the comparison are composed of the combination of RVs from CARMENES VIS, CARMENES NIR, HARPS, and SPIRou. The final adopted RV model in this paper is highlighted in bold.}\\
\label{tab:logZ}
\end{center}
\end{table*}

\subsubsection{The planet candidate AU\,Mic\,d}
\label{sec:pl_d}

\citet{Martioli2021}, \citet{Szabo2021}, \citet{Gilbert2022}, and \citet{zicher22} estimated TTVs of a few minutes. Later, \citet{Szabo2022} observed peak-to-peak TTVs of over 23 minutes combining photometric transits from TESS, CHEOPS, and Spitzer. The analysis of \citet{Wittrock2022}, using several ground-based transits, went in the same direction, indicating the presence of a third planet with an orbital period between AU\,Mic\,b and AU\,Mic\,c. Finally, \citet{Wittrock2023} presented a statistical 'validation' of a third non-transiting planet in the system with an orbital period of 12.73596 d and a mass estimated from the TTV of 1.1\,$\pm$\,0.5 M$_{\oplus}$. The expected Keplerian amplitude induced on the star by a $\sim$1 M$_{\oplus}$-mass planet at this orbital period is less than 1 m\,s$^{-1}$ and, therefore, beyond the reach of current RV detections in such active systems. 

We explored the presence of this potential planet and its influence on the amplitudes of the known transiting planets in our RV model (Sect.\ \ref{sec:rv_models}). The comparison between models with two and three planets is given in Table\,\ref{tab:pl_d}. The $\Delta \ln \mathcal{Z}$ values indicate that the model with two planets is favoured over the model with three planets. Therefore, there is no clear detection of a planet d in our RV data. Additionally, the amplitudes of the transiting planets b and c are not affected by the inclusion of this Earth-mass planet, for which we set an upper mass limit of $M_{p}^{[d]}\sin{i}$\,=\,8.6\,M$_{\oplus}$ at the 3$\sigma$ level of confidence. Our result is compatible with the mass upper limit of 11.4 M$_{\oplus}$ determined by \citet{Donati2023} and with the mass of 1.053\,$\pm$\,0.511 M$_{\oplus}$ determined using TTVs by \citet{Wittrock2023}.

\renewcommand{\arraystretch}{1.3} 
\begin{table}
\caption{Comparison between Keplerian model with two planets and Keplerian model with three planets.}\label{tab:pl_d}
\centering
\begin{tabular}{c c c c c}
\hline\hline
Planets & $K^b$[m\,s$^{-1}$] & $K^c$[m\,s$^{-1}$] & $K^d$[m\,s$^{-1}$] & $\Delta \ln \mathcal{Z}$\\
\hline
2 & 3.6$^{+1.1}_{-1.1}$ & 4.3$^{+1.0}_{-1.0}$ & ... & --2218.4\\
3 & 3.5$^{+1.1}_{-1.1}$ & 4.2$^{+1.0}_{-1.0}$ & 0.7$^{+0.8}_{-0.5}$ (3.1) & --2222.1\\
\hline
\end{tabular}
\tablefoot{The 99.7th percentile of the Keplerian amplitude for the AU\,Mic\,'d' candidate is provided in parentheses. The activity model used for this comparison was the GP model.}\\
\end{table}

\subsubsection{The planet candidate AU\,Mic\,e}
\label{sec:pl_e}

\begin{figure}
\includegraphics[width=9.cm]{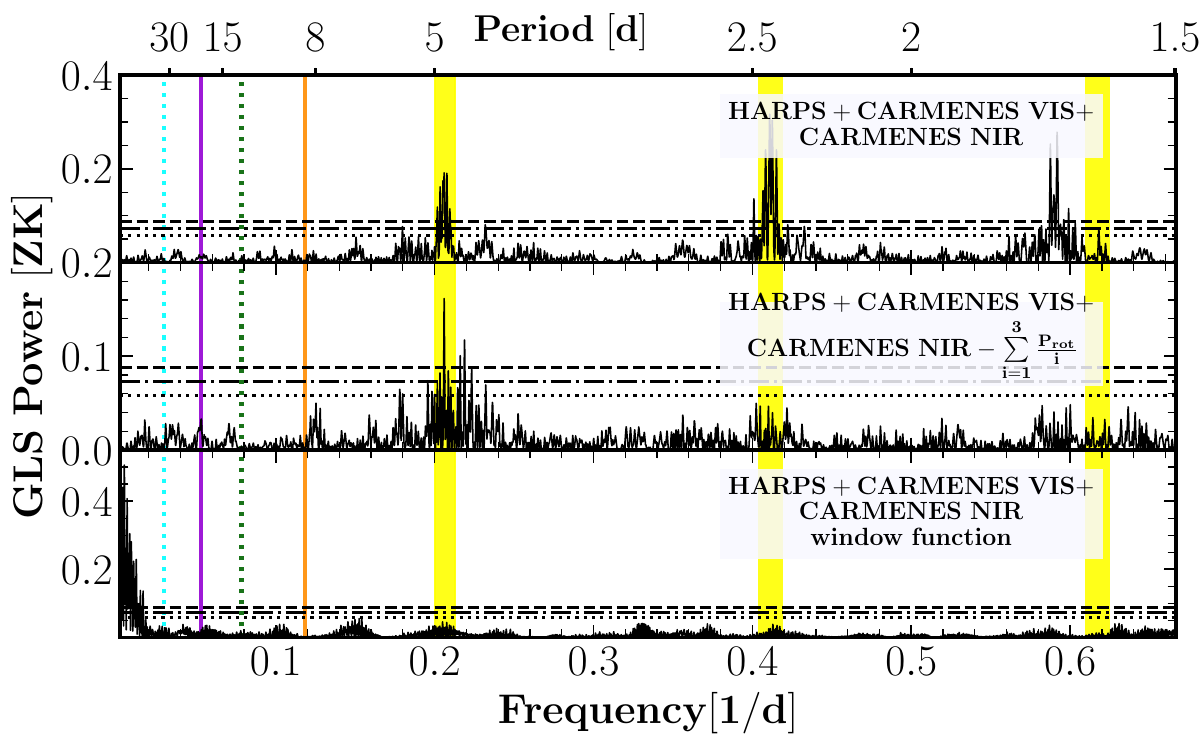}
\caption{GLS periodograms of combination of CARMENES VIS, CARMENES NIR, and HARPS datasets. The top, middle, and bottom panels show the GLS periodogram, the GLS periodogram of the residuals after applying the pre-whitening method, and the GLS periodogram of the window function, respectively. The stellar rotation period and its second and third harmonics are shown as vertical yellow bands, centred at 0.206\,d$^{-1}$ (4.9 d), 0.410\,d$^{-1}$ (2.5 d), and 0.617\,d$^{-1}$ (1.6 d). The vertical orange and purple lines indicate the orbital period of planets b and c. The vertical green and cyan dotted lines indicate the orbital period of candidates d and e, respectively. The dashed horizontal black lines corresponds to the FAP levels of 10\%, 1\%, and 0.1\% (from top to bottom).
\label{fig:GLS_33d}}
\end{figure}

\begin{figure}
\includegraphics[width=9.cm]{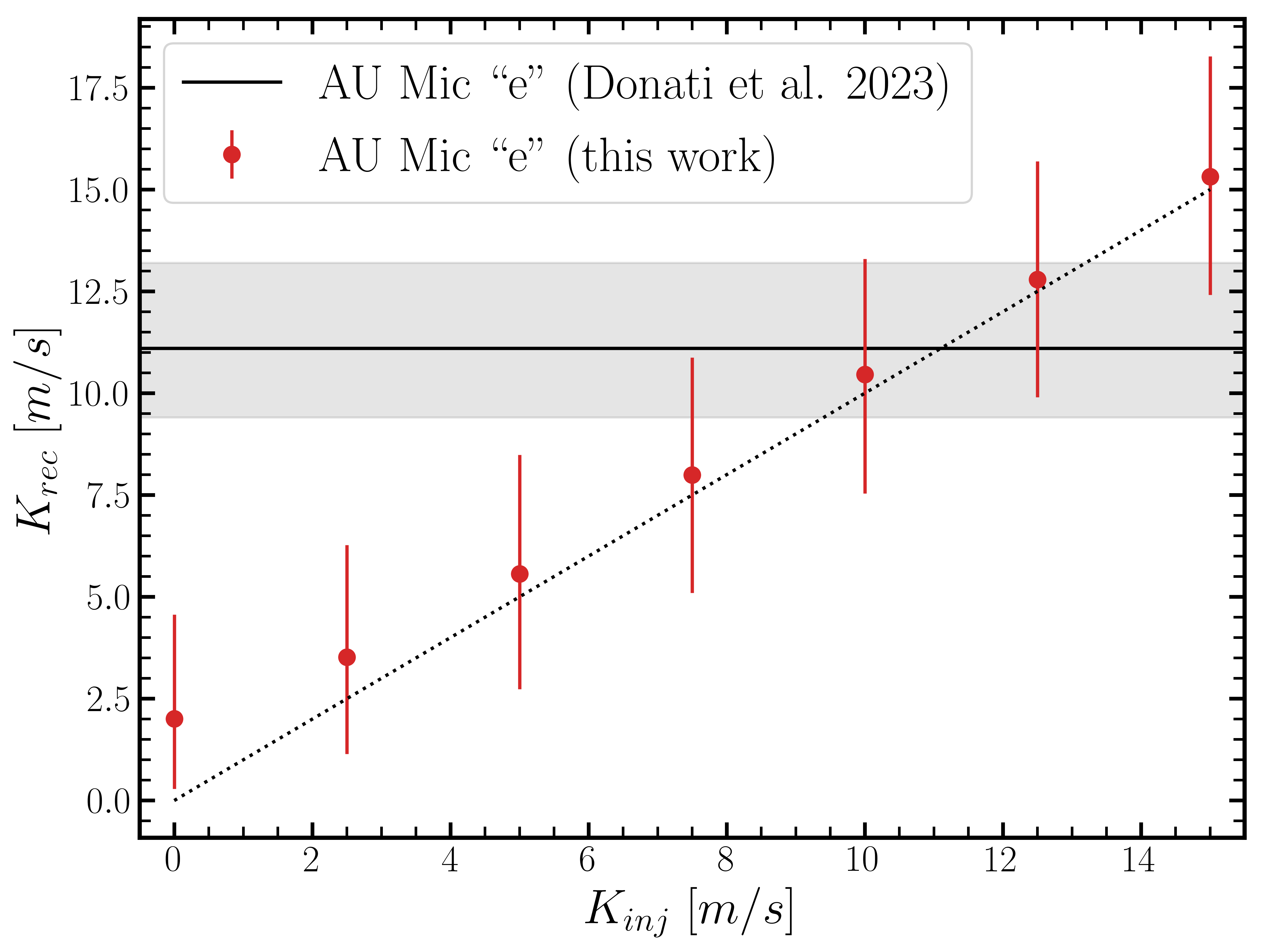}
\caption{Injection-recovery test for AU\,Mic\,e candidate (red dots). The horizontal black line and band represent the measurement and its 1$\sigma$ uncertainty obtained by \cite{Donati2023}.
\label{fig:inj-rec_d}}
\end{figure}

\cite{Donati2023} found a significant peak in the periodogram of the RV residuals in their analysis of the SPIRou data and proposed this signal as a new planet candidate, namely AU\,Mic\,'e'. This signal has a period of 33.4 d and an amplitude of 11.1$_{-1.7}^{+2.1}$ m\,s$^{-1}$ after fitting Keplerian models to planets b and c. We also observed this signal at 33.4 d, with two adjacent signals at 30.6 and 36.3 d in our analysis of the periodogram of the SPIRou data after subtracting the signals associated with the stellar rotation period through the pre-whitening method (Fig.\,\ref{fig:GLS_RV}, eleventh panel). In the periodogram of the window function, the maximum and significant peaks have periods of 28.9, 30.1, and 33.2 d, probably related with the monthly window. On the contrary, in the combination of the CARMENES VIS, CARMENES NIR, and HARPS data (Fig.\,\ref{fig:GLS_33d}, second panel) no significant signal is observed at $\sim$33 d. To test whether it would be possible to detect a similar signal in the datasets that do not include the SPIRou data, we performed an injection-recovery test (Fig.\,\ref{fig:inj-rec_d}). We injected circular Keplerians whose amplitudes vary from 0 to 15 m\,s$^{-1}$, and $P$ and $T_c$ are those obtained by \cite{Donati2023}. For the range of amplitudes similar to that obtained by \cite{Donati2023} (10--12.5 m\,s$^{-1}$), we recovered 10.5\,$\pm$\,2.9 and 12.8\,$\pm$\,2.9 m\,s$^{-1}$, respectively. Therefore, we can rule out amplitudes between 10 and 12.5 m\,s$^{-1}$ with a confidence level of between 3.5 and 4.4 $\sigma$ using the combination of CARMENES VIS, CARMENES NIR, and HARPS datasets. As a result, we questioned the Keplerian nature of this signal. In Appendix\,\ref{sec:ap_33d}, we present a study of how the period of the $\sim$33.4 d signal in the SPIRou data changes according to the different observation campaigns.

\subsection{Joint-fit analysis}
\label{sec:joint-fit}

\begin{figure*}
\includegraphics[width=1\linewidth]{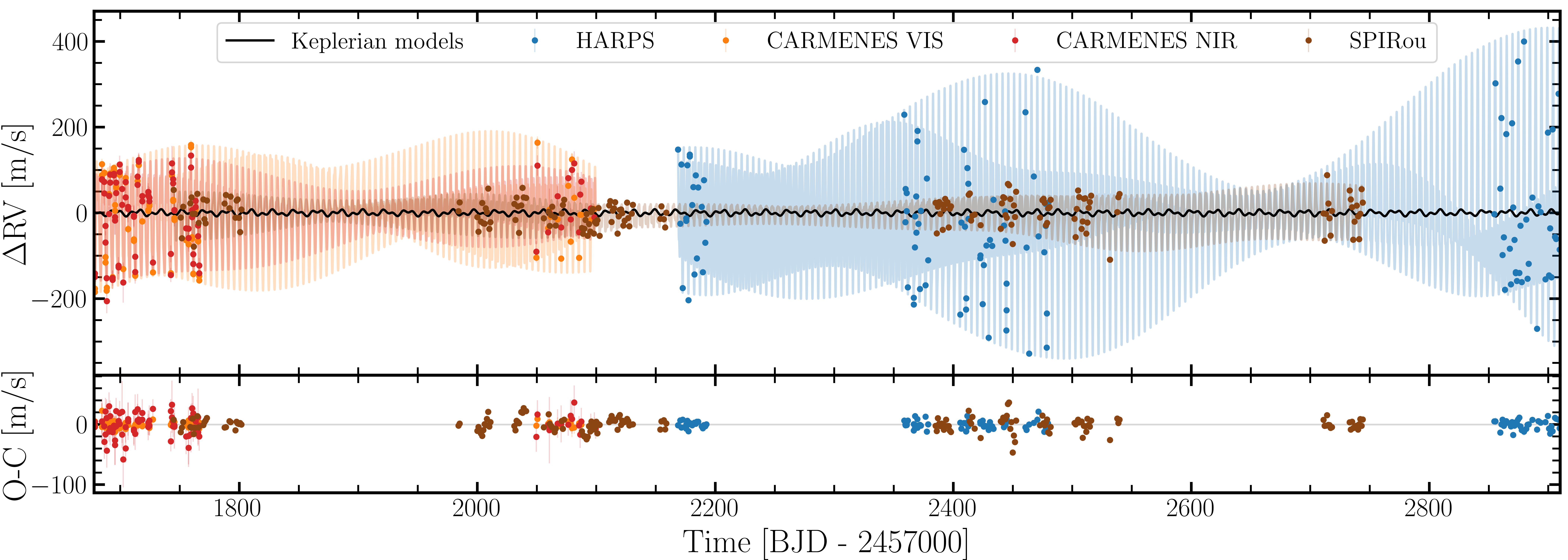}
\caption{CARMENES VIS, CARMENES NIR, HARPS, and SPIRou RV data for AU\,Mic. 
\textit{Top panel}: Individual activity models (coloured lines) and combined Keplerian models for the planets (black line). \textit{Bottom panel}: Residuals for the best-fit.
\label{fig:RV_curve}}
\end{figure*}

\begin{figure}
\includegraphics[width=1.0\linewidth]{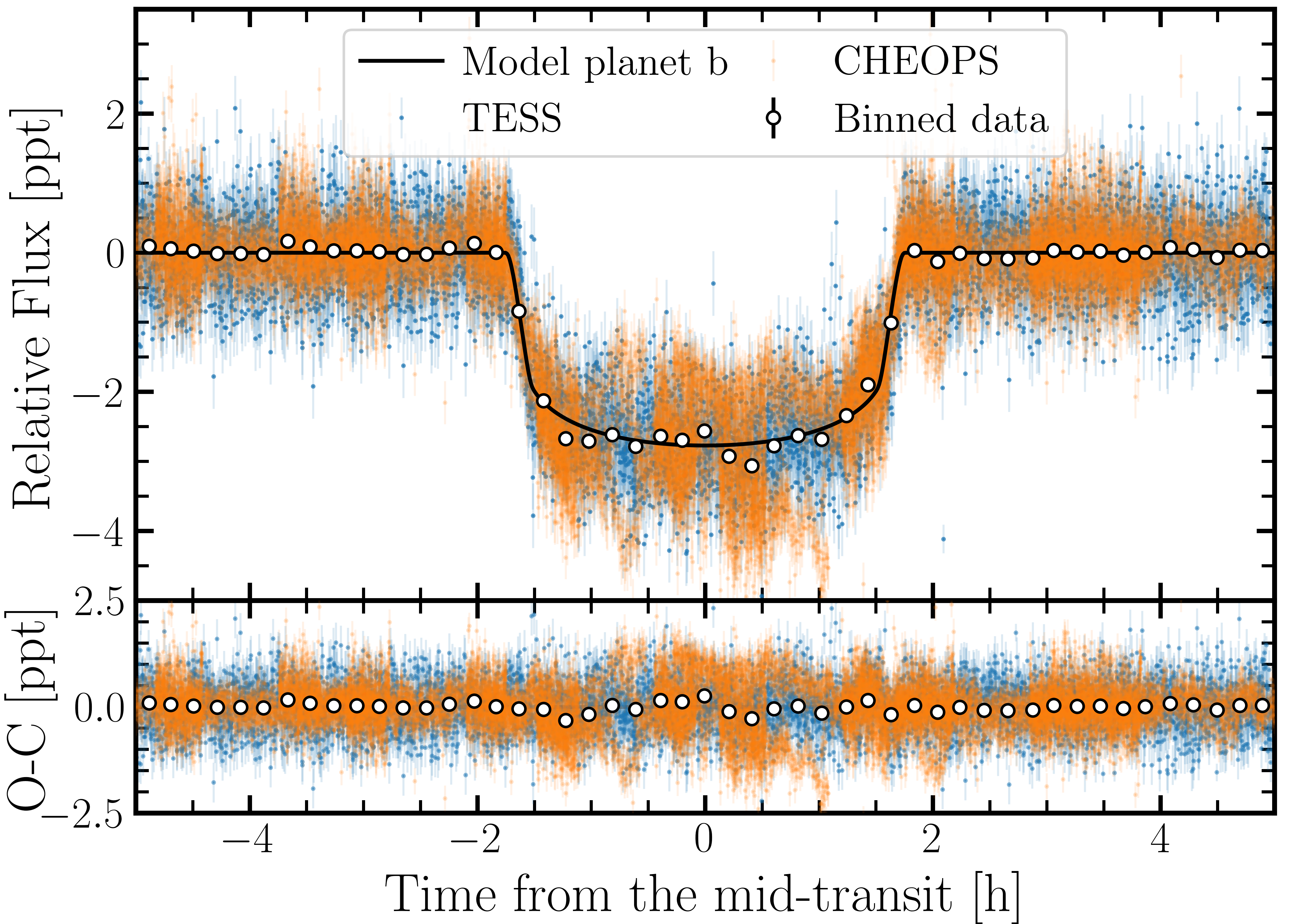}
\includegraphics[width=1.0\linewidth]{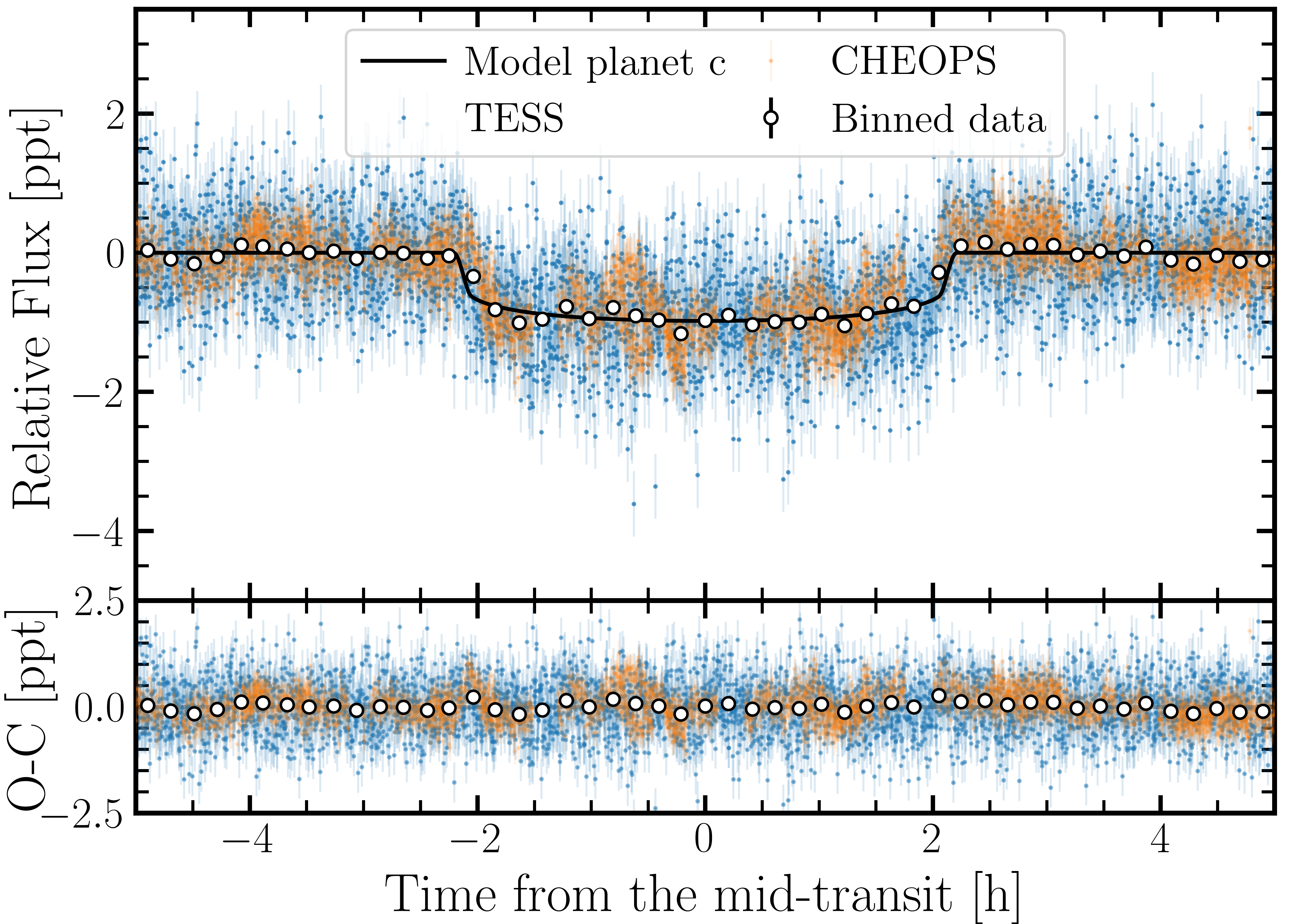}
\caption{Phase-folded light curves around transits of AU\,Mic\,b (top panel) and AU\,Mic\,c (bottom panel) for the alternative transit model. In each sub-panel, the TESS (blue dots) and CHEOPS data (orange dots) are shown, along with the binned data (white dots), the best TESS transit-fit model (black line) in the top, and the residuals from the best-fit in the bottom.
\label{fig:PH_folded}}
\end{figure}

\begin{figure*}
\includegraphics[width=0.49\linewidth]{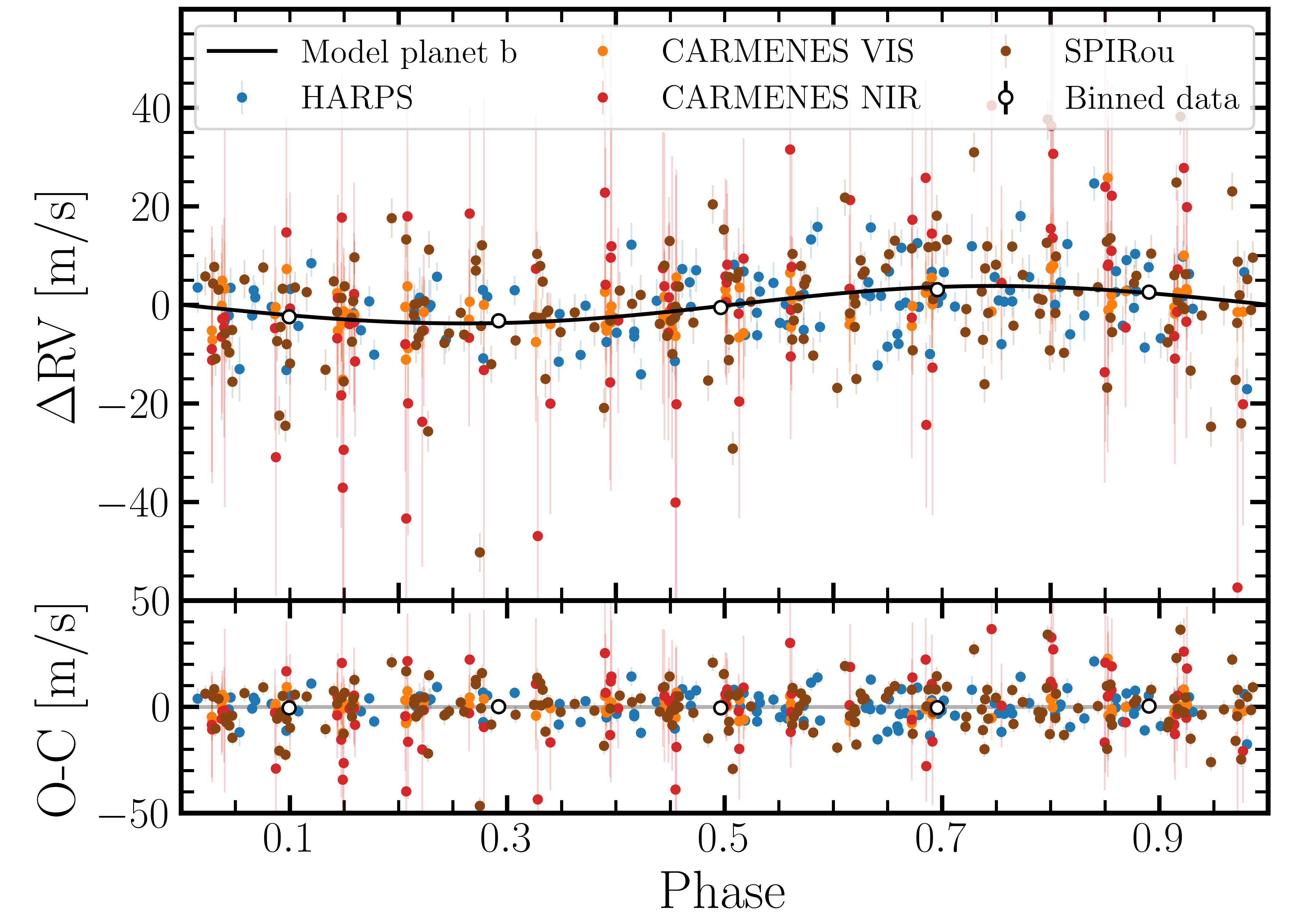}
\includegraphics[width=0.49\linewidth]{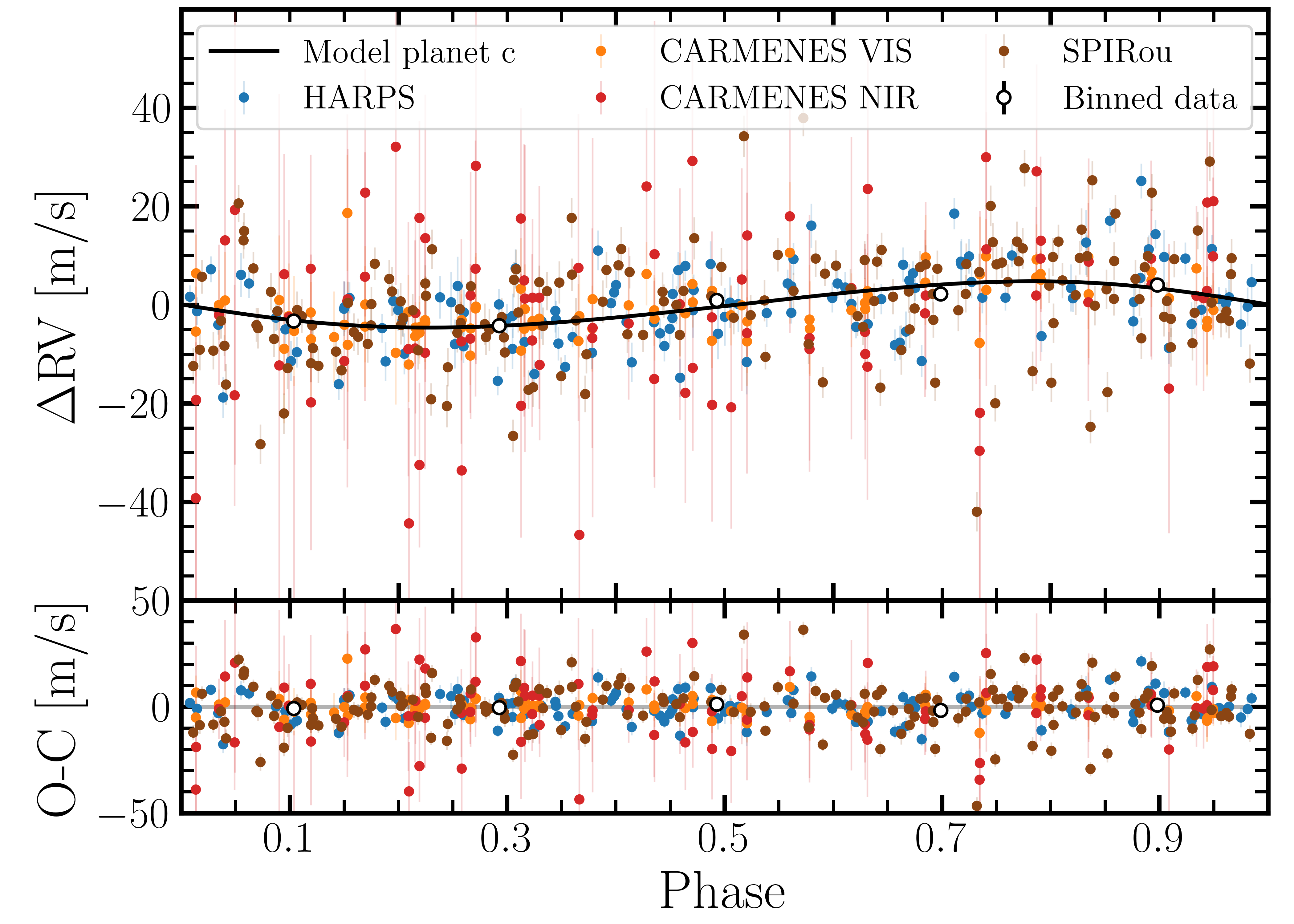}
\includegraphics[width=0.49\linewidth]{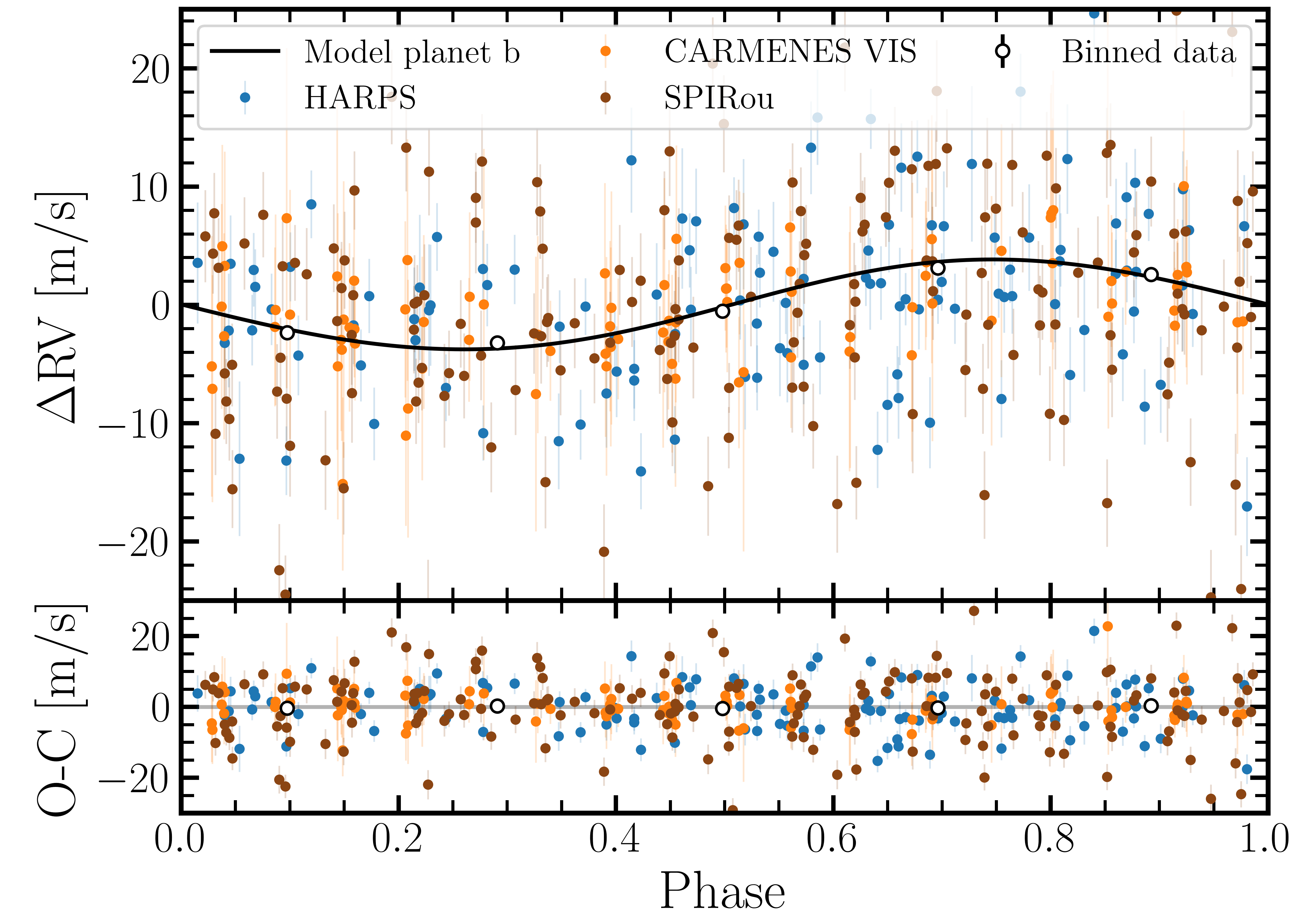}
\includegraphics[width=0.49\linewidth]{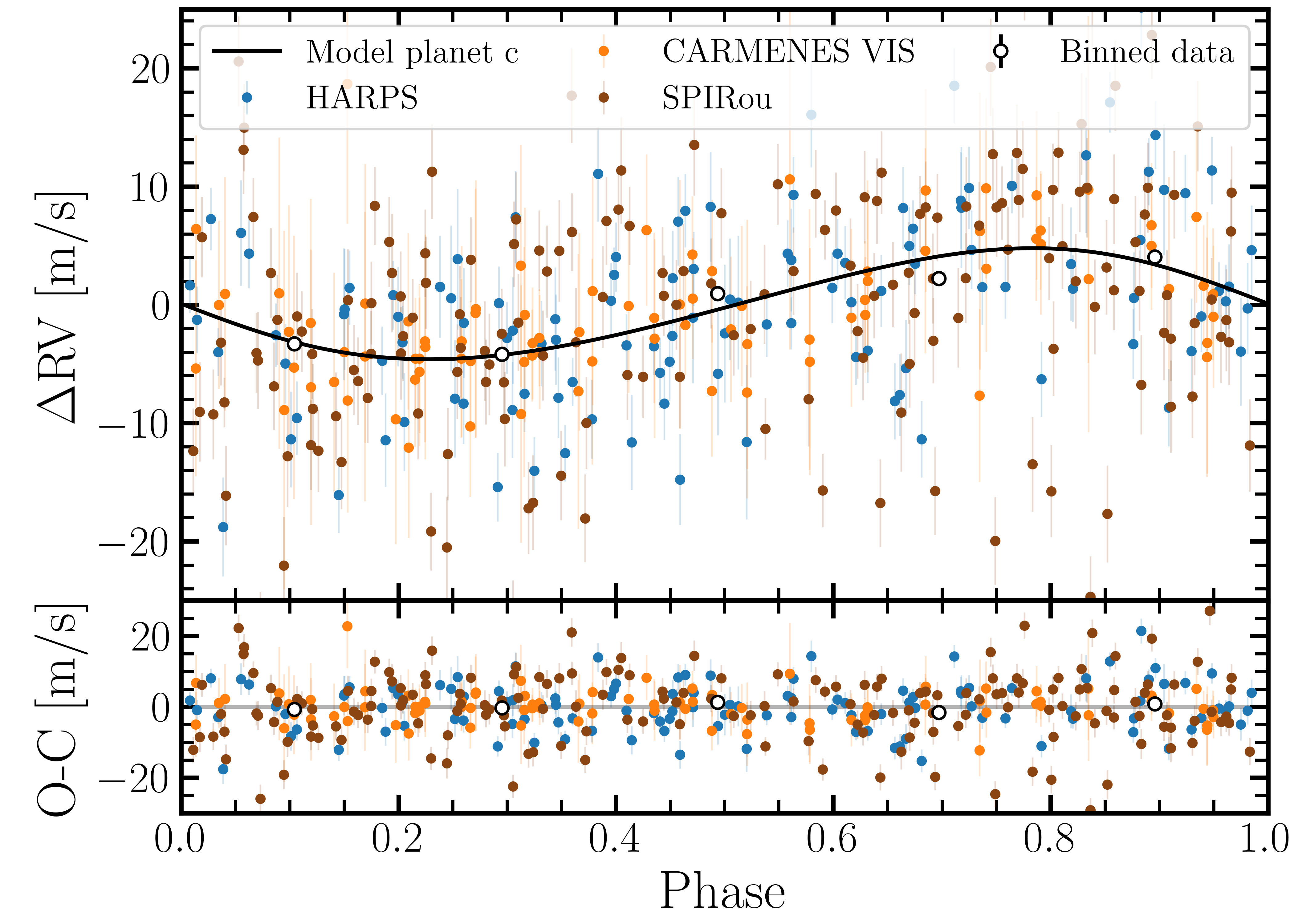}
\caption{RV phase-folded to periods of AU\,Mic\,b (left panels) and AU\,Mic\,c (right panels). The HARPS, CARMENES VIS, CARMENES NIR (only in the top panels), and SPIRou RV data are shown as blue, green, orange, and red dots, respectively. The binned data (white dots) and the Keplerian model of the joint fit (black line). The bottom panels show the residuals for the best fit. Due to the selected vertical scale, a few points are not shown in the figure.
\label{fig:RV_folded}}
\end{figure*}

We incorporated all datasets into a unified joint fit, combining the TESS and CHEOPS space-based photometry and the RV time series to derive more precise parameters for the planets in the AU\,Mic system. This global fit used a transit model for the TESS and CHEOPS datasets (phased-folded transits in Fig.\,\ref{fig:PH_folded}) and a Keplerian model for the CARMENES VIS, CARMENES NIR, HARPS, and SPIRou datasets (phase-folded RVs in Fig.\,\ref{fig:RV_folded}) to determine the planetary parameters. We shared the planetary parameters of $T_c$ and $P$ --both as normal priors-- among all the datasets. We set uniform priors for the $R_p/R_{\star}$ and $b$. We shared the RV amplitude $K$ with a uniform prior among all the RV datasets. We also analysed non-circular orbits with uniform priors for $\sqrt{e}\sin{\omega}$ and $\sqrt{e}\cos{\omega}$. These parameters represent a reparametrisation of $e$ and $\omega$ based on the approach introduced by \cite{anderson2011}. Moreover, we included a stellar activity model via GPs to model both the photometric TESS light curves using the dSHO kernel (Fig.\,\ref{fig:LC_TESS}) and the RV dataset with a QP kernel (Fig.\,\ref{fig:RV_curve}). The hyperparameter $\eta_P$ was shared between TESS and the RV datasets, while $\eta_L$ and $\eta_\omega$ were shared among the RV datasets, all of them with uniform priors. However, normal priors were assigned to the covariance amplitude hyperparameters, centred on the standard deviation of each dataset. We refer the reader to Table \ref{tab:joint-fit} for a detailed overview of the prior and posterior results of the joint fit.

\begin{table*}
\caption{Prior and posterior parameters of joint fit for AU\,Mic\,b and c.}\label{tab:joint-fit}
\centering
\begin{tabular}{lccc}
\hline\hline
Parameter & Prior & Posterior ($e$\,$=$\,0) & Posterior ($e$, $\omega$ free)\\
\hline
$T_{c}^{b}$[BJD] & $\mathcal{N}$(2458330.387, 0.05) & 2458330.350825$^{+0.000497}_{-0.000502}$ & 2458330.350878$^{+0.000535}_{-0.000503}$\\
$P^{b}$[d] & $\mathcal{N}$(8.463, 0.05) & 8.463446$^{+0.000005}_{-0.000005}$ & 8.463446$^{+0.000005}_{-0.000005}$\\
$R_{p}^{b}/R_{\star}$ & $\mathcal{U}$(0, 1) & 0.0495$^{+0.0001}_{-0.0001}$ & 0.0499$^{+0.0004}_{-0.0004}$\\
$b^{b}$ & $\mathcal{U}$(0, 1) & 0.451$^{+0.002}_{-0.002}$ & 0.502$^{+0.044}_{-0.048}$\\
$K^{b}$[m s$^{-1}$] & $\mathcal{U}$(0, 50) & 3.712$^{+1.066}_{-1.063}$ & 3.800$^{+1.093}_{-1.095}$\\
$(\sqrt{e}\sin\omega)^{b}$ & $\mathcal{U}$(-1, 1) & ... & --0.138$^{+0.110}_{-0.082}$\\
$(\sqrt{e}\cos\omega)^{b}$ & $\mathcal{U}$(-1, 1) & ... & 0.074$^{+0.243}_{-0.284}$\\
\hline
$T_{c}^{c}$[BJD] & $\mathcal{N}$(2458342.224, 0.05) & 2458342.223487$^{+0.001080}_{-0.001116}$ & 2458342.223325$^{+0.001085}_{-0.001105}$\\
$P^{c}$[d] & $\mathcal{N}$(18.859, 0.05) & 18.859018$^{+0.000023}_{-0.000022}$ & 18.859023$^{+0.000022}_{-0.000024}$\\
$R_{p}^{c}/R_{\star}$ & $\mathcal{U}$(0, 1) & 0.0300$^{+0.0003}_{-0.0003}$ & 0.0291$^{+0.0005}_{-0.0003}$\\
$b^{c}$ & $\mathcal{U}$(0, 1) & 0.544$^{+0.007}_{-0.006}$ & 0.259$^{+0.203}_{-0.178}$\\
$K^{c}$[m s$^{-1}$] & $\mathcal{U}$(0, 50) & 4.284$^{+0.993}_{-0.985}$ & 4.696$^{+1.031}_{-1.054}$\\
$(\sqrt{e}\sin\omega)^{c}$ & $\mathcal{U}$(-1, 1) & ... & 0.293$^{+0.088}_{-0.164}$\\
$(\sqrt{e}\cos\omega)^{c}$ & $\mathcal{U}$(-1, 1) & ... & 0.071$^{+0.333}_{-0.349}$\\
\hline
$\gamma_{\mathrm{TESS}}$[ppt] & $\mathcal{U}$(--3$\sigma_{\mathrm{TESS}}$, 3$\sigma_{\mathrm{TESS}}$) & 6.5$^{+3.2}_{-3.2}$ & 6.6$^{+3.0}_{-3.1}$\\
$\gamma_{\mathrm{CHEOPS}}$[ppt] & $\mathcal{U}$(--3$\sigma_{\mathrm{CHEOPS}}$, 3$\sigma_{\mathrm{CHEOPS}}$) & --0.1$^{+0.1}_{-0.1}$ & --0.1$^{+0.1}_{-0.1}$\\
$\gamma_{\mathrm{HARPS}}$[m s$^{-1}$] & $\mathcal{U}$(--3$\sigma_{\mathrm{HARPS}}$, 3$\sigma_{\mathrm{HARPS}}$) & 27.0$^{+43.2}_{-42.7}$ & 29.7$^{+43.1}_{-43.0}$\\
$\gamma_{\mathrm{CARMENES\ VIS}}$[m s$^{-1}$] & $\mathcal{U}$(--3$\sigma_{\mathrm{CARMENES\ VIS}}$, 3$\sigma_{\mathrm{CARMENES\ VIS}}$) & 11.0$^{+35.2}_{-34.6}$ & 10.6$^{+36.2}_{-34.3}$\\
$\gamma_{\mathrm{CARMENES\ NIR}}$[m s$^{-1}$] & $\mathcal{U}$(--3$\sigma_{\mathrm{CARMENES\ NIR}}$, 3$\sigma_{\mathrm{CARMENES\ NIR}}$) & --10.6$^{+26.6}_{-27.1}$ & --10.9$^{+27.2}_{-27.8}$\\
$\gamma_{\mathrm{SPIRou}}$[m s$^{-1}$] & $\mathcal{U}$(--3$\sigma_{\mathrm{SPIRou}}$, 3$\sigma_{\mathrm{SPIRou}}$) & --2.1$^{+7.4}_{-7.4}$ & --2.3$^{+7.5}_{-7.7}$\\
$\sigma_{\mathrm{jit,TESS,120}}$[ppt] & $\mathcal{U}$(0, 3$\sigma_{\mathrm{TESS}}$) & 0.27$^{+0.01}_{-0.01}$ & 0.27$^{+0.01}_{-0.01}$\\
$\sigma_{\mathrm{jit,TESS,20}}$[ppt] & $\mathcal{U}$(0, 3$\sigma_{\mathrm{TESS}}$) & 0.42$^{+0.01}_{-0.01}$ & 0.42$^{+0.01}_{-0.01}$\\
$\sigma_{\mathrm{jit,CHEOPS,15}}$[ppt] & $\mathcal{U}$(0, 3$\sigma_{\mathrm{CHEOPS}}$) & 0.34$^{+0.01}_{-0.01}$ & 0.34$^{+0.01}_{-0.01}$\\
$\sigma_{\mathrm{jit,CHEOPS,3}}$[ppt] & $\mathcal{U}$(0, 3$\sigma_{\mathrm{CHEOPS}}$) & 0.54$^{+0.01}_{-0.01}$ & 0.54$^{+0.01}_{-0.01}$\\
$\sigma_{\mathrm{HARPS}}$[m s$^{-1}$] & $\mathcal{U}$(0, 3$\sigma_{\mathrm{HARPS}}$) & 10.2$^{+1.6}_{-1.4}$ & 10.1$^{+1.6}_{-1.3}$\\
$\sigma_{\mathrm{CARMENES\ VIS}}$[m s$^{-1}$] & $\mathcal{U}$(0, 3$\sigma_{\mathrm{CARMENES\ VIS}}$) & 1.1$^{+1.2}_{-0.8}$ & 1.1$^{+1.3}_{-0.8}$\\
$\sigma_{\mathrm{CARMENES\ NIR}}$[m s$^{-1}$] & $\mathcal{U}$(0, 3$\sigma_{\mathrm{CARMENES\ NIR}}$) & 3.4$^{+3.7}_{-2.4}$ & 3.5$^{+3.7}_{-2.4}$\\
$\sigma_{\mathrm{SPIRou}}$[m s$^{-1}$] & $\mathcal{U}$(0, 3$\sigma_{\mathrm{SPIRou}}$) & 13.0$^{+1.1}_{-1.0}$ & 13.0$^{+1.1}_{-1.1}$\\
$q_{1,\mathrm{TESS}}$ & $\mathcal{N}$(0.291, 0.005) & 0.295$^{+0.005}_{-0.005}$ & 0.295$^{+0.005}_{-0.005}$\\
$q_{2,\mathrm{TESS}}$ & $\mathcal{N}$(0.298, 0.005) & 0.300$^{+0.005}_{-0.005}$ & 0.300$^{+0.005}_{-0.005}$\\
$q_{1,\mathrm{CHEOPS}}$ & $\mathcal{N}$(0.423, 0.005) & 0.422$^{+0.005}_{-0.005}$ & 0.421$^{+0.005}_{-0.005}$\\
$q_{2,\mathrm{CHEOPS}}$ & $\mathcal{N}$(0.326, 0.005) & 0.320$^{+0.005}_{-0.005}$ & 0.319$^{+0.005}_{-0.005}$\\
$\eta_{\sigma_1, \mathrm{TESS}}$[ppt] & $\mathcal{N}$($\sigma_{\mathrm{TESS}}$, 4) & 10.8$^{+3.4}_{-3.2}$ & 10.8$^{+3.5}_{-3.1}$\\
$\eta_{\sigma_2, \mathrm{TESS}}$[ppt] & $\mathcal{N}$($\sigma_{\mathrm{TESS}}$, 4) & 47.4$^{+1.3}_{-1.3}$ & 47.4$^{+1.2}_{-1.3}$\\
$\eta_{\sigma, \mathrm{HARPS}}$[m s$^{-1}$] & $\mathcal{N}$($\sigma_{\mathrm{HARPS}}$, 35) & 158.5$^{+17.4}_{-15.4}$ & 159.1$^{+17.5}_{-15.6}$\\
$\eta_{\sigma, \mathrm{CARMENES\ VIS}}$[m s$^{-1}$] & $\mathcal{N}$($\sigma_{\mathrm{CARMENES\ VIS}}$, 25) & 107.1$^{+13.7}_{-12.1}$ & 106.7$^{+13.4}_{-12.2}$\\
$\eta_{\sigma, \mathrm{CARMENES\ NIR}}$[m s$^{-1}$] & $\mathcal{N}$($\sigma_{\mathrm{CARMENES\ NIR}}$, 20) & 77.6$^{+11.5}_{-10.1}$ & 78.0$^{+12.0}_{-9.6}$\\
$\eta_{\sigma, \mathrm{SPIRou}}$[m s$^{-1}$] & $\mathcal{N}$($\sigma_{\mathrm{SPIRou}}$, 8) & 31.6$^{+3.6}_{-3.2}$ & 31.8$^{+3.5}_{-3.1}$\\
$\eta_{L_1, \mathrm{TESS}}$[d] & $\mathcal{U}$($P_\mathrm{rot}$, 1000) & 547.3$^{+309.9}_{-327.7}$ & 546.3$^{+314.2}_{-318.0}$\\
$\eta_{L_2, \mathrm{TESS}}$[d] & $\mathcal{U}$($P_\mathrm{rot}$, 1000) & 5.0$^{+0.1}_{-0.1}$ & 5.0$^{+0.1}_{-0.1}$\\
$\eta_{L, \mathrm{RV}}$[d] & $\mathcal{U}$($P_\mathrm{rot}$, 2500) & 89.0$^{+9.3}_{-8.4}$ & 89.8$^{+9.4}_{-8.6}$\\
$\eta_{\omega, \mathrm{RV}}$ & $\mathcal{U}$(0.1, 1.0) & 0.30$^{+0.02}_{-0.02}$ & 0.30$^{+0.02}_{-0.02}$\\
$\eta_{P_{\mathrm{rot}}}$[d] & $\mathcal{U}$(1, 10) & 4.860$^{+0.002}_{-0.002}$ & 4.860$^{+0.002}_{-0.002}$\\
\hline
$\Delta \ln \mathcal{Z}$ & ... & --102199.4 & --102187.4\\
\hline
\end{tabular}
\tablefoot{The prior label of $\mathcal{N}$ and $\mathcal{U}$ represent normal and uniform distributions, respectively.}\\
\end{table*}

\begin{table}
\caption{Derived parameters of joint fit for AU\,Mic\,b and c.}\label{tab:joint-fit_der}
\centering
\begin{tabular}{l@{\hskip 0.01in}c@{\hskip 0.04in}c}
\hline\hline
Parameter & Posterior ($e$\,$=$\,0) & Posterior ($e$, $\omega$ free)\\
\hline
$\delta^{b}$ [ppt] & 2.45$^{+0.01}_{-0.01}$ & 2.49$^{+0.04}_{-0.04}$\\
$R_{p}^{b}$ [R$_{\oplus}$] & 4.75$^{+0.29}_{-0.29}$ & 4.79$^{+0.29}_{-0.29}$\\
$a^{b}/R_{\star}$ & 17.51$^{+1.12}_{-1.24}$ & 17.51$^{+1.12}_{-1.24}$\\
$a^{b}$[AU] & 0.070$^{+0.006}_{-0.007}$ & 0.070$^{+0.006}_{-0.007}$\\
$i^{b}$ [$^\circ$] & 88.53$^{+0.09}_{-0.10}$ & 88.39$^{+0.20}_{-0.23}$\\
$e^{b}$ & ... & 0.07$^{+0.09}_{-0.04}$\\
$\omega^{b}$ [rad] & ... & --0.91$^{+0.75}_{-1.62}$\\
$M_p^{b}$ [M$_{\oplus}$] & 8.79$^{+2.55}_{-2.60}$ & 8.99$^{+2.61}_{-2.67}$\\
$\rho_{p}^{b}$ [g\,cm$^{-3}$] & 0.48$^{+0.16}_{-0.17}$ & 0.49$^{+0.16}_{-0.16}$\\
$T_{\mathrm{eq}}^{b}$(A\,$=$\,0) [K] & 554.8$^{+31.6}_{-32.0}$ & 554.8$^{+31.6}_{-32.0}$\\
$T_{\mathrm{eq}}^{b}$(A\,$=$\,0.6) [K] & 441.2$^{+25.2}_{-25.5}$ & 441.2$^{+25.2}_{-25.5}$\\
$t_{12}^{b}$ [h] & 0.20$^{+0.01}_{-0.01}$ & 0.22$^{+0.02}_{-0.01}$\\
$t_{14}^{b}$ [h] & 3.30$^{+0.23}_{-0.21}$ & 3.24$^{+0.23}_{-0.21}$\\
\hline
$\delta^{c}$ [ppt] & 0.90$^{+0.01}_{-0.01}$ & 0.85$^{+0.03}_{-0.02}$\\
$R_{p}^{c}$ [R$_{\oplus}$] & 2.87$^{+0.18}_{-0.18}$ & 2.79$^{+0.18}_{-0.17}$\\
$a^{c}/R_{\star}$ & 29.87$^{+1.91}_{-2.11}$ & 29.87$^{+1.91}_{-2.11}$\\
$a^{c}$[AU] & 0.119$^{+0.011}_{-0.011}$ & 0.119$^{+0.011}_{-0.011}$\\
$i^{c}$ [$^\circ$] & 88.96$^{+0.07}_{-0.07}$ & 89.46$^{+0.43}_{-0.38}$\\
$e^{c}$ & ... & 0.18$^{+0.08}_{-0.07}$\\
$\omega^{c}$ [rad] & ... & 1.27$^{+1.06}_{-0.86}$\\
$M_p^{c}$ [M$_{\oplus}$] & 13.25$^{+3.12}_{-3.19}$ & 14.46$^{+3.24}_{-3.42}$\\
$\rho_{p}^{c}$ [g\,cm$^{-3}$] & 3.28$^{+0.98}_{-0.99}$ & 3.90$^{+1.14}_{-1.17}$\\
$T_{\mathrm{eq}}^{c}$(A\,$=$\,0) [K] & 424.7$^{+24.2}_{-24.5}$ & 424.7$^{+24.2}_{-24.5}$\\
$T_{\mathrm{eq}}^{c}$(A\,$=$\,0.6) [K] & 337.7$^{+19.3}_{-19.5}$ & 337.7$^{+19.3}_{-19.5}$\\
$t_{12}^{c}$ [h] & 0.17$^{+0.01}_{-0.01}$ & 0.13$^{+0.01}_{-0.01}$\\
$t_{14}^{c}$ [h] & 4.05$^{+0.29}_{-0.29}$ & 4.29$^{+0.30}_{-0.27}$\\
\hline
$u_{1,\mathrm{TESS}}$ & 0.325$^{+0.006}_{-0.006}$ & 0.326$^{+0.006}_{-0.006}$\\
$u_{2,\mathrm{TESS}}$ & 0.217$^{+0.006}_{-0.006}$ & 0.218$^{+0.006}_{-0.006}$\\
$u_{1,\mathrm{CHEOPS}}$ & 0.415$^{+0.007}_{-0.007}$ & 0.415$^{+0.007}_{-0.007}$\\
$u_{2,\mathrm{CHEOPS}}$ & 0.234$^{+0.007}_{-0.007}$ & 0.234$^{+0.006}_{-0.007}$\\
\hline
\end{tabular}
\end{table}

\section{Discussion}
\label{sec:disc}

\subsection{Time evolution and wavelength chromaticity of stellar activity}
\label{sec:chr}

We have already observed in photometry (Fig.\,\ref{fig:LC_TESS}) that the two main contributions to amplitude variations are the stellar activity due to flares (randomly distributed) and the stellar activity due to spots on the surface of the star (quasi-periodically distributed due to the rotation of the star). As the star rotates, the cool dark star spots modify the shape of the absorption spectral lines and, thus, the RV calculated from them \citep{Reiners2010, Jeffers2022}. Since the contrast between the temperature of the spots and the temperature of the photosphere decreases at redder wavelengths, these variations in the line profile are lower, and, therefore, the amplitude variations of the RV in NIR wavelengths are smaller than in the VIS.

\begin{figure}
\includegraphics[width=1\linewidth]{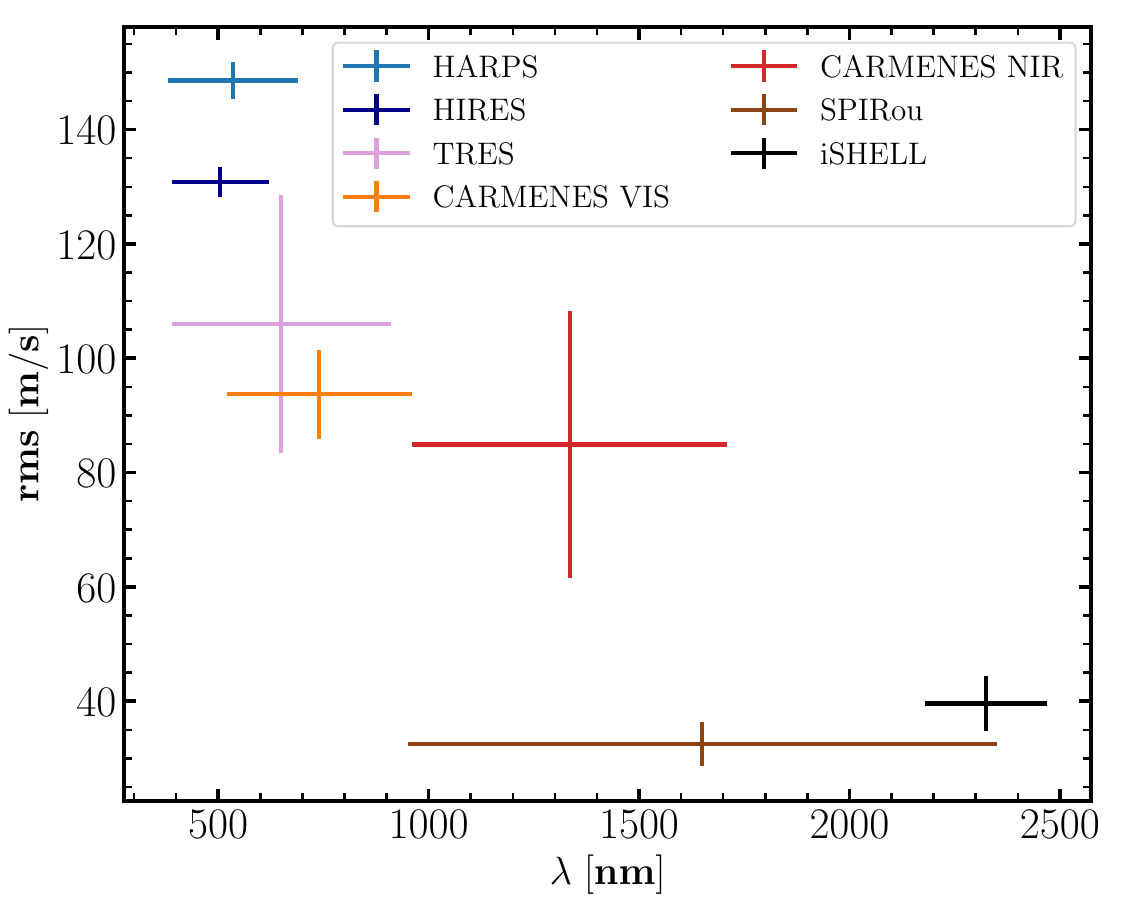}
\includegraphics[width=1\linewidth]{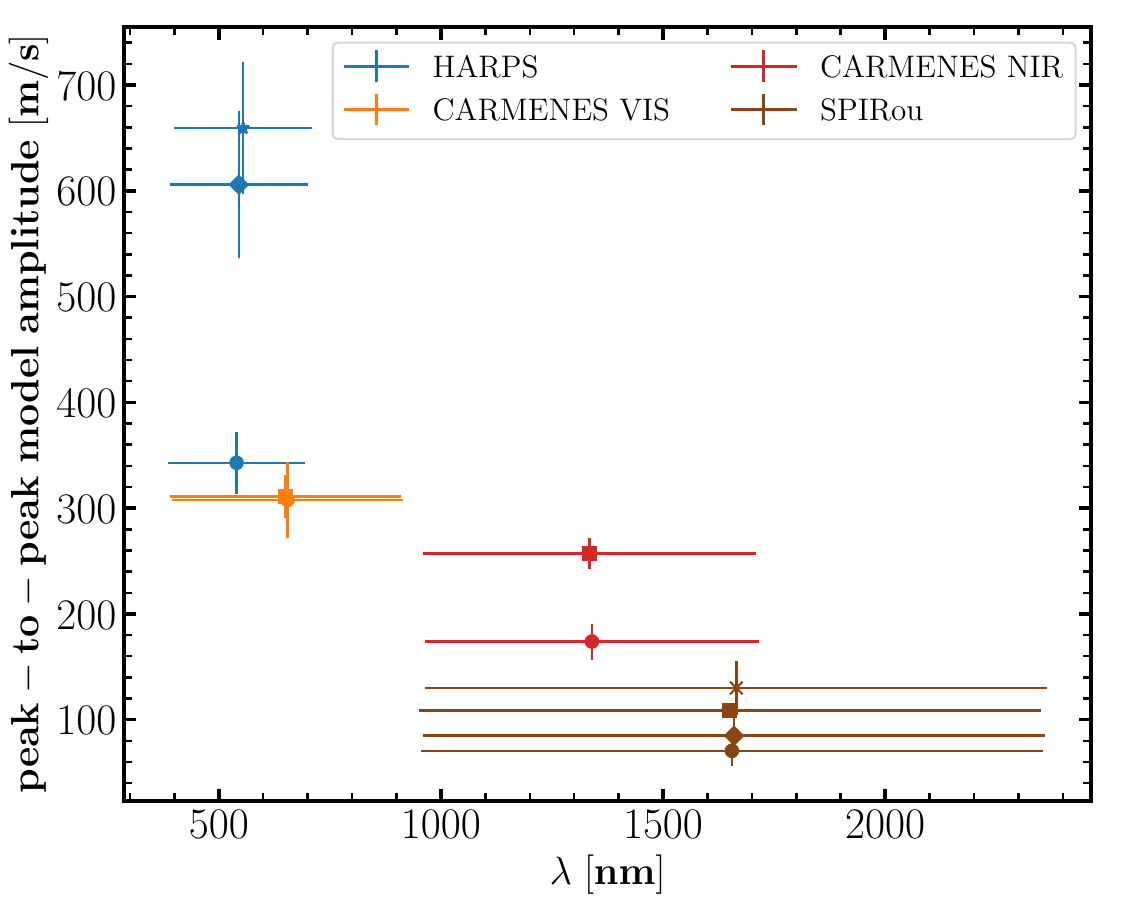}
\caption{Induced stellar activity as function of wavelength measured as rms (top panel) and as peak-to-peak amplitude (bottom panel). The symbols represent the different observation campaigns of the star: C1 squares, C2 dots, C3 diamonds, C4 cross, and C5 asterisk. }
\label{fig:chr}
\end{figure}

Thanks to the long temporal coverage and the different instrument used in the spectroscopic observations of AU\,Mic, we can study the level variations of the stellar activity in RV as a function of wavelength and time. In Fig.\,\ref{fig:chr} (upper panel), the rms of the RV in the different datasets considered in Table\,\ref{tab:RVdatasets} is shown. It is lower at longer wavelengths (SPIRou and iSHELL) and 3--4 times higher at shorter wavelengths (HARPS, HIRES, TRES, CARMENES VIS). To study how the amplitude of activity varies with time, we used the GP activity model of the joint fit for each RV dataset, which is represented in Fig.\,\ref{fig:RV_curve}. For each stellar rotation period, we calculated the peak-to-peak amplitude by taking the difference between the maximum and minimum. Then, we calculated the median and dispersion of those amplitudes. Additionally, we sub-divided the sample into five 'campaigns', whose baselines are listed in Table\,\ref{tab:camp}. The results are shown in Fig.\,\ref{fig:chr} (bottom panel). The figure shows that stellar activity is clearly lower at redder wavelengths. The rms of the RVs is a factor of 3--4 times lower at 1700-2500\,nm than at $\sim$ 500\,nm wavelengths, which can be crucial for measuring the Keplerian signals due to the presence of the planets. Furthermore, if we compare the activity variation between campaigns, for example between campaigns C2 and C3 (represented by the dot and diamond symbols, respectively) of HARPS and SPIRou, we can see that the HARPS variations increase by 80\% (from $\sim$340 to $\sim$600 m s$^{-1}$) from one campaign to another, while in the same SPIRou campaigns they increase by 20\% (from $\sim$70 to $\sim$85 m s$^{-1}$). Therefore, stellar activity is also less variable at longer wavelengths from one campaign to another. This shows that the sweet spot for observation to measure the masses of planets around this type of star would be in the $H$ and $K$ bands.

\begin{table*}
\caption{Dates and number of observations per instrument in each campaign.}\label{tab:camp}
\centering
\begin{tabular}{c|c c c c c}
\hline\hline
Dates [BJD] & 2458650--2458805 & 2458980--2459200 & 2459350--2459550 & 2459700--2459750 & 2459850--2459925\\
Campaign name & C1 & C2 & C3 & C4 & C5\\
\hline
CARMENES VIS & 69 & 17 & 0 & 0 & 0\\
CARMENES NIR & 71 & 16 & 0 & 0 & 0\\
HARPS & 0 & 21 & 58 & 0 & 38 \\
SPIRou & 25 & 71 & 61 & 19 & 0\\
\hline
\hline
\end{tabular}
\end{table*}

\subsection{Comparison of transit depths}
\label{sec:tr_depth}

\begin{figure}
\includegraphics[width=1\linewidth]{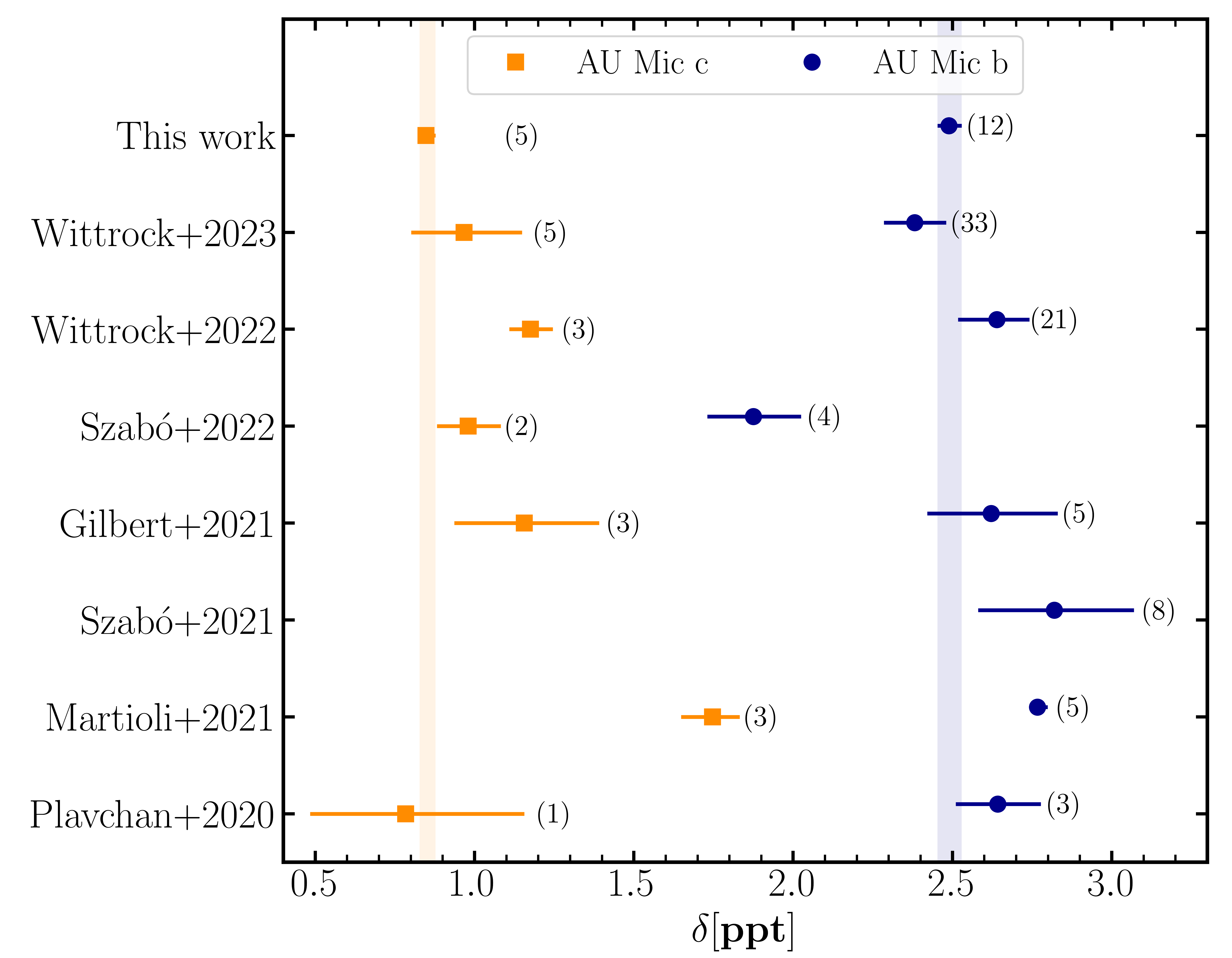}
\caption{Transit depths of AU\,Mic\,b and c in the literature compared to our determinations. The number in parentheses indicates the number of transits used to calculate the transit depth. The coloured bands represent the transit depth measurements of this work.}
\label{fig:depth_comparison}
\end{figure}

We compared our measurements of transit depths of planets AU\,Mic\,b and c with those of the literature in Fig.\,\ref{fig:depth_comparison}. The transit depths shown are the direct measurements obtained from the light-curve analysis and, in contrast to the planetary radii, they are independent of the stellar radius employed by the different authors. All of these publications (except \citealp{Szabo2022}) use, at least, the TESS photometry. Some of them use only TESS photometry, others use TESS and Spitzer photometry, TESS and CHEOPS photometry, and/or ground-based photometry. In addition, stellar activity is often treated differently, especially in the case of the TESS light curves. Some authors have modelled stellar activity due to rotation with GP or polynomial functions, while some of them have modelled flares or masked them. Despite the different datasets or models used, all the results are consistent with our results within 2$\sigma,$ and most of them within 1$\sigma$. The exceptions are the larger transit depths with very small error bars for both planets determined by \citet{Martioli2021} and the rather shallow depth for planet b found by \citet{Szabo2022}. A slight decrease in the radii of both planets is observed when the most recent CHEOPS measurements (four transits of planet b and two transits of planet c) are introduced. In previous works \citep{Martioli2021, Szabo2022}, it has already been discussed how stellar activity can affect the measurement of the transit depth. Flares that occur during, just after, or just before the transit may affect the continuum and, therefore, the depth; flares during ingress or egress hamper the determination of the transit duration, while flares during the total eclipse affect the depth. Furthermore, the modelling of stellar activity can also alter the transit depth as it depends on the intrinsic activity of the star. For example, the peak-to-peak variation is lower in the second TESS sector than in the first, which might have an indirect effect on the transit depth measurement through the GP modelling.

\begin{figure}
\includegraphics[width=1\linewidth]{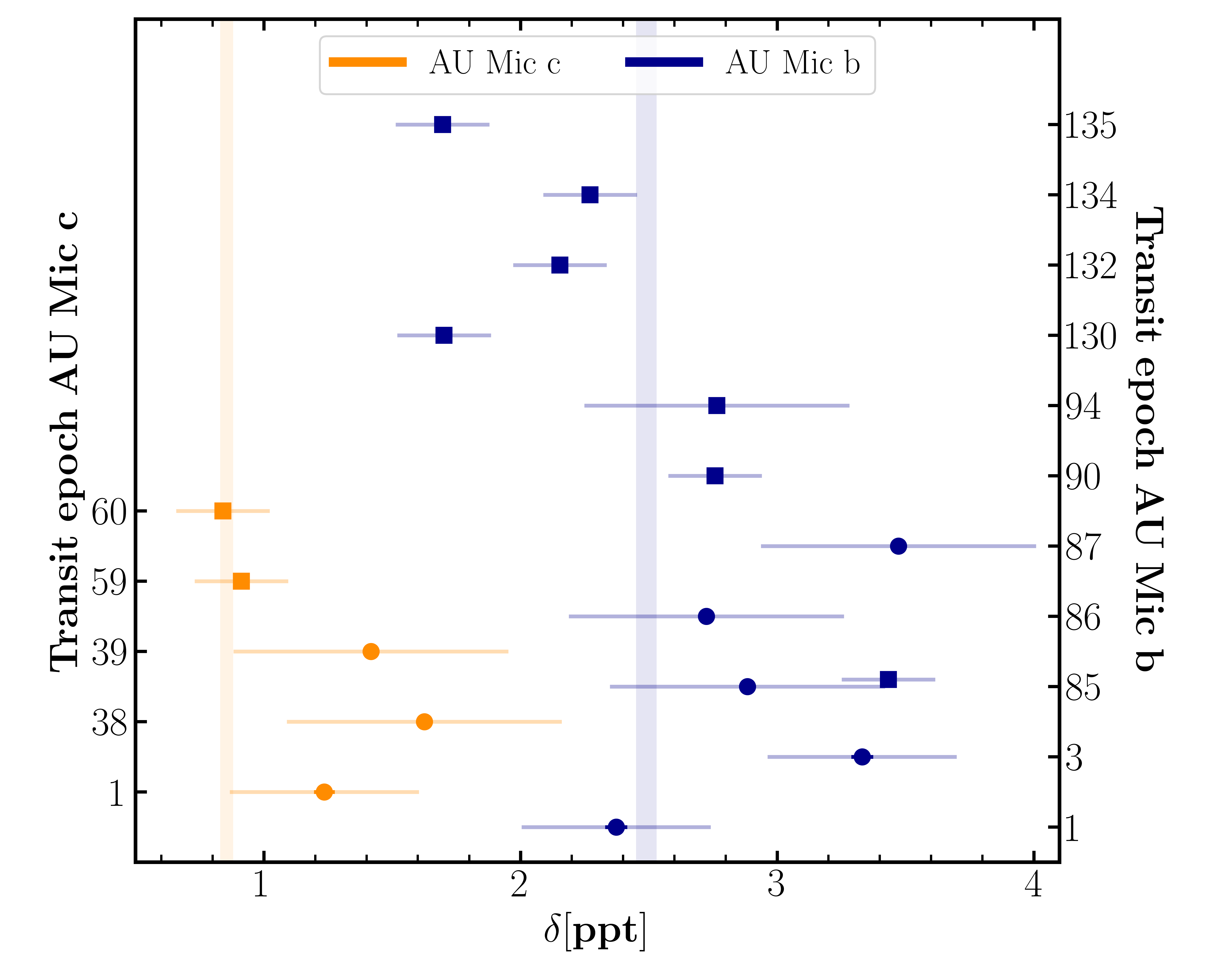}
\caption{Transit depth determinations of AU\,Mic\,b (blue) and c (orange) for individual transits. The dot and square symbols represent the TESS and CHEOPS data, respectively. The vertical bands represent our transit depth determination from the joint fit. The opacity in the error bars of each measurement indicates the 1$\sigma$ error of the posterior distribution (higher opacity) and the error of the jitter term (less opacity). The transit with epoch 85 was observed simultaneously with TESS and CHEOPS and is slightly shifted in the vertical axis for clarity.}
\label{fig:depth_ind_comparison}
\end{figure}

In addition, to study how the individual transit depth changes, we created an alternative transit model (Sect.\,\ref{sec:tr}) where each transit has an independent depth parameter (Table\,\ref{tab:tr_depth}). Figure \ref{fig:depth_ind_comparison} shows a similar scheme to Fig.\,\ref{fig:depth_comparison}, where each point now represents an individual transit. The circle symbols correspond to TESS data and the squares to CHEOPS data. The error bars represented in the figure include the uncertainty in the $\delta$ parameter (higher opacity) and the jitter term added in quadrature to the error bars (lower opacity). The vertical bands represent the results of our joint fit. We observed variations between individual transits that are larger than the error bar associated with the model fit. However, they are within 1--2$\sigma$ if we take into account the model jitter. Figures \ref{fig:TTVs_b} and \ref{fig:TTVs_c} show the transit-to-transit data with its models. The high stellar activity of AU\,Mic and the intermittent cadence of the CHEOPS photometry make it difficult to visualise where the minimum of each transit is (when the eclipse is total) and, therefore, to correctly measure the radius of the planets. While in Sect.\,\ref{sec:analysis} we describe how we modelled all flares affecting the transits, the observed noise level during the transit remains somewhat higher than out of the transit (Fig.\,\ref{fig:PH_folded}). If we observe the individual transits and their residuals in Fig.\,\ref{fig:TTVs_b} and Fig.\,\ref{fig:TTVs_c}, we can see some artefacts that we do not recognise as flares (in particular transit epochs 1, 3, 85, 87, 90, 130, 132, and 134 for AU\,Mic\,b and 38, 39, 59, and 60 for AU\,Mic\,c). The shape of these artefacts is similar to that found by \cite{Dai2018}. We think that given the high stellar activity that AU\,Mic shows, its surface must have several active regions of spots and faculae that also evolve over time. These variations may be a consequence of the shadow of the planet crossing these regions during the transit of the planet over the stellar surface. The weighted means of all the individual transit depths of AU\,Mic\,b and AU\,Mic\,c are 2.61\,$\pm$\,0.17 ppt and 1.11\,$\pm$\,0.13 ppt, respectively. These values are slightly higher and have larger uncertainties than those obtained in the joint fit analysis ($\delta^b$\,=\,2.49\,$\pm$\,0.04 ppt, $\delta^c$\,=\,0.85\,$\pm$\,0.03 ppt). However, these depths are translated into 4.92\,$\pm$\,0.34 R$_{\oplus}$ for AU\,Mic\,b and 3.20\,$\pm$\,0.27 R$_{\oplus}$ for AU\,Mic\,c, which is consistent within the error bars with those of the joint fit, and the precision level is similar. This is because the dominant source of error in the radius of the planet is the determination of the radius of the star. Therefore, to obtain a precise measurement of planetary radii of planets transiting stars as young and active as AU\,Mic, a proper modelling of the stellar activity generated by spots and flares during transits is necessary, together with a more precise measurement of the star's radius.

\subsection{Previous planet mass measurements}

\begin{figure}
\includegraphics[width=1\linewidth]{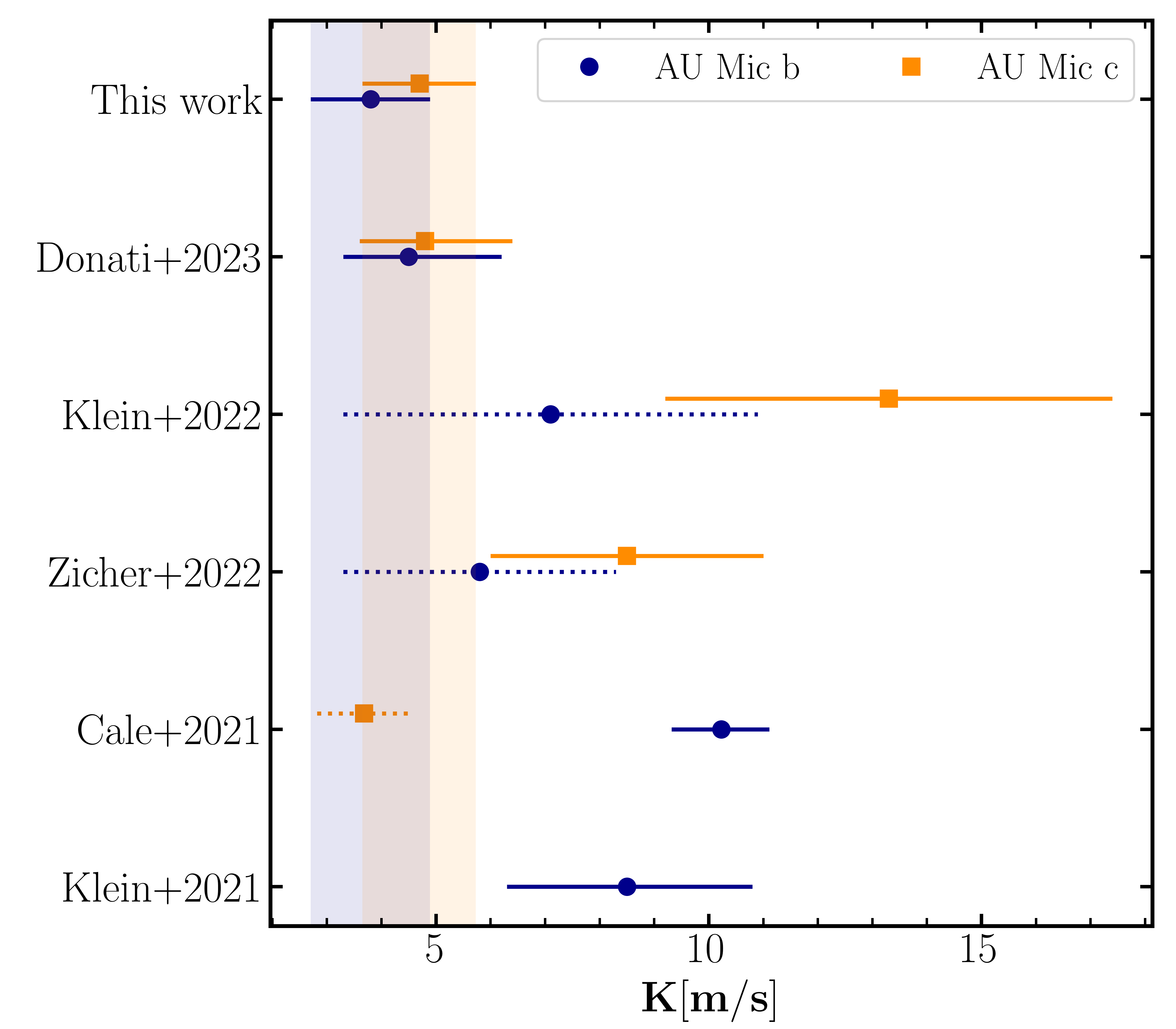}
\caption{Keplerian amplitudes of AU\,Mic\,b and c in the literature compared to our determinations. The dotted lines represent the measurements with a less than 3$\sigma$ confidence level. The coloured bands represent the Keplerian amplitude measurements of this work.}
\label{fig:kep_comparison}
\end{figure}

As for transits, we compiled different induced Keplerian amplitudes of the planets AU\,Mic\,b and c published in the literature in Fig.\,\ref{fig:kep_comparison} and compare them with each other and with our results. We used this value because, on the contrary to the planet masses, it is independent of stellar parameters. We note that all the previous mass measurements come from an RV-only analysis, except ours, which comes from a joint photometric-RV analysis. To understand the observed differences, we summarise the main features of each work that make use of different RV datasets below, except for \citet{zicher22} and \citet{klein22}. In addition, all studies but \cite{klein22} model stellar activity with GPs. Chronologically, \citet{klein21} used 28 RVs collected with SPIRou, a circular Keplerian model for planet b and a GP activity model with a QP 
kernel. In their final model, the planetary parameters $T_c$ and $P$, as well as the hyperparameters $\eta_{\omega}$, $\eta_{L}$, and the jitter term ($\sigma_{\mathrm{jit, RV}}$) are fixed values. \citet{Cale2021} used an RV dataset collected with several spectrographs, including HIRES, TRES, CARMENES VIS, CARMENES NIR, SPIRou, and iSHELL. They used a Keplerian with non-zero eccentricity for AU\,Mic\,b ($e$\,=\,0.19) and a circular orbit for AU\,Mic\,c. To model the activity, they proposed a novel method based on the wavelength dependence of stellar activity. This is modelled with a GP and a QP kernel but not in each dataset separately, but their model assumes a combined dataset where each dataset has a different amplitude hyperparameter; this is done so that a correlation between all the data is maintained. As in \cite{klein21}, the parameters of $T_c$, $P$, $\eta_{\omega}$, $\eta_{L}$, and jitter values are fixed. \cite{zicher22} used a single dataset of 85 HARPS spectra, employing a Keplerian model with non-zero eccentricity for both planets. They modelled the stellar activity with a multi-dimensional GP \citep{Rajpaul2015, Barragan2022}, meaning a more general case of GP where the correlation between different datasets is taken into account. In their particular case, they used activity indicators as tracers of stellar activity and included a term of the first derivative of the GP to take into account the variation of stellar activity with time. All the parameters of their fit are free. \citet{klein22} used the same HARPS dataset as \citet{zicher22} but measured the Keplerian amplitude of AU\,Mic\,c using the Doppler imaging technique. The stellar activity of AU\,Mic is high enough to produce noticeable distortions of the spectral lines. These authors measured the amplitude of AU\,Mic\,c simultaneously fitting a model of activity in the line profile and a shift in the lines potentially due to the presence of the planet. Finally, \citet{Donati2023} used $\sim$180 SPIRou epochs to infer the mass of AU\,Mic\,b and c, assuming circular orbits. They modelled the activity with GP with a QP kernel and fixed the planetary parameters $T_c$ and $P$ in their model.

Overall, the previous published mass measurements are based on notably different datasets, circular or eccentric Keplerian orbits, techniques to model stellar activity, and types or ranges of priors used in the fits. Although many determinations are non-detections and only upper limits can be set (detections with less than 3$\sigma$ are represented as dotted lines in the Fig.\,\ref{fig:kep_comparison}), all the results for planet b are consistent with our result within 1$\sigma$, except for that of \cite{Cale2021}, which is within 2$\sigma,$ and \cite{klein21}, which is beyond 2$\sigma$. For planet c, our results are consistent within 1$\sigma$ with the results of \cite{Cale2021} and \cite{Donati2023}, and within 2$\sigma$ with the results of \cite{klein22} and \cite{zicher22}. 

\subsection{Planet characterisation}

We characterise the young transiting planetary system composed of AU\,Mic\,b and AU\,Mic\,c. The bayesian evidence shows that the model including eccentric orbits is clearly favoured over the circular orbit model; thus, we adopted the model with non-zero eccentricity as our final model. For the inner planet, AU\,Mic\,b, we infer a radius of $R_p^b$=\,4.79\,$\pm$\,0.29 R$_\oplus$ and a mass of $M_p^b$=\,9.0\,$\pm$\,2.7 M$_\oplus$ with a 3.4$\sigma$ detection. For AU\,Mic\,c, we derive a radius of $R_p^c$=\,2.79\,$\pm$\,0.18 R$_\oplus$ and a mass of $M_p^c$=\,14.5\,$\pm$\,3.4 M$_\oplus$ at a 4.3$\sigma$ level of confidence. From these results, we derive bulk densities of $\rho^{c}$\,=\,0.49\,$\pm$\,0.16 g\,cm$^{-3}$ and $\rho^{b}$\,=\,3.90\,$\pm$\,1.17 g\,cm$^{-3}$ for AU\,Mic\,b and c, respectively. These densities are slightly smaller than those of Saturn and higher than those of Neptune in the case of AU\,Mic\,b and AU\,Mic\,c, respectively, which could indicate differences in their compositions.

\subsection{Mass-radius diagram}

\begin{figure*}
\includegraphics[width=1\linewidth]{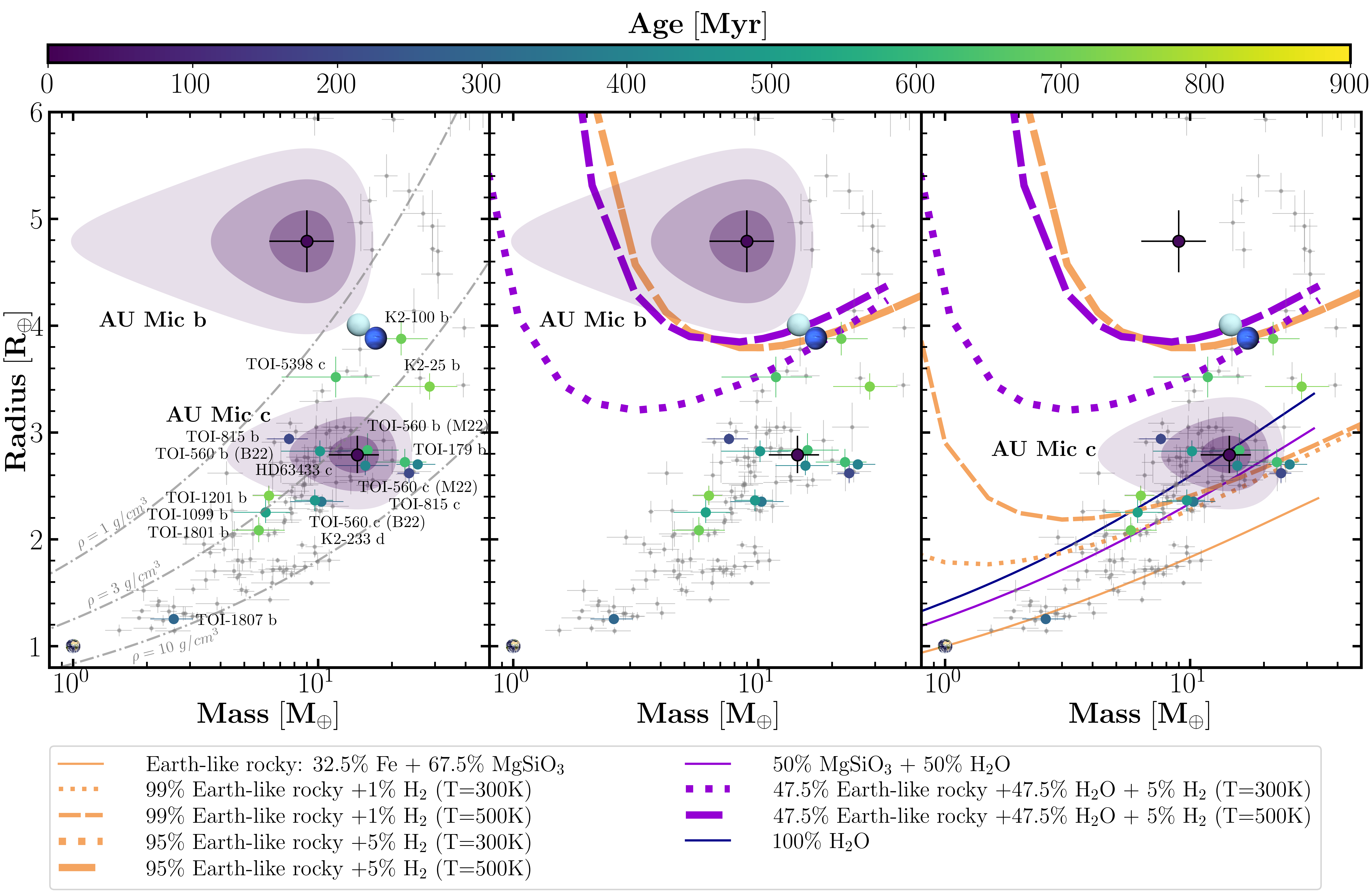}
\caption{Mass-radius diagram for AU\,Mic\,b and c, together with all known exoplanets (grey dots) with a precision level that is better than 8\% in radius (through transit) and 20\% in mass (from RV). The population of young transiting exoplanets ($<$ 900 Myr) with measured masses is plotted as coloured dots according to their ages. The uncertainties of AU\,Mic\,b and c are shown as coloured shaded regions indicating 1, 2, and 3$\sigma$ confidence regions. In the left panel, the iso-density lines are displayed as dashed grey lines. In the remaining panels where the population of exoplanets is plotted along AU\,Mic\,b and c, coloured lines indicate different composition models without gas and with a gas envelope from \cite{Zeng2019PNAS..116.9723Z}. The dotted and dashed lines show the models with temperatures of 300 and 500\,K, respectively, of the corresponding specific entropy at 100 bar level in the gas envelope. The Earth, Uranus, and Neptune are also depicted as references. We include B22 and M22 as references because different results have been published for the same planet, and this refers to \cite{barr22} and \cite{elmufti2023}, respectively.
\label{fig:RM}}
\end{figure*}

The precise measurements of radius and mass presented in this work allow us to compare the AU\,Mic planetary system with known exoplanets from the Extrasolar Planets Encyclopedia\footnote{\url{http://exoplanet.eu/}} and other young systems. Figure\,\ref{fig:RM} shows the mass-radius distribution of all exoplanets with measured radius and mass (grey dots) in the range of 1--6 R$_\oplus$ and 1--50 M$_\oplus,$ with radius uncertainties better than 8\% from the transit method and mass uncertainties better than 20\%. In addition, the coloured dots indicate young planets and their ages. Our values of radius and mass obtained in the joint-fit for AU\,Mic\,b and c are represented as coloured regions (according to their age) and opacity indicating the level of confidence (1, 2, or 3$\sigma$). In the left panel, the planets are not uniformly distributed throughout the diagram, but rather seem to follow a sequence within the iso-density lines between values of 1 and 10 g\,cm$^{-3}$; although, for masses under 5 M$_\oplus$, this sequence is narrowed to density values between 3 and 10 g\,cm$^{-3}$. So far, all young planets with measured radii and masses are found on the sequence of old field planets. However, none of these young systems has an age below 100 Myr. While AU\,Mic\,c appears to fit the old field planet sequence, AU\,Mic\,b does not fit in the sequence, which indicates that it has an extended envelope (3$\sigma$ off the sequence), has lost a considerable part of its mass (2$\sigma$ off the sequence), or both. Given the extremely young age of the star ($\sim$20\,Myr), we argue that it may be due to the formation or evolution mechanisms that are still occurring in the planet. The middle panel shows the composition models of \cite{Zeng2019PNAS..116.9723Z} for rocky, and rocky plus water compositions, with or without an H$_2$ envelope. AU\,Mic\,b is consistent with an H$_2$ envelope larger than 5\% in mass, regardless of whether its internal composition is mainly rocky or rock plus water. On the other hand, if AU\,Mic\,c has mainly a rocky core, its envelope may contain between 1\% and 5\% of H$_2$ in mass. If its composition contains water, AU\,Mic\,c could either be a 100\% water planet without envelope, or a planet with less water content but with an atmosphere of up to 5\% of H$_2$ in mass.

According to the models of \citet{Venturini20}, gaseous (and giant) planets form in regions far from the star where there is still gas. In contrast, in the regions closer to the star, the planets have a composition of iron and silicates and may have a thin envelope. In addition, beyond the ice line, rocky and gaseous planets can also be composed of water. This formation occurs on a timescale of hundreds of thousands or a few million years. Observationally, planets have been found in short orbits (within the ice line) with water in their composition. Therefore, these planets may have migrated from more distant orbits. Migration is expected to occur before the dissipation of the disc ($\sim$5--10 Myr; \citealp{Williams2011}). The youngest planet with a measured radius in a short orbit is K2-33\,b \citep[5--10\,Myr;][]{Trevor2016, Mann2016}, with a radius slightly larger than Neptune, which fits with this hypothesis. Therefore, we suggest that AU\,Mic\,b and c formed in outer regions rich in gas and/or water, and then migrated inwards. In their new orbit, the planets receive more radiation from the star, and mechanisms such as photo-evaporation \citep{owen17} can be especially efficient. If the planet is sufficiently massive ($\sim$1 M$_{\mathrm{Jup}}$), it can retain its envelope. However, for lower masses, as in the case of AU\,Mic\,b and c, they can lose part or all of their envelopes. The timescales of processes such as photo-evaporation are estimated to be from tens to 100 Myr, and therefore they fit with the age of AU\,Mic. It is possible that AU\,Mic\,b still retains part of its original envelope and that due to photo-evaporation it will lose it in the next tens of millions of years, dropping down in the diagram to a position closer to the rest of older small planets, as suggested by \cite{zicher22}. On the other hand, this same process may be occurring in AU\,Mic\,c. However, it is possible that its original envelope was not as large as that of AU\,Mic\,b and that due to its more distant orbit, the photo-evaporation mechanism is not as efficient. In this context, measuring whether both planets are losing H or He from their atmospheres and in what quantity is key to understanding the formation and evolution of AU\,Mic.

Another point is that our inferred planetary equilibrium temperature place both AU\,Mic\,b and AU\,Mic\,c in the region closer to its host star than the anticipated inner boundary of the habitable zone \citep{Kopparapu2014}. The low bulk density of AU\,Mic\,b may indicate that the planet has undergone an instellation-induced runaway greenhouse effect \citep[e.g.][]{Kasting1988, Nakajima1992}, in which case it may still possess an extended steam atmosphere \citep{Turbet2020, Dorn2021}. This hypothesis is supported by the prolonged durations of runaway greenhouse phases expected for planets orbiting M dwarfs \citep{Luger2015} in combination with AU\,Mic's young age. On the other hand, the higher bulk density of AU\,Mic\,c may indicate that it already lost its water to space \citep[e.g.][]{Hamano2015} or, being less irradiated, never entered a runaway greenhouse. While these scenarios are challenging when it comes to testing for individual planets and without deep atmospheric characterisation, the runaway greenhouse transition is expected to manifest as a habitable zone inner edge discontinuity in the radius/density distribution of small exoplanets \citep{Turbet2019, Schlecker2024}. The AU\,Mic system adds to the limited pool of planets suitable for investigating this demographic feature.

\subsection{Mass-loss rate}
\label{mlr}

AU\,Mic has been widely observed at high energies. We analysed high-resolution spectra taken with XMM-Newton/RGS, EUVE, FUSE, and HST/STIS to perform a very detailed model of the corona and transition region of AU\,Mic \citep[][updated in Sanz-Forcada et al., in prep.]{cha15}. This model was used to create a synthetic SED in the whole XUV spectral range (5--920\,\AA), which is used to evaluate the stellar irradiation of the planets orbiting AU\,Mic. The stellar luminosity in the different bands is $\log L_{\rm X}$\,=\,29.44\,$\pm$\,0.07 (5--100\,\AA), $\log L_{\rm EUV,H}$\,=\,29.4\,$\pm$\,0.2 (100--920\,\AA), and $\log L_{\rm EUV,He}$\,=\,29.2\,$\pm$\,0.2 (100--504\,\AA), in erg\,s$^{-1}$. We calculated the expected mass-loss rate in the planet atmospheres, assuming an H-based atmosphere, and an energy-limited approach, following \citet{Sanz2011} and references therein. In the case of AU\,Mic\,b, a mass loss rate of $1.4\times10^{12}$\,g\,s$^{-1}$ or 7.5\,M$_{\oplus}$ Gyr$^{-1}$ is expected, while AU\,Mic\,c would lose its atmosphere at a rate of 0.4\,M$_{\oplus}$ Gyr$^{-1}$. Given the spectral type of AU\,Mic and its age, it is expected that the X-ray (and thus XUV) emission remains at the same level for $\sim$1\,Gyr. Thus, it is likely that AU\,Mic\,b will lose most of its atmosphere along the next Gyr due to photo-evaporation, moving down the diagram in Fig.\,\ref{fig:RM} until reaching the rocky or water composition line and fitting into the field planet sequence. In contrast, AU\,Mic\,c will suffer less and should remain in the same position in the mass-radius diagram.

We also explored whether these planets are good candidates to observe the \ion{He}{i}~10830 triplet in their atmospheres. The formation of this line is linked to the stellar ionising (i.e.\ $<$ 504\,\AA{}) irradiation at the planet. We calculate a flux in the 5--504\,\AA{} range of 3.6\,$\times$\,$10^4$ and 12\,$\times$\,$10^3$ erg\,s$^{-1}$\,cm$^{-2}$ for planets b and c, respectively. This flux is higher than in some planets with positive detections \citep{flo19}, although more challenging for planets of this size. \citet{Hirano2020} and \citet{Allart2023} observed the region of the \ion{He}{i} triplet with the IRD, Keck/NIRSPEC, and SPIRou spectrographs. However, given the observational conditions and stellar activity, they could only establish upper limits on the detection of \ion{He}{i} triplet of 4.4 and 3.7 m\AA{} (with a 99\% confidence level) for IRD and NIRSPEC \citep{Hirano2020} and a 0.26\% limit at 3$\sigma$ in the absorption with SPIRou \citep{Allart2023}.

\section{Conclusions}
\label{sec:concl}

In this study, we focused on the planetary system around the M star AU\,Mic in the young $\beta$\,Pictoris moving group. We conducted an analysis using publicly available photometric and spectroscopic datasets and revised the stellar parameters of AU\,Mic. Our analysis integrates TESS and CHEOPS light curves and more than 400 high-resolution spectra from the CARMENES VIS, CARMENES NIR, HARPS, and SPIRou instruments, as well as iSHELL and TRES. The high stellar activity present in this young star was modelled with GP. In addition, a chromatic analysis of the RV activity level indicates that in this star, stellar activity and its variation with time are lower at redder wavelengths (H and K bands) than at bluer wavelengths. For stars as active as AU\,Mic, we recommend obtaining the largest possible number of transits, with continuous sampling and with sufficient pre- and post-transit duration to model the activity without influencing the transit depth.

For the innermost planet AU\,Mic\,b, we derived a radius of $R_p^b$=\,4.79\,$\pm$\,0.29 R$_\oplus$, a mass of $M_p^b$=\,9.0\,$\pm$\,2.7 M$_\oplus$, and a bulk density of $\rho^{b}$\,=\,0.49\,$\pm$\,0.16 g\,cm$^{-3}$. For AU\,Mic\,c, we inferred a radius of $R_p^c$=\,2.79\,$\pm$\,0.18 R$_\oplus$, a mass of $M_p^c$=\,14.5\,$\pm$\,3.4 M$_\oplus,$ and a bulk density of $\rho^{c}$\,=\,3.90\,$\pm$\,1.17 g\,cm$^{-3}$. We set an upper limit on the mass at a 3$\sigma$ level of confidence of $M_{p}^{[d]}\sin{i}$\,=\,8.6\,M$_{\oplus}$ for the candidate proposed by TTVs, AU\,Mic\,d. We were not able to detect the signal of planet candidate AU\,Mic\,e in HARPS and CARMENES data, and we argue that this could be a false positive in the SPIRou data.

The star AU\,Mic\,c has a radius and mass similar to other exoplanets, with a composition of up to 5\% H$_2$ by mass, according to the models of \citet{Zeng2019PNAS..116.9723Z}. On the other hand, AU\,Mic\,b seems to be outside the planet sequence, which could indicate an extended envelope due to either a composition with more than 5\% H$_2$ by mass or an ongoing runaway greenhouse. This, together with the activity of the host star and the proximity of the planet, indicates that AU\,Mic\,b will probably lose its planetary envelope in the next tens or hundreds of millions of years due to the mechanism of photo-evaporation. This investigation advances our understanding of planetary properties, providing valuable insights into the field of planet formation and evolution around young stars.

\begin{acknowledgements}
CARMENES is an instrument at the Centro Astron\'omico Hispano en Andaluc\'ia (CAHA) at Calar Alto (Almer\'{\i}a, Spain), operated jointly by the Junta de Andaluc\'ia and the Instituto de Astrof\'isica de Andaluc\'ia (CSIC). CARMENES was funded by the Max-Planck-Gesellschaft (MPG), the Consejo Superior de Investigaciones Cient\'{\i}ficas (CSIC), the Ministerio de Econom\'ia y Competitividad (MINECO) and the European Regional Development Fund (ERDF) through projects FICTS-2011-02, ICTS-2017-07-CAHA-4, and CAHA16-CE-3978, and the members of the CARMENES Consortium (Max-Planck-Institut f\"ur Astronomie, Instituto de Astrof\'{\i}sica de Andaluc\'{\i}a, Landessternwarte K\"onigstuhl, Institut de Ci\`encies de l'Espai, Institut f\"ur Astrophysik G\"ottingen, Universidad Complutense de Madrid, Th\"uringer Landessternwarte Tautenburg, Instituto de Astrof\'{\i}sica de Canarias, Hamburger Sternwarte, Centro de Astrobiolog\'{\i}a and Centro Astron\'omico Hispano-Alem\'an), with additional contributions by the MINECO, the Deutsche Forschungsgemeinschaft (DFG) through the Major Research Instrumentation Programme and Research Unit FOR2544 ''Blue Planets around Red Stars'', the Klaus Tschira Stiftung, the states of Baden-W\"urttemberg and Niedersachsen, and by the Junta de Andaluc\'{\i}a.
Based on observations obtained at the Canada-France-Hawaii Telescope (CFHT) which is operated from the summit of Maunakea by the National Research Council of Canada, the Institut National des Sciences de l'Univers of the Centre National de la Recherche Scientifique of France, and the University of Hawaii. The observations at the Canada-France-Hawaii Telescope were performed with care and respect from the summit of Maunakea which is a significant cultural and historic site.
Based on observations obtained with SPIRou, an international project led by Institut de Recherche en Astrophysique et Plan\'etologie, Toulouse, France. 
The authors wish to recognize and acknowledge the very significant cultural role and reverence that the summit of Maunakea has always had within the Native Hawaiian community. We are most fortunate to have the opportunity to conduct observations from this mountain.
This paper includes data collected by the TESS mission. Funding for the TESS mission is provided by the NASA Explorer Program. We acknowledge the use of public TOI Release data from pipelines at the TESS Science Office and at the TESS Science Processing Operations Center. Resources supporting this work were provided by the NASA High-End Computing (HEC) Program through the NASA Advanced Supercomputing (NAS) Division at Ames Research Center for the production of the SPOC data products. This research has made use of the Exoplanet Follow-up Observation Program website, which is operated by the California Institute of Technology, under contract with the National Aeronautics and Space Administration under the Exoplanet Exploration Program.
CHEOPS is an ESA mission in partnership with Switzerland with important contributions to the payload and the ground segment from Austria, Belgium, France, Germany, Hungary, Italy, Portugal, Spain, Sweden, and the United Kingdom. The CHEOPS Consortium would like to gratefully acknowledge the support received by all the agencies, offices, universities, and industries involved. Their flexibility and willingness to explore new approaches were essential to the success of this mission. CHEOPS data analysed in this article will be made available in the CHEOPS mission archive (\url{https://cheops.unige.ch/archive_browser/}).
This work has made use of data from the European Space Agency (ESA) mission {\it Gaia} (\url{https://www.cosmos.esa.int/gaia}), processed by the {\it Gaia} Data Processing and Analysis Consortium (DPAC, \url{https://www.cosmos.esa.int/web/gaia/dpac/consortium}). Funding for the DPAC has been provided by national institutions, in particular the institutions participating in the {\it Gaia} Multilateral Agreement.
We acknowledge financial support from the Agencia Estatal de Investigaci\'on (AEI/10.13039/501100011033) of the Ministerio de Ciencia e Innovaci\'on and the ERDF "A way of making Europe" through projects
PID2022-137241NB-C4[1:4],    
PID2021-125627OB-C31,        
PID2019-109522GB-C5[1:4]
and the Centre of Excellence "Severo Ochoa" and "Mar\'ia de Maeztu" awards to the Instituto de Astrof\'isica de Canarias (CEX2019-000920-S), Instituto de Astrof\'isica de Andaluc\'ia (CEX2021-001131-S) and Institut de Ci\`encies de l'Espai (CEX2020-001058-M).

This work was also funded by the Generalitat de Catalunya/CERCA programme, the DFG through grant HA3279/14-1, the University of La Laguna and EU Next Generation funds through the Margarita Salas Fellowship from the Spanish Ministerio de Universidades through grant UNI/551/2021-May~26, and the European Research Council under the European Union's Horizon 2020 research and innovation programme through grant 865624~GPRV. The results reported herein benefitted from collaborations and/or information exchange within NASA’s Nexus for Exoplanet System Science (NExSS) research coordination network sponsored by NASA’s Science Mission Directorate under Agreement No. 80NSSC21K0593 for the program "Alien Earths".
\end{acknowledgements}
%
%

\bibliographystyle{aa} 
\bibliography{biblio}

\clearpage
\onecolumn

\begin{appendix}

\section{GLS periodograms of SPIRou data}
\label{sec:ap_33d}
\begin{figure*}
\begin{center}
\includegraphics[width=0.8\linewidth]{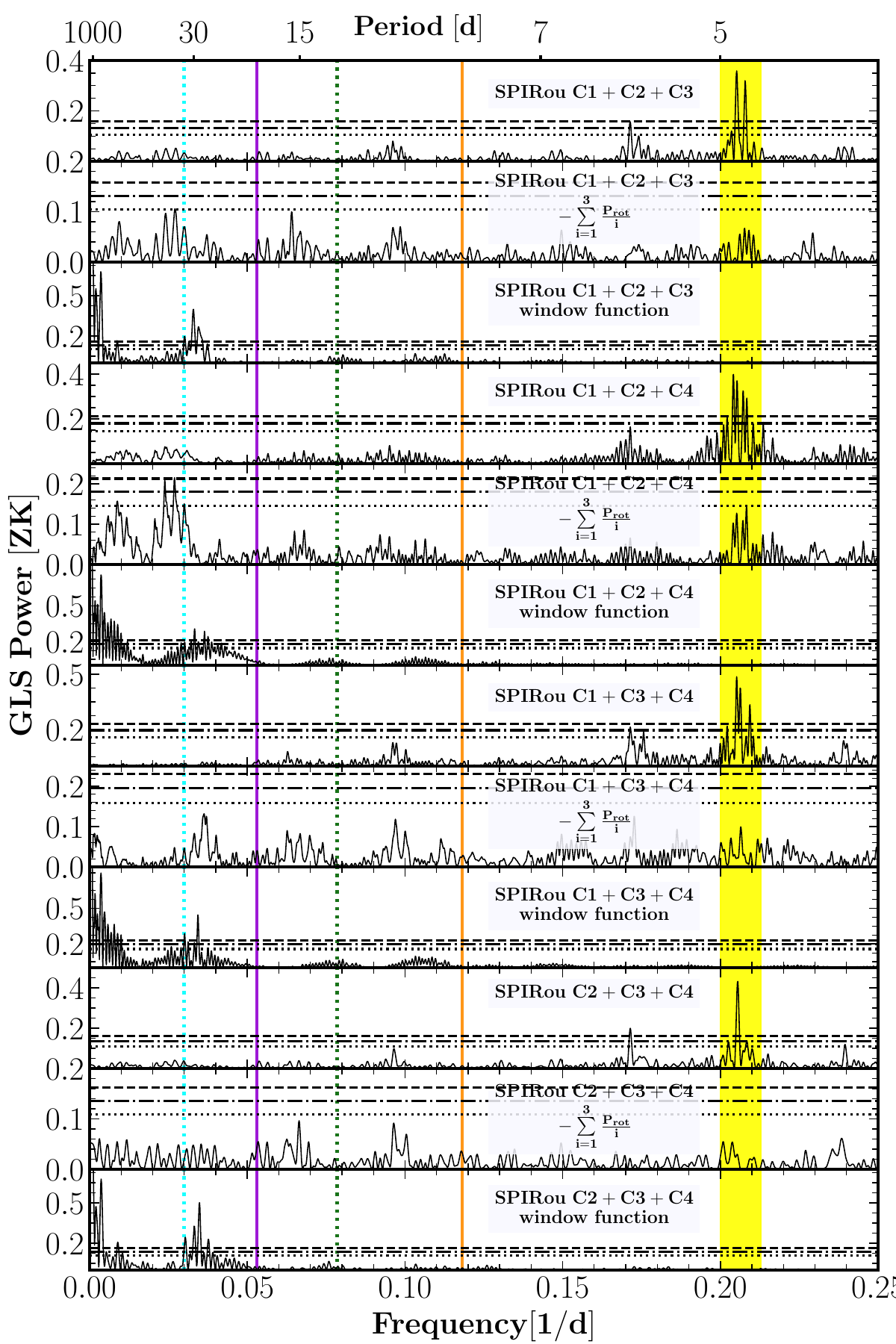}
\caption{GLS periodograms for the SPIRou seasonal datasets. For each dataset, the GLS periodogram, GLS periodogram of the residuals after applying the pre-whitening, and GLS periodogram of the window function are shown. The stellar rotation period and its second and third harmonics are shown as vertical yellow bands, centred at 0.206\,d$^{-1}$ (4.9 d), 0.410\,d$^{-1}$ (2.5 d), and 0.617\,d$^{-1}$ (1.6 d). The vertical orange and purple lines indicate the orbital periods of planets b and c. The vertical green and cyan dotted lines indicate the orbital periods of the candidates d and e, respectively. The dashed horizontal black lines of each panel correspond to the FAP levels of 10\%, 1\%, and 0.1\% (from top to bottom).
\label{fig:GLS_RV_SPIRou}}
\end{center}
\end{figure*}

With the aim of investigating the 33.4 d signal in depth, we performed a seasonal analysis of the SPIRou observations. SPIRou has observed AU\,Mic in campaigns C1, C2, C3 and C4 (see Table\,\ref{tab:camp} for details), covering a total baseline around 1000 d. To study the stability of the signal against the dataset, we chose the data from three SPIRou campaigns and their corresponding permutations (C1 + C2 + C3, C1 + C2 + C4, C1 + C3 + C4, and C2 + C3 + C4). Following the same procedure as in Sect.\,\ref{sec:rv_GLS}, the Fig.\,\ref{fig:GLS_RV_SPIRou} shows the GLS periodograms of the RVs (panels one, four, seven and eleven, from top to bottom), of the RVs after subtracting the stellar activity using pre-whitening (panels two, five, eight and twelve, from top to bottom) and of the window functions (panels three, six, nine and thirteen, from top to bottom). In the panels where the activity has been subtracted, it is observed that the most significant signals correspond to periods of 37.0 (panel two), 37.2 (panel five), 27.5 (panel eight) and 15.1 (panel eleven) d. Therefore, we concluded that the signal may not be of planetary nature.

\section{Chromatic transit analysis}

\begin{figure*}
\begin{center}
\includegraphics[width=0.8\linewidth]{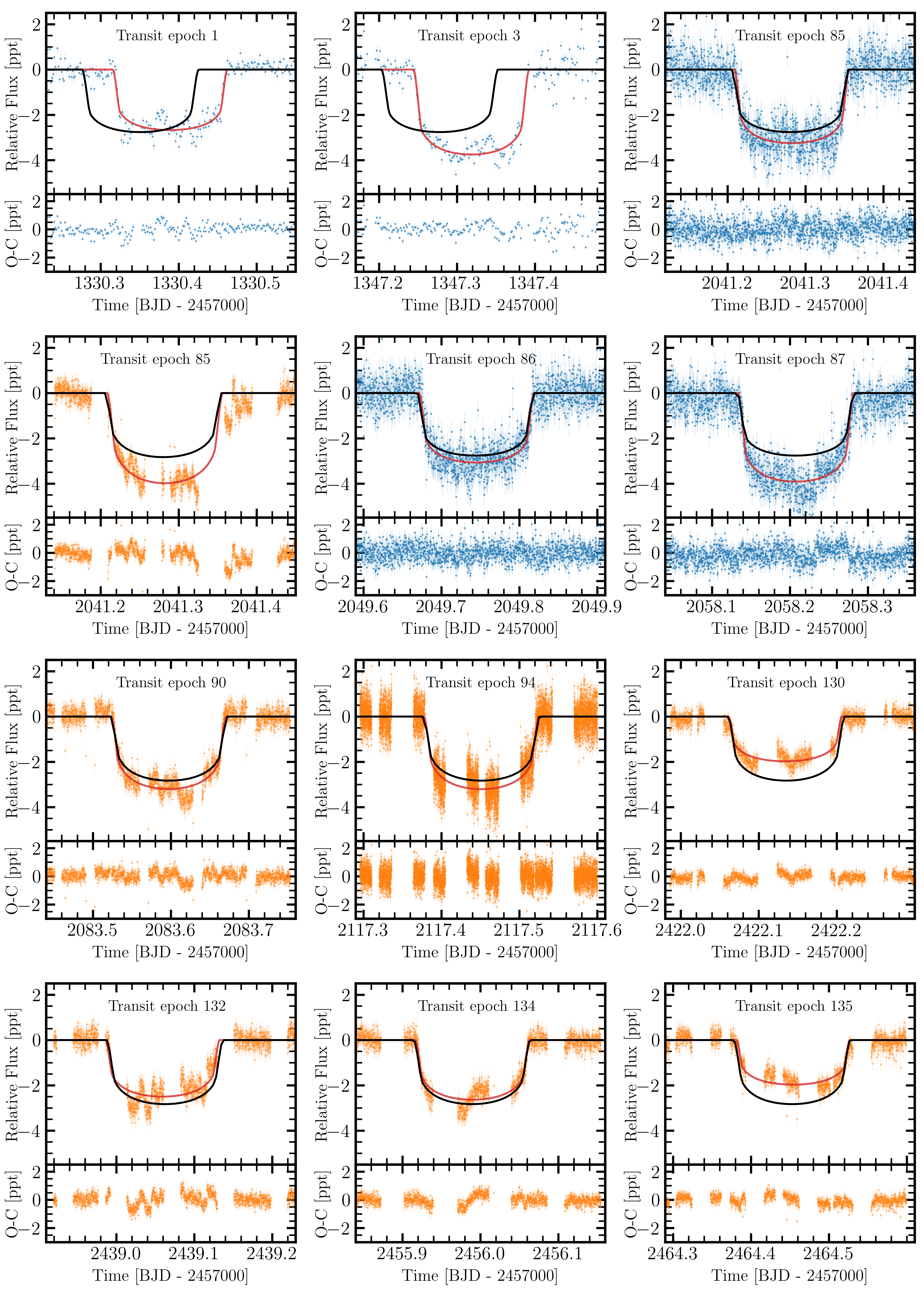}
\caption{Individual transits for AU\,Mic\,b. The blue and orange dots represent the flattened transit photometry from TESS and CHEOPS, respectively. The lines represent the transit model from the joint fit (black line) and from the alternative transit model (red line).
\label{fig:TTVs_b}}
\end{center}
\end{figure*}
\begin{figure*}
\begin{center}
\includegraphics[width=0.8\linewidth]{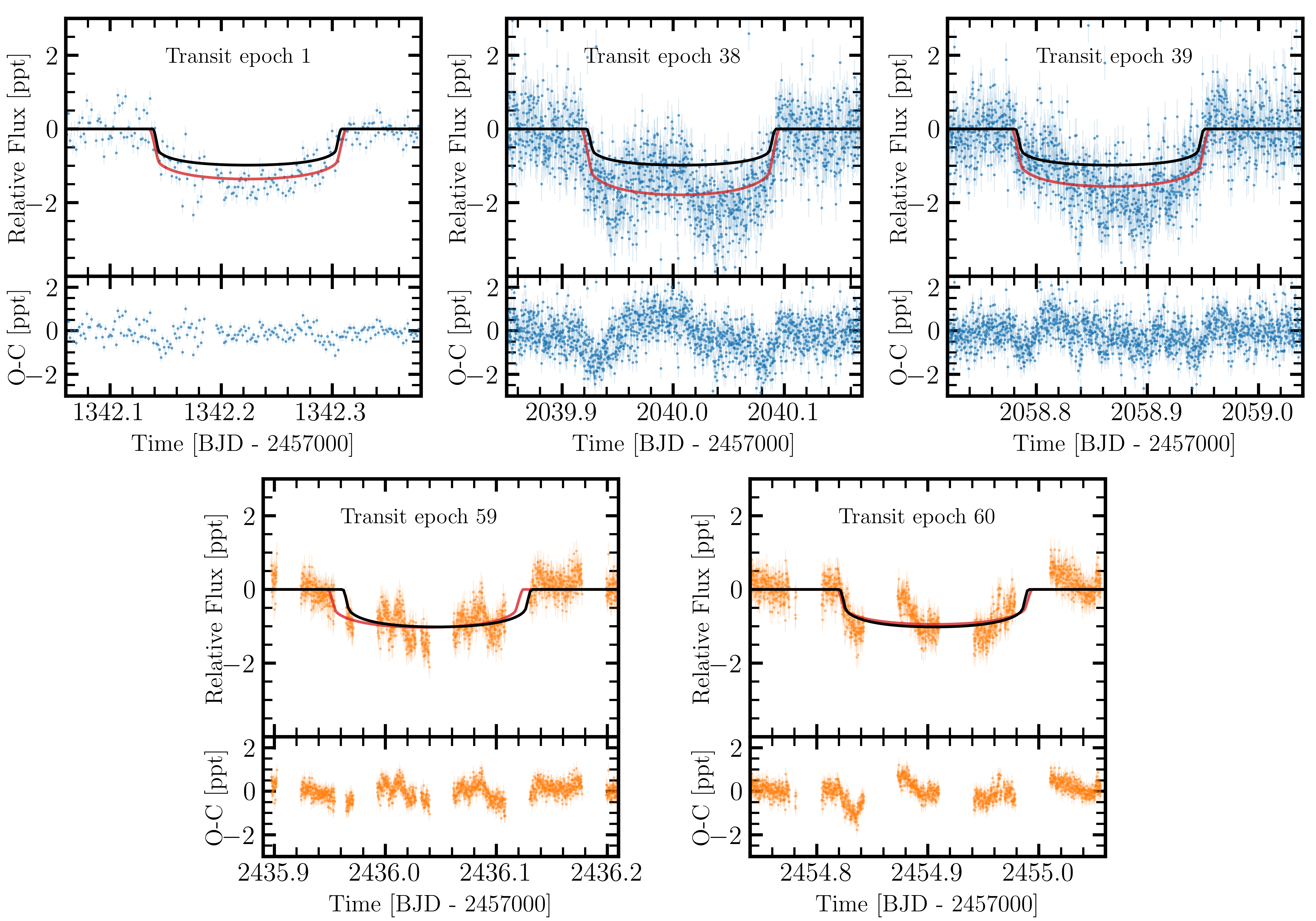}
\caption{Individual transits for AU\,Mic\,c. The blue and orange dots represent the flattened transit photometry from TESS and CHEOPS, respectively. The lines represent the transit model from the joint fit (black line) and from the alternative transit model (red line).
\label{fig:TTVs_c}}
\end{center}
\end{figure*}

\begin{table*}
\begin{center}
\caption{Transit-fit parameters for the alternative transit model for AU\,Mic\,b and c.}\label{tab:tr_depth}
\begin{tabular}{lcc}
\hline\hline
Parameter & Prior & Posterior ($e$\,$=$\,0)\\
\hline
$R_{p}^{b}/R_{\star}$$_{\mathrm{TESS,1}}$ & $\mathcal{U}$(0, 0.1) & 0.0486$^{+0.0005}_{-0.0005}$\\
$R_{p}^{b}/R_{\star}$$_{\mathrm{TESS,3}}$ & $\mathcal{U}$(0, 0.1) & 0.0563$^{+0.0005}_{-0.0005}$\\
$R_{p}^{b}/R_{\star}$$_{\mathrm{TESS,85}}$ & $\mathcal{U}$(0, 0.1) & 0.0540$^{+0.0003}_{-0.0003}$\\
$R_{p}^{b}/R_{\star}$$_{\mathrm{TESS,86}}$ & $\mathcal{U}$(0, 0.1) & 0.0525$^{+0.0003}_{-0.0003}$\\
$R_{p}^{b}/R_{\star}$$_{\mathrm{TESS,87}}$ & $\mathcal{U}$(0, 0.1) & 0.0569$^{+0.0003}_{-0.0003}$\\
$R_{p}^{b}/R_{\star}$$_{\mathrm{CHEOPS,85}}$ & $\mathcal{U}$(0, 0.1) & 0.0588$^{+0.0001}_{-0.0001}$\\
$R_{p}^{b}/R_{\star}$$_{\mathrm{CHEOPS,90}}$ & $\mathcal{U}$(0, 0.1) & 0.0528$^{+0.0001}_{-0.0001}$\\
$R_{p}^{b}/R_{\star}$$_{\mathrm{CHEOPS,94}}$ & $\mathcal{U}$(0, 0.1) & 0.0526$^{+0.0001}_{-0.0001}$\\
$R_{p}^{b}/R_{\star}$$_{\mathrm{CHEOPS,130}}$ & $\mathcal{U}$(0, 0.1) & 0.0406$^{+0.0002}_{-0.0002}$\\
$R_{p}^{b}/R_{\star}$$_{\mathrm{CHEOPS,132}}$ & $\mathcal{U}$(0, 0.1) & 0.0460$^{+0.0002}_{-0.0002}$\\
$R_{p}^{b}/R_{\star}$$_{\mathrm{CHEOPS,134}}$ & $\mathcal{U}$(0, 0.1) & 0.0480$^{+0.0002}_{-0.0002}$\\
$R_{p}^{b}/R_{\star}$$_{\mathrm{CHEOPS,135}}$ & $\mathcal{U}$(0, 0.1) & 0.0415$^{+0.0002}_{-0.0002}$\\
$T_{c,\mathrm{TESS,1}}^{b}$[BJD] & $\mathcal{N}$(2458330.4, 0.1) & 2458330.38900$^{+0.00068}_{-0.00068}$\\
$T_{c,\mathrm{TESS,3}}^{b}$[BJD] & $\mathcal{N}$(2458347.3, 0.1)& 2458347.31777$^{+0.00050}_{-0.00050}$\\
$T_{c,\mathrm{TESS,CHEOPS,85}}^{b}$[BJD] & $\mathcal{N}$(2459041.3, 0.1)& 2459041.28213$^{+0.00022}_{-0.00022}$\\
$T_{c,\mathrm{TESS,86}}^{b}$[BJD] & $\mathcal{N}$(2459049.8, 0.1)& 2459049.74498$^{+0.00030}_{-0.00030}$\\
$T_{c,\mathrm{TESS,87}}^{b}$[BJD] & $\mathcal{N}$(2459058.2, 0.1)& 2459058.19806$^{+0.00055}_{-0.00049}$\\
$T_{c,\mathrm{CHEOPS,90}}^{b}$[BJD] & $\mathcal{N}$(2459083.6, 0.1)& 2459083.59710$^{+0.00014}_{-0.00014}$\\
$T_{c,\mathrm{CHEOPS,94}}^{b}$[BJD] & $\mathcal{N}$(2459117.5, 0.1)& 2459117.45137$^{+0.00018}_{-0.00018}$\\
$T_{c,\mathrm{CHEOPS,130}}^{b}$[BJD] & $\mathcal{N}$(2459422.1, 0.1)& 2459422.13341$^{+0.00034}_{-0.00035}$\\
$T_{c,\mathrm{CHEOPS,132}}^{b}$[BJD] & $\mathcal{N}$(2459439.1, 0.1)& 2459439.05936$^{+0.00023}_{-0.00023}$\\
$T_{c,\mathrm{CHEOPS,134}}^{b}$[BJD] & $\mathcal{N}$(2459456.0, 0.1)& 2459455.98972$^{+0.00017}_{-0.00017}$\\
$T_{c,\mathrm{CHEOPS,135}}^{b}$[BJD] & $\mathcal{N}$(2459464.5, 0.1)& 2459464.45412$^{+0.00025}_{-0.00025}$\\
$P^{b}$[d] & fixed & 8.463446\\
$b^{b}$ & $\mathcal{U}$(0, 1) & 0.45$^{+0.01}_{-0.01}$\\
$R_{p}^{c}/R_{\star\mathrm{TESS,1}}$ & $\mathcal{U}$(0, 0.1) & 0.0367$^{+0.0006}_{-0.0006}$\\
$R_{p}^{c}/R_{\star\mathrm{TESS,38}}$ & $\mathcal{U}$(0, 0.1) & 0.0443$^{+0.0008}_{-0.0009}$\\
$R_{p}^{c}/R_{\star\mathrm{TESS,39}}$ & $\mathcal{U}$(0, 0.1) & 0.0369$^{+0.0005}_{-0.0005}$\\
$R_{p}^{c}/R_{\star\mathrm{CHEOPS,59}}$ & $\mathcal{U}$(0, 0.1) & 0.0315$^{+0.0003}_{-0.0003}$\\
$R_{p}^{c}/R_{\star\mathrm{CHEOPS,60}}$ & $\mathcal{U}$(0, 0.1) & 0.0302$^{+0.0003}_{-0.0003}$\\
$T_{c,\mathrm{TESS,1}}^{c}$[BJD] & $\mathcal{N}$(2458342.2, 0.1)& 2458342.22052$^{+0.00173}_{-0.00143}$\\
$T_{c,\mathrm{TESS,38}}^{c}$[BJD] & $\mathcal{N}$(2459040.0, 0.1)& 2459040.00385$^{+0.00078}_{-0.00082}$\\
$T_{c,\mathrm{TESS,39}}^{c}$[BJD] & $\mathcal{N}$(2459058.9, 0.1)& 2459058.87122$^{+0.00099}_{-0.00102}$\\
$T_{c,\mathrm{CHEOPS,59}}^{c}$[BJD] & $\mathcal{N}$(2459436.0, 0.1)& 2459436.03801$^{+0.00100}_{-0.00122}$\\
$T_{c,\mathrm{CHEOPS,60}}^{c}$[BJD] & $\mathcal{N}$(2459454.9, 0.1)& 2459454.89591$^{+0.00048}_{-0.00047}$\\
$P^{c}$[d] & fixed & 18.859023 \\
$b^{c}$ & $\mathcal{U}$(0, 1) & 0.65$^{+0.01}_{-0.01}$\\
\hline
$\gamma_{\mathrm{TESS}}$[ppt] & $\mathcal{U}$(-3$\sigma_{\mathrm{TESS}}$, 3$\sigma_{\mathrm{TESS}}$) & 0.01$^{+0.01}_{-0.01}$\\
$\sigma_{\mathrm{jit,TESS,120}}$[ppt] & $\mathcal{U}$(0, 3$\sigma_{\mathrm{TESS,120}}$) & 0.39$^{+0.01}_{-0.01}$\\
$\sigma_{\mathrm{jit,TESS,20}}$[ppt] & $\mathcal{U}$(0, 3$\sigma_{\mathrm{TESS,30}}$) & 0.52$^{+0.01}_{-0.01}$\\
$q_{1,\mathrm{TESS}}$ & $\mathcal{N}$(0.29, 0.01) & 0.30$^{+0.01}_{-0.01}$\\
$q_{2,\mathrm{TESS}}$ & $\mathcal{N}$(0.31, 0.01) & 0.30$^{+0.01}_{-0.01}$\\
$\gamma_{\mathrm{CHEOPS}}$[ppt] & $\mathcal{U}$(-3$\sigma_{\mathrm{CHEOPS}}$, 3$\sigma_{\mathrm{CHEOPS}}$) & --0.03$^{+0.01}_{-0.01}$\\
$\sigma_{\mathrm{jit,CHEOPS,15}}$[ppt] & $\mathcal{U}$(0, 3$\sigma_{\mathrm{CHEOPS,15}}$) &0.19$^{+0.01}_{-0.01}$\\
$\sigma_{\mathrm{jit,CHEOPS,3}}$[ppt] & $\mathcal{U}$(0, 3$\sigma_{\mathrm{CHEOPS,3}}$) &0.52$^{+0.01}_{-0.01}$\\
$q_{1,\mathrm{CHEOPS}}$ & $\mathcal{N}$(0.42, 0.01) & 0.42$^{+0.01}_{-0.01}$\\
$q_{2,\mathrm{CHEOPS}}$ & $\mathcal{N}$(0.33, 0.01) & 0.32$^{+0.01}_{-0.01}$\\
\hline
\end{tabular}
\end{center}
\tablefoot{The prior label of $\mathcal{N}$ and $\mathcal{U}$ represent normal and uniform distributions, respectively. In the planetary parameters, the subscripts represent the instrument and the epoch of the transit. In the $\sigma_{\mathrm{jit}}$ parameters, the subscripts represent the instrument and the exposure time of the data.}\\

\end{table*}

\section{Radial velocity data}

\onecolumn

\begin{table*}
\begin{center}
\caption{RV data of CARMENES VIS.}\label{tab:rv_CARMV}
\begin{tabular}{crr}
\hline\hline
Time & \multicolumn{1}{c}{RV} & \multicolumn{1}{c}{$\sigma$} \\ 

[BJD] & [m\,s$^{-1}$] & [m\,s$^{-1}$] \\ 
\hline
2458678.5657 & --165.72 & 14.66\\ 
2458678.5706 & --173.70 & 12.94\\ 
2458684.5660 & 124.66 & 11.80\\ 
2458684.5707 & 98.79 & 9.47\\ 
2458686.5460 & 103.43 & 6.72\\ 
2458686.5508 & 104.53 & 8.05\\ 
2458687.5767 & --92.52 & 10.22\\ 
2458687.5812 & --105.12 & 10.89\\ 
2458688.5772 & --170.57 & 11.62\\ 
2458688.5915 & --163.92 & 8.34\\ 
... & ... & ... \\
\hline
\end{tabular}
\end{center}
\tablefoot{Only the first 10 measurements are shown. The complete table can be found in the online version of the manuscript and in the CDS.}
\end{table*}

\begin{table*}
\begin{center}
\caption{RV data of CARMENES NIR.}\label{tab:rv_CARMN}
\begin{tabular}{crr}
\hline\hline
Time & \multicolumn{1}{c}{RV} & \multicolumn{1}{c}{$\sigma$} \\ 

[BJD] & [m\,s$^{-1}$] & [m\,s$^{-1}$] \\ 
\hline
2458678.5655 & --159.99 & 20.09\\ 
2458678.5703 & --152.17 & 20.90\\ 
2458684.5661 & 66.53 & 37.85\\ 
2458684.5706 & 67.62 & 30.44\\ 
2458686.5461 & 61.48 & 20.78\\ 
2458686.5510 & 34.75 & 28.11\\ 
2458687.5759 & --41.17 & 22.37\\ 
2458687.5809 & --79.91 & 26.76\\ 
2458688.5774 & --162.83 & 31.13\\ 
2458688.5914 & --216.65 & 28.35\\ 
... & ... & ... \\
\hline
\end{tabular}
\end{center}
\tablefoot{Only the first 10 measurements are shown. The complete table can be found in the online version of the manuscript and in the CDS.}
\end{table*}

\begin{table*}
\begin{center}
\caption{RV data of HARPS.}\label{tab:rv_HARPS}
\begin{tabular}{crr}
\hline\hline
Time & \multicolumn{1}{c}{RV} & \multicolumn{1}{c}{$\sigma$} \\ 

[BJD] & [m\,s$^{-1}$] & [m\,s$^{-1}$] \\ 
\hline
2459168.6006 & 177.72 & 4.79\\ 
2459170.5810 & 4.51 & 3.26\\ 
2459171.5437 & 142.80 & 4.57\\ 
2459172.5353 & --145.72 & 2.62\\ 
2459175.5082 & 11.01 & 2.89\\ 
2459175.5630 & 17.11 & 3.95\\ 
2459176.5088 & 140.86 & 4.31\\ 
2459177.5126 & --173.93 & 2.90\\ 
2459178.5080 & 165.65 & 4.37\\ 
2459178.5192 & 161.28 & 3.82\\ 
... & ... & ... \\
\hline
\end{tabular}
\end{center}
\tablefoot{Only the first 10 measurements are shown. The complete table can be found in the online version of the manuscript and in the CDS.}
\end{table*}

\begin{table*}
\begin{center}
\caption{RV data of SPIRou.}\label{tab:rv_SPIRou}
\begin{tabular}{crr}
\hline\hline
Time & \multicolumn{1}{c}{RV} & \multicolumn{1}{c}{$\sigma$} \\ 

[BJD] & [m\,s$^{-1}$] & [m\,s$^{-1}$] \\ 
\hline
2458744.8220 & 51.25 & 4.30\\ 
2458750.7550 & --13.15 & 5.00\\ 
2458751.7461 & --61.25 & 4.20\\ 
2458752.7906 & 12.15 & 4.30\\ 
2458758.7296 & 20.25 & 3.70\\ 
2458759.8061 & 34.15 & 3.70\\ 
2458760.7286 & --4.05 & 3.80\\ 
2458761.7313 & --80.85 & 4.30\\ 
2458762.7323 & --4.45 & 4.00\\ 
2458764.7579 & 42.15 & 4.50\\ 
... & ... & ... \\
\hline
\end{tabular}
\end{center}
\tablefoot{Only the first 10 measurements are shown. The complete table can be found in the online version of the manuscript and in the CDS.}
\end{table*}

\end{appendix}

\end{document}